\documentclass[journal,11pt,draftclsnofoot,onecolumn]{IEEEtran}

\usepackage{latexsym}
\usepackage{amsfonts}
\usepackage{graphicx} 
\usepackage{subfigure}

\usepackage{amssymb}
\usepackage{natbib}
\usepackage{enumerate}
\usepackage{amsmath}

\usepackage[usenames]{color}

\usepackage{supertabular}

\usepackage{enumerate}
\usepackage{verbatim}

\newtheorem{Teorema}{\bf Theorem}

\newtheorem{Corolario}{\bf Corollary}
\newtheorem{Proposicion}{\bf Proposition}
\newtheorem{Demo}{\bf Proof}
\newtheorem{Ex}{\bf Example}

\newcommand{\dfn}{\triangleq}

\newcommand{\Prob}{\mbox{Prob}}

\def\BibTeX{{\rm B\kern-.05em{\sc i\kern-.025em b}\kern-.08em
    T\kern-.1667em\lower.7ex\hbox{E}\kern-.125emX}}

\begin{document}

%
\title{On the Generalized Ratio of Uniforms as a Combination of Transformed Rejection and Extended Inverse of Density Sampling}
\author{Luca Martino$^\dagger$, David Luengo$^\ddagger$, Joaqu\'{\i}n M\'{\i}guez$^\dagger$\\
$^\dagger$Department of Signal Theory and Communications, Universidad Carlos III de Madrid.\\
Avenida de la Universidad 30, 28911 Legan\'es, Madrid, Spain.\\
$^\ddagger$Department of Circuits and Systems Engineering, Universidad Polit\'ecnica de Madrid.\\
Carretera de Valencia Km. 7, 28031 Madrid, Spain.\\
E-mail: {\tt luca@tsc.uc3m.es, david.luengo@upm.es, jmiguez@tsc.uc3m.es}}
\maketitle 

\vspace{-1.5cm}

\begin{abstract}
In this work we investigate the relationship among three classical sampling techniques: the inverse of density (Khintchine's theorem), the transformed rejection (TR) and the generalized ratio of uniforms (GRoU).
Given a monotonic probability density function (PDF), we show that the transformed area obtained using the generalized ratio of uniforms method can be found equivalently by applying the transformed rejection sampling approach to the inverse function of the target density.
Then we provide an extension of the classical inverse of density idea, showing that it is completely equivalent to the GRoU method for monotonic densities.
Although we concentrate on monotonic probability density functions (PDFs), we also discuss how the results presented here can be extended to any non-monotonic PDF that can be decomposed into a collection of intervals where it is monotonically increasing or decreasing. In this general case, we show the connections with transformations of certain random variables and the generalized inverse PDF with the GRoU technique. Finally, we also introduce a GRoU technique to handle unbounded target densities.
\end{abstract}

\begin{keywords}
Transformed rejection sampling; inverse of density method; Khintchine's theorem; generalized ratio of uniforms technique; vertical density representation.
\end{keywords}

\IEEEpeerreviewmaketitle

\section{Introduction}
\label{sec:introduction}

Monte Carlo (MC) methods are often used for the implementation of optimal Bayesian estimators in many practical applications, ranging from statistical physics \citep{Rosenbluth55,Siepmann92} to nuclear medicine \citep{Ljungberg98} and statistical signal processing \citep{Djuric03,MartinoSigPro10,Fitzgerald96}. 
Many Monte Carlo techniques have been proposed for solving this kind of problems either sequentially (SMC methods, also known as particle filters), making use of Markov chains (MCMC methods) or otherwise \citep{Fitzgerald01,Gilks95bo,Fitzgerald96}.
Sampling techniques (see e.g. \citep{Devroye86, Gentle04, Hormann03} for a review) are the core of Monte Carlo simulations, since all of them rely on the efficient generation of samples from some proposal PDF \citep{Liu04b, Robert04}.
Many sampling algorithms have been proposed, but the problem of drawing samples efficiently from a generic distribution is far from trivial and many open questions still remain.

In this paper we investigate the relationship between three classical sampling techniques: 
\begin{itemize}
\item the {\em inverse-of-density} technique \citep{Devroye86,Khintchine38},
\item  the {\em transformed rejection sampling} \citep{VonNeumann51,Wallace76}, also known as ``almost exact inversion''  \citep[Chapter 3]{Devroye86} and ``exact approximation method'' \citep{Marsaglia84}, 
\item and  the {\em ratio-of-uniforms} method \citep{Kinderman77,Wakefield91}.
\end{itemize}
We present new connections among them, useful to design more efficient sampling techniques. 
Although in the sequel we concentrate mainly on monotonic PDFs, we also discuss the relationships among these techniques in more general cases, especially in the last two sections and the Appendix.

The first method considered, the \emph{inverse-of-density} (IoD) technique \citep[Chapter 4]{Devroye86}, \citep{Devroye84,Isii58,Jones02,Khintchine38} (often known as \emph{Khintchine's theorem} \citep[pp. 157-159]{Feller71}, \citep{Bryson82,Chaubey10,Jones02,Khintchine38,Olshen70,Shepp62}, both for monotonic PDFs and for symmetric unimodal PDFs), is a classical sampling technique.
Given a monotonic target PDF, $p_0(x) = K p(x)$ (also denoted often as $p_0(x) \propto p(x)$, omitting the normalization constant, $K>0$), this method provides a closed-form relationship between the samples from the PDF defined by the unnormalized inverse density, $p^{-1}(y)$, and the desired samples, distributed according to the normalized PDF, $p_0(x)$.
Hence, if we are able to draw samples easily from $p^{-1}(y)$, then it is straightforward to generate samples from the target PDF by using the IoD approach.
Clearly, the practical applicability of the IoD method depends on the feasibility of drawing samples from the inverse density $p^{-1}(y)$.

The IoD method can be easily extended to non-monotonic densities (see e.g. \citep{Devroye86,Jones02}) both unidimensional and multidimensional \citep{Bryson82,DeSilva78}. Moreover, the IoD presents several relationships (see e.g. \citep{Jones02}) with  {\it vertical density representation} (VDR) \citep{Fang01,Kotz97,Troutt96,KOZUBOWSKI02,Troutt91,Troutt93,Troutt04}, especially with the so-called {\it second type} VDR  \citep{Fang01} , \citep[Chapter 3]{Troutt04}.

The second tackled method is \emph{transformed rejection sampling} (TRS) \citep{Devroye86,Marsaglia84,Wallace76}. The rejection sampling (RS) is another standard Monte Carlo technique that use a a simpler proposal distribution,$\pi(x)$ to generate samples and, then, to accept or discard them according to a ratio between the target and proposal densities $\frac{p(x)}{L\pi(x)}$ (where $L\pi(x)\geq p(x)$).
Hence, the fundamental figure of merit of a rejection sampler is the \emph{mean acceptance rate} (i.e. the expected number of accepted samples out of the total number of proposed candidates).

The most favorable scenario for using the RS algorithm occurs when the unnormalized target PDF, $p(x)$, is bounded with bounded domain.
In this case, the proposal PDF $\pi(x)$ can be a uniform density (the easiest possible proposal), and calculating the bound $L$ for the ratio $p(x)/\pi(x)$ is equivalent to finding an upper bound for the unnormalized target PDF, $p(x)$, which is in general a much easier task \citep{Devroye86,Hormann03}.
Indeed, in this scenario several sophisticated and efficient acceptance/rejection methods that achieve a high acceptance rate have been devised: adaptive schemes \citep{Gilks92,MartinoStatCo10}, strips techniques \citep[Chapter 5]{Hormann03}, \citep{Ahrens93,Ahrens95, Marsaglia00}, patchwork algorithms \citep{Kemp90,Stadlober99}, etc.

However, in general the target $p(x)$ can be  unbounded or with an infinite support and the choice of a good proposal PDF becomes more critical (see, for instance \cite{MartinoStatCo10-2}).   
In order to overcome this problem, different methods have been proposed to transform the region corresponding to the area below $p(x)$ into an alternative bounded region.
A straightforward solution from a theoretical point of view is the transformed rejection sampling (TRS) \citep{Botts11,Hormann93,Hormann94,Marsaglia84,Wallace76}, which is based on finding a suitable invertible transformation, $f(x): \mathcal{D}_X \rightarrow  \mathcal{D}_Z$, such that the region below $p(x)$ is transformed into an appropriate bounded set. 
Making use of this transformation we can define a random variable (RV) $Z = f(X)$, with unnormalized PDF $\rho(z) = p(f^{-1}(z)) \times |\dot{f}^{-1}(z)|$ and $\dot{f}^{-1}(z)$ denoting the derivative of $f^{-1}(z)$, draw samples $\{z^{(1)}, \ldots, z^{(N)}\}$ from $\rho(z)$, and convert them into samples from the target PDF, $\{x^{(1)},\ \ldots,\ x^{(N)}\} = \{f^{-1}(z^{(1)}),\ \ldots,\ f^{-1}(z^{(N)})\}$ by inverting the transformation.

Obviously, attaining a {\em bounded} PDF $\rho(z)$ requires imposing some restrictions on the transformation $f(x)$ that depend on the unnormalized target PDF, $p(x)$ \citep{Hormann94,Wallace76}.
Furthermore, the performance of the TRS approach depends critically on a suitable choice of the transformation function, $f(x)$.
Indeed, if  $f(x)$ is chosen to be similar to the cumulative distribution function (CDF) of the target RV, $F_X(x)$, the PDF of the transformed RV, $\rho(z)$, becomes flatter and closer to a uniform PDF and higher acceptance rates can be achieved.
In particular, if $f(x) = F_X(x)$, then $\rho(z)$ is the uniform density in $[0,1]$, implying that we can easily draw samples from it without any rejection and justifying the fact that this technique is sometimes also called \emph{almost exact inversion method} \citep{Devroye86}.

Another approach to work with bounded region is the so-called \emph{ratio-of-uniforms} (RoU) technique \citep{Devroye86, Kinderman77} (the third technique that we address here).
The RoU ensures that, given a pair or independent RVs, $(V,U)$, uniformly distributed inside $\mathcal{A}_r = \{(v,u) \in \mathbb{R}^2: 0 \le u \le \sqrt{p(v/u)}\}$, then $x=v/u$ is distributed exactly according to the target PDF, $p_0(x)$.
Hence, in the cases of interest (i.e. when the region $\mathcal{A}_r$ is bounded) the RoU provides us with a bidimensional region, $\mathcal{A}_r$, such that drawing samples from the univariate target density is equivalent to drawing samples uniformly inside $\mathcal{A}_r$, which can be done efficiently by means of rejection sampling schemes  \citep{MartinoLuengoLetter1_12,Leydold00,Leydold03,Perez08}.
Unfortunately the region $\mathcal{A}_r$ provided by RoU is only bounded when the tails of the target density decay faster than $1/x^2$, which is not always fulfilled for the PDFs of interest.

Consequently, several generalizations of the RoU method have been proposed in the literature (see e.g. \citep{Jones96,Wakefield91} and more related materials that can be found in \citep{Barbu82,Curtiss41,Dieter89,Marsaglia65,Perez08,Stefanescu87,Vaduva82}).
The most popular of those extensions is the so called \emph{generalized ratio-of-uniforms} (GRoU) \citep{Wakefield91}, which shows that $x=v/\dot{g}(u)$ is distributed according to the target PDF, $p_0(x)$, when the random vector $(V,U)$ is uniformly distributed inside the region $\mathcal{A}_g = \{(v,u) \in \mathbb{R}^2: 0 \le u \le g^{-1}(c p(v/\dot{g}(u)))\}$, with $c > 0$ being a constant term and $g(u)$ a strictly increasing differentiable function on $\mathbb{R}^+$ such that $g(0)=0$.

These two techniques (TRS and GRoU) have been introduced separately in the literature and their connection has not been explored as far as we know.
The primary goal of this paper is showing that there is a close relationship between both approaches.
Indeed, one of the main results in this work is proving that the transformed region attained using the GRoU technique \citep{Wakefield91} can also be obtained applying the transformed rejection approach \citep{Wallace76} to the unnormalized inverse PDF, $p^{-1}(y)$, for monotonic target PDFs, $p_0(x) \propto p(x)$.
Moreover, we introduce an {\em extended} version of the standard inverse-of-density method \citep[Chapter 4]{Devroye86}, \citep{Jones02,Khintchine38}, which is strictly related to the GRoU method and show that the GRoU sampling technique coincides with this extended version of the inverse-of-density method.  

Considering a monotonic unnormalized target PDF, $p(x)$, in this work we show that the region $\mathcal{A}_g$ defined by the GRoU can be obtained transforming an RV $Y$ with unnormalized inverse PDF $p^{-1}(y)$, and that the relationship between the points in this region $\mathcal{A}_g$ and the samples drawn from $p(x)$ is provided by the novel extended version of the IoD method, introduced here.
Hence, as a conclusion we can assert that, for monotonic PDFs, the GRoU can be seen as a combination of the transformed rejection sampling method applied to the unnormalized inverse PDF, $p^{-1}(y)$, and an extended inverse-of-density technique.
%
%
%
%
%
%
We also investigate the connections among TRS, IoD and GRoU for generic non-monotonic target PDFs.
%
%
Finally, taking advantage of the previous considerations we introduce a GRoU technique to handle unbounded target distributions.

The rest of the paper is organized as follows.
In Section \ref{sec:background} we provide some important considerations about the notation, we formulate the fundamental theorem of simulation (which is the basis for all the sampling methods discussed), and briefly describe the standard inverse-of-density and rejection sampling techniques, thus providing the background for the rest of the paper.
Then, Sections \ref{sec:TR} and \ref{sec:GRoU} provide a detailed description of the two sampling methods compared, transformed rejection sampling and the generalized ratio-of-uniforms respectively, focusing on the different possible situations that may be found and particularly on the conditions required for obtaining finite sampling regions.
This is followed by Section \ref{sec:ExtIoD}, where we introduce an extension of the inverse-of-density method, and Section \ref{sec:RoUvsTRS}, which provides the main result of the paper: the relationship between the ratio-of-uniforms, transformed rejection and the inverse-of-density methods.
Section \ref{sec:further} provides some further considerations about the different approaches considered, whereas Section \ref{sec:GenPdfs} discusses their extension to non-monotonic PDFs. Section \ref{sec:GRoUforUnbounded} is devoted to design a GRoU for unbounded distributions, using the previous considerations and observations.
Finally, the conclusions and the appendix close the paper.

\section{Background}
\label{sec:background}

\subsection{Important consideration about the notation}

In the sequel we always work with proper but unnormalized PDFs, meaning that integrating them over their whole domain results in a finite positive constant, but not necessarily equal to one.
As an example, consider the normalized target PDF, $p_0(x) = K p(x)$, with $K > 0$ denoting the normalization constant.
All the subsequent methods will be formulated in terms of $p(x)$, which is the unnormalized target PDF, since
\begin{equation}
	\int_{\mathcal{D}_X}p(x) dx = \frac{1}{K},
\end{equation}
with $K > 0$, but $K \neq 1$ in general. Hence, the integral is finite but not necessarily equal to one.

Furthermore, in order to get rid of the normalization constant, we will also work with the unnormalized inverse target PDF, $p^{-1}(y)$, for monotonic target PDFs or its generalized version, $p_G^{-1}(y)$, for non-monotonic PDFs.
Note that the normalized inverse target PDF, $p_0^{-1}(y)$, can no longer be obtained from the unnormalized inverse target PDF, $p^{-1}(y)$, simply multiplying by a normalization constant.
A scaling of the independent variable, $y$, must be performed instead in order to attain $p_0^{-1}(y) = p^{-1}(y/K)$.
We remark also that $Kp^{-1}(y) \ne p_0^{-1}(y) = p^{-1}(y/K)$, i.e. the normalized version of the unnormalized inverse target PDF will be different from the normalized inverse target PDF in general.
This is due to the fact that the support of $p_0^{-1}(y)$ will usually be different from the support of $p^{-1}(y)$, due to the scaling of the independent variable, $y$, performed on $p^{-1}(y)$ in order to obtain $p_0^{-1}(y)$.
Finally, note also that, given a sample $y'$ from $Kp^{-1}(y)$ we can easily obtain samples from the normalized inverse target RV, $p_0^{-1}(y)$, since $y'/K \sim p_0^{-1}(y)$.
All these issues are clearly illustrated in the following example.

\begin{Ex}
Consider a half Gaussian random variable with the following PDF:
\begin{equation}
	p_0(x) = \sqrt{\frac{2}{\pi\sigma^2}} \exp\left(-\frac{x^2}{2\sigma^2}\right),
\label{eq:halfGaussian}
\end{equation}
for $0 \le x < \infty$.
The half Gaussian PDF given by \eqref{eq:halfGaussian} is bounded, $0 \le p_0(x) \le \sqrt{2/(\pi\sigma^2)}$, with bounded support, $\mathcal{D}_X = \mathbb{R}^{+} = [0,\ \infty)$, and we can easily identify the corresponding unnormalized PDF,
\begin{equation}
	p(x) = \exp\left(-\frac{x^2}{2\sigma^2}\right),
\label{eq:halfGaussianUnnormalized}
\end{equation}
for $0 \le x < \infty$, and the normalization constant,
\begin{equation}
	K = \sqrt{\frac{2}{\pi\sigma^2}}.
\label{eq:halfGaussianNormalization}
\end{equation}
The functional inverse of \eqref{eq:halfGaussian} (i.e. the normalized inverse taget PDF) is given by
\begin{equation}
	p_0^{-1}(y) = \sqrt{\sigma^2 \log\frac{2}{\pi \sigma^2 y^2}},
\label{eq:inverseHalfGaussianNormalized}
\end{equation}
for $0 < y \le \sqrt{2/(\pi\sigma^2)}$, and $\log$ indicating the natural logarithm.
Note that \eqref{eq:inverseHalfGaussianNormalized} defines a proper normalized PDF, since $p_0^{-1}(y) \ge 0$ for any value of $y$ and
\begin{equation}
	\int_{0}^{\sqrt{2/(\pi\sigma^2)}}{p_0^{-1}(y) \textrm{d}y} = 1.
\end{equation}
Furthermore, it is an unbounded PDF, since $\lim_{y \to 0}\ p_0^{-1}(y) = \infty$, but with a bounded support, $\mathcal{D}_Y = (0,\ \sqrt{2/(\pi\sigma^2)}]$.
Similarly, the unnormalized inverse target PDF is given by
\begin{equation}
	p^{-1}(y) = \sqrt{-2\sigma^2 \textrm{log}\ y},
\label{eq:inverseHalfGaussianUnnormalized}
\end{equation}
for $0 < y \le 1$, which is also a proper unbounded PDF with a bounded support, $\mathcal{D}_Y = (0,\ 1]$.
We note that the normalized version of \eqref{eq:inverseHalfGaussianUnnormalized}, $Kp^{-1}(y) = 2\sqrt{-\frac{1}{\pi} \textrm{log}\ y}$ for $0 < y \le 1$, is clearly different from the normalized inverse target PDF, $p_0^{-1}(y) = p^{-1}(y/K)$ for $0 < y \le \sqrt{2/(\pi\sigma^2)}$, given by \eqref{eq:inverseHalfGaussianNormalized}.
Finally, we also notice that, given $y' \sim Kp^{-1}(y)$, then $y'/K = y' \sqrt{\pi\sigma^2/2}$ is distributed as $p^{-1}(y/K) = p_0^{-1}(y)$, as discussed before.
\end{Ex}

In order to conclude this section, it is important to remark that all the discussions and algorithms shown below do not require the knowledge of the normalization constant.
Hence, we can work with unnormalized PDFs without any loss of generality, since all the results attained in the sequel can also be formulated using normalized PDFs, although the notation becomes more cumbersome.
This is a standard approach, followed not only by the RS and TRS methods, but also by most other standard sampling algorithms, like the RoU or the GROU.
Therefore, in the sequel we will use $X \sim p(x)$ and $Y \sim p^{-1}(y)$ to indicate that the PDFs of the RVs $X$ and $Y$ are proportional to the unnormalized target and inverse target RVs, even though $p(x)$ and $p^{-1}(y)$ will not be normalized in general.
See the Appendix for a detailed revision of the notation used throughout the paper.

\subsection{Fundamental theorem of simulation}
\label{sec:FToS}

Many Monte Carlo techniques (inverse of density method, rejection sampling, slice sampling, etc.) are based on a simple result, known as the fundamental theorem of simulation, that we enunciate in the sequel.

\vspace*{6pt}
\begin{Teorema}
Drawing samples from a unidimensional random variable $X$ with probability density function $p_0(x) = K p(x)$, where $K > 0$ is a constant, is equivalent to sampling uniformly inside the bidimensional region
\begin{equation}
	\mathcal{A}_0 = \{(x,y)\in \mathbb{R}^2: \ 0 \leq y \leq p(x)\}.
\label{eq:A0}
\end{equation}
\label{theo:sampling}
\end{Teorema}
\begin{Demo}
	Straightforward. See \citep[Chapter 2]{Robert04}.
\end{Demo}
\vspace*{6pt}
                        


\begin{figure}[htb]
\centering 
  \includegraphics[width=6cm]{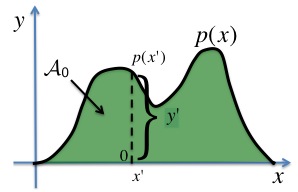}
  \caption{The region $\mathcal{A}_0$ corresponding to the area below the unnormalized target PDF, $p(x)$.}
\label{fig:A0}
\end{figure}

Hence, according to Theorem \ref{theo:sampling}, if the pair of random variables $(X, Y)$ is uniformly distributed inside the region $\mathcal{A}_0$, which corresponds to the area below $p(x)$, then the PDF of $X$ is proportional to $p(x)$, whereas the random variable $Y$ plays the role of an auxiliary variable.
Many Monte Carlo techniques make use of this theorem explicitly to simulate jointly the random variables $(X,Y)$, discarding $Y$ and considering only $X$, which is a univariate random variable marginally distributed according to the unnormalized target PDF, $p(x)$ \citep{Robert04}.
Figure \ref{fig:A0} depicts an example of an unnormalized target PDF, $p(x)$, and the region $\mathcal{A}_0$ delimited by it.
The two methods described in the sequel, inverse of density and rejection sampling, are clear examples of how this simple idea can be applied in practice to design Monte Carlo sampling algorithms. 

\subsection{Inverse of density method for monotone PDFs}
\label{sec:IoD}

In this section we present the inverse of density (IoD) method \citep[Chapter 4]{Devroye86}, \citep{Jones02}, often known as Khintchine's theorem  \citep[pp. 157-159]{Feller71}, \citep{Bryson82,DeSilva78,Isii58,Khintchine38,Olshen70}, both for monotonic and for symmetric unimodal densities \citep{Chaubey10,Shepp62}.
Note once more that, although we concentrate here on monotonic PDFs, this result can be easily extended to generic PDFs, as tackled in Section \ref{sec:GenericINVofDEN} and also shown in \citep{Devroye86}.
The standard formulation for the IoD method is the following.
Let us consider a monotonic unnormalized target PDF, $p(x)$, and denote by $p^{-1}(y)$ the corresponding inverse function of the unnormalized target density.

\begin{figure}[htb]
	\centering 
  \subfigure[]{\label{fig:InvPdf1}\includegraphics[width=6.2cm]{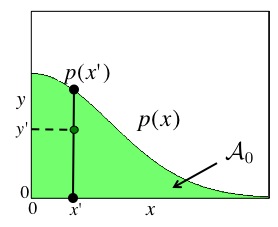}}
  \subfigure[]{\label{fig:InvPdf2}\includegraphics[width=6.2cm]{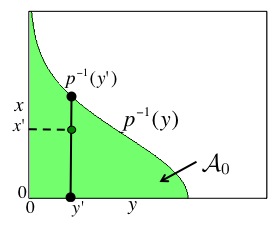}}
  \caption{Two ways of drawing a random point $(x',y')$ uniformly inside the region $\mathcal{A}_0$.
  	\textbf{(a)} Draw first $x'$ from $p(x)$ and then $y'\sim \mathcal{U}([0,p(x')])$. 
  	\textbf{(b)} Draw first $y'$ from $p^{-1}(y)$ and then $x'\sim \mathcal{U}([0,p^{-1}(y')])$.}
\label{fig:InvPdf}
\end{figure}

The fundamental idea underlying the IoD approach is noticing that $p^{-1}(y)$ can also be used to describe $\mathcal{A}_0$, as illustrated graphically in Figure \ref{fig:InvPdf}.
Consequently, the region associated to $p(x)$,
\begin{equation}
	\mathcal{A}_0 = \{(x,y)\in \mathbb{R}^2: \  0\leq y \leq p(x)\},
\label{eq:AzeroDef1}
\end{equation}
shown in Figure \ref{fig:InvPdf1}, can be expressed alternatively in terms of the inverse PDF as
\begin{equation}
	\mathcal{A}_0=\{(y,x)\in \mathbb{R}^2: \  0\leq x \leq p^{-1}(y)\},
\label{eq:AzeroDef2}
\end{equation}
as depicted in Figure \ref{fig:InvPdf2}.
Therefore, we can proceed in two alternative ways in order to generate samples $(x',y')$ uniformly distributed inside $\mathcal{A}_0$: 
\begin{enumerate}
	\item Draw first $x'$ from $p(x)$ and then $y'$ uniformly in the interval $[0,p(x')]$, i.e. $y' \sim \mathcal{U}([0,p(x')])$,
		as shown in Figure \ref{fig:InvPdf1}.\footnote{Noting that the samples $y'$ generated in this way are distributed according to
		$p^{-1}(y)$, we remark that this method can always be used to generate samples from the generalized unnormalized inverse PDF,
		$p_G^{-1}(y)$, even	when $p(x)$ is non-monotonic. However, in this case the geometric interpretation of this generalized inverse
		PDF becomes more complicated, since its definition may not be straightforward, as shown in Section \ref{sec:GenPdfs}.}
	\item Draw first $y'$ from $p^{-1}(y)$ and then $x'$ uniformly in the interval $[0,p^{-1}(y')]$, i.e.
		$x' \sim \mathcal{U}([0,p^{-1}(y')])$, as shown in Figure \ref{fig:InvPdf2}.
\end{enumerate}
Both procedures allow us to generate points $(x',y')$ uniformly distributed inside the region $\mathcal{A}_0$.
Moreover, from the fundamental theorem of simulation, the first coordinate $x'$ is distributed according to the unnormalized target PDF, $p(x)$, whereas the PDF of the second coordinate $y'$ is proportional to the unnormalized inverse PDF, $p^{-1}(y)$.
Hence, the key idea of the inverse of density method is that, whenever we are able to draw samples $y'$ from $p^{-1}(y)$ more easily than samples $x'$ from $p(x)$, we can use the second procedure to generate samples $x'$ from $p(x)$ more efficiently.

Note that generating a sample $x'$ uniformly inside the interval $[0,a]$, i.e. $x'\sim\mathcal{U}([0,a])$, is equivalent to drawing a sample $w'$ uniformly inside $[0,1]$ and then multiplying it by $a$, i.e. $x' = w' a$. Thus, given a known value $y'$, drawing a sample $x'$ uniformly inside the interval $[0,p^{-1}(y')]$, i.e. $x'\sim\mathcal{U}([0, p^{-1}(y')])$, is equivalent to generating a sample $w'$ uniformly inside $[0,1]$ and then taking
\begin{equation}
	x' = w' p^{-1}(y'),
\label{eq:IoD1}
\end{equation}
which is the expression frequently provided for the IoD method.
%
We also remark that, for a proper monotonic unnormalized density $p(x)$, $p^{-1}(x)$ is also a proper monotonic unnormalized PDF, obtained simply through functional inversion.

Obviously, the interest in using this technique depends on the feasibility of drawing samples from the unnormalized inverse PDF, $p^{-1}(y)$, more easily than from the unnormalized target PDF, $p(x)$, as already mentioned.
The following example shows a practical application where the IoD method provides a clear advantage over the direct generation of a random variable.
\begin{Ex}
Assume that we need to draw samples from
\begin{equation}
p_0(x) = p(x) = \sqrt{\log\left(\frac{2}{\pi x^2}\right)}, \mbox{  } \mbox{   } \mbox{ with } \mbox{   }  \mbox{   }  0\leq x \leq \sqrt{\frac{2}{\pi}}.
\end{equation}
Since $p^{-1}(y) = \sqrt{\frac{2}{\pi}\exp(-y^2)} = 2\frac{1}{\sqrt{2\pi}} \exp(-y^2/2)$ is the half Gaussian PDF used in the previous example, we can easily draw $y'$ from $p^{-1}(y)$, then $w'$ from a uniform PDF inside $[0,1]$, and finally obtain a sample $x' = w' p^{-1}(y')$, which is distributed according to the target PDF, $p(x)$.
\end{Ex}

Finally, we notice that it is possible to find this method in other forms related to \emph{vertical density representation} in the literature \citep{Fang01,Jones02,Khintchine38, Troutt04}.
Indeed, let us consider a random variable $Y$ which follows a strictly decreasing unnormalized PDF, $p^{-1}(y)$.
Then, the random variable $\widetilde{U} = p^{-1}(Y)$ is distributed as 
\begin{equation}
\label{eq:vertPDF}
	q(\tilde{u}) = -p^{-1}(p(\tilde{u}))\frac{dp(\tilde{u})}{d\tilde{u}} = -\tilde{u}\frac{dp(\tilde{u})}{d\tilde{u}},
\end{equation}
and this unnormalized PDF, $q(\tilde{u})$, is called the \emph{vertical density} w.r.t. $p^{-1}(y)$.
Making use of this result, the inverse of density method, summarized by equation \eqref{eq:IoD1}, can be expressed alternatively in this way: given $w' \sim \mathcal{U}([0,1])$ and $\tilde{u}' \sim q(\tilde{u})$, then the sample
\begin{equation}
	x' = w' \tilde{u}' = w' p^{-1}(y'),
\label{SecondFormIoD}
\end{equation}
is distributed as $p(x)$ provided that $y'$ is a sample from $p^{-1}(y)$.
The relationship in Eq. (\ref{SecondFormIoD}) is usually known as Khintchine's theorem.

\subsection{Rejection sampling}
\label{sec:RS}
\begin{figure}[htb]
	\centering 
		\centerline{
      \includegraphics[width=6cm]{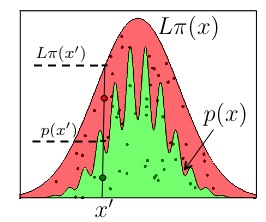}
     }
	\caption{Graphical description of the RS procedure. The green region corresponds to the acceptance region $\mathcal{A}_0$, as defined
		by \eqref{eq:A0}, whereas the red region indicates the rejection region, $\mathcal{A}_0^c$, located between the functions
		$L\pi(x)$ and $p(x)$, as defined by \eqref{eq:exclusionRegion}, and $x'$ denotes a sample drawn from the proposal PDF, $\pi(x)$,
		for the two possible situations that can occur: $x' \in \mathcal{A}_0$ (filled dark green circle) and $x' \in \mathcal{A}_0^c$
		(filled dark red circle).}
  \label{fig:RSfig}
\end{figure}

Another technique that clearly applies the simple idea exposed in Section \ref{sec:FToS} is rejection sampling.
Rejection sampling (RS) is a universal method for drawing independent samples from an unnormalized target density, $p(x)$, known up to a proportionality constant $K >0$.
Let $\pi(x)$ be a (possibly unnormalized) proposal PDF and $L$ an upper bound for the ratio $p(x)/\pi(x)$, i.e.
\begin{equation}
L\geq \frac{p(x)}{\pi(x)}.
\label{eq:ratioRS}
\end{equation}
RS works by generating samples from the proposal PDF, $\pi(x)$, and accepting or rejecting them on the basis of this ratio.
The standard RS algorithm can be outlined as follows.
\begin{enumerate}
	\item Draw $x' \sim \pi(x)$ and $w' \sim \mathcal{U}([0,1])$.
	\item If $w' \leq \frac{p(x')}{L\pi(x')}$, then $x'$ is accepted. Otherwise, $x'$ is discarded.
	\item Repeat steps 1--2 until as many samples as required have been obtained from the target PDF.
\end{enumerate}
Alternatively, the procedure undertaken by the RS method can also be summarized in the following equivalent way that remarks its close connection to the fundamental theorem of simulation.
\begin{enumerate}
	\item Draw $x'$ from $\pi(x)$.
	\item Generate $y'$ uniformly inside the interval $[0,L\pi(x')]$, i.e. $y' \sim \mathcal{U}([0,L\pi(x')])$.
	\item If the point $(x',y')$ belongs to $\mathcal{A}_0$, the region corresponding to the area below the unnormalized target PDF 
		$p(x)$ as defined by \eqref{eq:A0}, the sample $x'$ is accepted.
	\item Otherwise, i.e. whenever the point $(x',y')$ falls inside the region located between the functions $L\pi(x)$ and $p(x)$, the
		sample $x'$ is rejected.
	\item Repeat steps 1--4 until as many samples as required have been obtained from the target PDF.
\end{enumerate}
Figure \ref{fig:RSfig} provides a graphical representation of the rejection sampling technique.
Here, the green region corresponds to $\mathcal{A}_0$ as defined by \eqref{eq:A0}, the region associated to the target PDF inside which we want to sample uniformly (i.e. the \emph{acceptance region}), whereas the red region indicates the region located between the functions $L\pi(x)$ and $p(x)$, where we do not want our samples to lie (i.e. the \emph{rejection region}).
Defining
\begin{equation}
	\mathcal{A}_{\pi} = \{(x,y) \in \mathbb{R}^2: 0 \le y \le L \pi(x)\},
\label{eq:proposalRegion}
\end{equation}
this rejection or exclusion region is given by the set-theoretic difference or relative complement of $\mathcal{A}_0$ inside $\mathcal{A}_{\pi}$:
\begin{equation}
	\mathcal{A}_0^c = \mathcal{A}_{\pi} \setminus \mathcal{A}_0 = \{(x,y) \in \mathbb{R}^2: p(x) < y \le L \pi(x)\}.
\label{eq:exclusionRegion}
\end{equation}
Now, the RS algorithm proceeds by drawing first a sample from the proposal PDF, $x' \sim \pi(x)$, and then a second sample from a uniform distribution, $y' \sim \mathcal{U}([0,L\pi(x')])$.
If the point $(x',y')$ belongs to $\mathcal{A}_0$ (green region), as it happens for the point indicated by a filled dark green circle in Figure \ref{fig:RSfig}, the sample $x'$ is accepted.
Otherwise, whenever the point $(x',y')$ belongs to $\mathcal{A}_0^c$ (red region), as it happens for the point indicated by a filled dark red circle in Figure \ref{fig:RSfig}, it is discarded.
Note that, since $y' \sim \mathcal{U}([0,L\pi(x')])$ can be expressed alternatively as $y' = L\pi(x')w'$ with $w' \sim \mathcal{U}([0,1])$, the previous conditions are equivalent to accepting $x'$ whenever $y' = L\pi(x')w' \le p(x')$, which happens if and only if $(x',y')$ belongs to the green region, and rejecting $x'$ otherwise, i.e. whenever $y' = L\pi(x')w' > p(x')$, which happens if and only if $(x',y')$ belongs to the red region. This is equivalent to the condition shown in step 2 of the first formulation, demonstrating the equivalence between both descriptions of the RS algorithm.

The fundamental figure of merit of a rejection sampler is the \emph{mean acceptance rate}, i.e. the expected number of accepted samples out of the total number of proposed candidates, which is given by
\begin{equation}
	P_a = \frac{|\mathcal{A}_0|}{|\mathcal{A}_{\pi}|} = \frac{|\mathcal{A}_0|}{|\mathcal{A}_0|+|\mathcal{A}_0^c|} = 1 - \frac{|\mathcal{A}_0^c|}{|\mathcal{A}_{\pi}|},
\label{eq:Pa}
\end{equation}
where $|\mathcal{C}|$ denotes the Lebesgue measure of set $\mathcal{C}$, and the last two expressions arise from the fact that, since $\mathcal{A}_{\pi} = \mathcal{A}_0 \cup \mathcal{A}_0^c$ and $\mathcal{A}_0 \cap \mathcal{A}_0^c = \emptyset$, then $|\mathcal{A}_{\pi}| = |\mathcal{A}_0| + |\mathcal{A}_0^c|$.
Hence, from \eqref{eq:Pa} we notice that finding a tight overbounding function $L \pi(x)$ as close as possible to $p(x)$, i.e. making $|\mathcal{A}_0^c|$ as small as possible, is crucial for the good performance of a rejection sampling algorithm.

The most favourable scenario to use the RS algorithm occurs when $p(x)$ is bounded with bounded domain.
In this case, the proposal PDF, $\pi(x)$, can be chosen as a uniform density (the easiest possible proposal), and the calculation of the bound $L$ for the ratio $p(x)/\pi(x)$ is converted into the problem of finding an upper bound for the unnormalized target PDF, $p(x)$, which is in general a much easier task.
Indeed, in this scenario the performance of the rejection sampler can be easily improved using adaptive schemes \citep{Gilks92,MartinoStatCo10} or strip methods \citep[Chapter 5]{Hormann03}, \citep{Ahrens93,Ahrens95,Devroye84,Hormann02,Marsaglia00} among other techniques.
Unfortunately, when $p(x)$ is unbounded or its domain is infinite, the proposal $\pi(x)$ cannot be a uniform density and, in general, it is not straightforward to design a good proposal PDF (i.e. a proposal from which samples can be easily drawn and with a shape as close as possible to the shape of the target PDF) inside an infinite domain \citep{Devroye86,Gorur08rev,Hormann03,MartinoStatCo10-2}.
 
\begin{figure}[htb]
\centerline{
	\subfigure[]{\includegraphics[width=4.6cm]{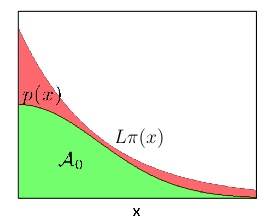}}	
	\subfigure[]{\includegraphics[width=4.6cm]{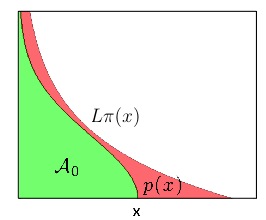}}	
	\subfigure[]{\includegraphics[width=4.6cm]{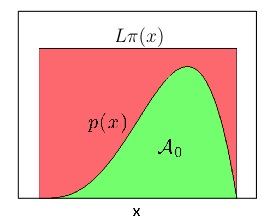}}	
}	
\caption{Three possible cases of unnormalized target density, $p(x)$, with three possible overbounding functions $L\pi(x)$: \textbf{(a)} bounded with infinite domain,  \textbf{(b)} unbounded inside a finite domain and \textbf{(c)} bounded with a finite domain. Only in the last case it is possible to use a uniform distribution as a proposal PDF, $\pi(x)$.} 
\label{fig:3cases}
\end{figure}

Figure \ref{fig:3cases} illustrates the three possible cases considered in the sequel: bounded PDF with an infinite support, Figure \ref{fig:3cases}(a), unbounded PDF with a finite support, Figure \ref{fig:3cases}(b), and bounded PDF with a finite support, Figure \ref{fig:3cases}(c).
In fact, there exists a fourth possible scenario: an unbounded PDF with an infinite support. However, since we can consider this case as a combination of the first two cases shown in Figure \ref{fig:3cases}(a) and Figure \ref{fig:3cases}(b), it will only be briefly discussed.
The next two sections  are devoted to describing methods that deal with these problematic situations by transforming $p(x)$ and embedding it inside a finite region. First, Section \ref{sec:TR} describes the transformed rejection (TR) sampling approach, and then Section \ref{sec:GRoU} describes the generalized ratio of uniforms (GRoU) technique.

\section{Transformed rejection method}
\label{sec:TR}

As already discussed in Section \ref{sec:RS}, the simplest scenario for the RS algorithm occurs when the target density is bounded with bounded support, since a uniform PDF can be used as proposal density, as suggested by several authors (see e.g. \citep{Botts11,Devroye86,Hormann93,Hormann94,Marsaglia84,Wallace76}).
Therefore, an interesting and very active line of research is trying to find a suitable invertible transformation of the target RV that allows us to apply RS to a bounded PDF defined inside a finite domain, where we can use the uniform or some other simple proposal.
Namely, our goal is finding a transformation that converts a PDF of the type displayed in Figure \ref{fig:3cases}(a) or Figure \ref{fig:3cases}(b) into a PDF of the type depicted in Figure \ref{fig:3cases}(c).

Conceptually, we can distinguish two cases: a bounded target PDF defined inside an unbounded domain, as in Figure \ref{fig:3cases}(a), and an unbounded target PDF with bounded support, as in Figure \ref{fig:3cases}(b).
The third case, unbounded target PDF with unbounded support, can be dealt with as a combination of the other two cases.
Moreover, we can tackle the problem by applying a transformation directly to an RV distributed according to the target PDF, $X \sim p(x)$, or to an RV that follows the inverse target PDF, $Y \sim p^{-1}(y)$.
Hence, taking into account all the possibilities, in the sequel we have to consider six different situations:
\begin{enumerate}[A.]
	\item Applying a suitable invertible transformation to an RV $X \sim p(x)$, obtaining $Z = f(X) \sim \rho(z)$.
		\begin{enumerate}[1)]
			\item When $p(x)$ is bounded with unbounded domain.
			\item When $p(x)$ is unbounded but has a finite support.
			\item When $p(x)$ is unbounded and with an infinite support.
		\end{enumerate}
	\item Applying an appropriate invertible transformation to an RV $Y \sim p^{-1}(y)$, obtaining
		$\widetilde{U} = h(Y) \sim q(\tilde{u})$.
		\begin{enumerate}[1)]
			\item When $p(x)$ is bounded with unbounded domain, implying that $p^{-1}(y)$ is unbounded but with bounded support.
			\item When $p(x)$ is unbounded but has a finite support, implying that $p^{-1}(y)$ is bounded but has an infinite support.
			\item When both $p(x)$ and $p^{-1}(y)$ are unbounded and with an infinite support.
		\end{enumerate}
\end{enumerate}

Finally, before discussing in detail all these cases, it is important to remark that we can always generate samples distributed according to the target PDF from samples of the transformed RVs.
On the one hand, when an invertible transformation $Z = f(X)$ is applied directly to the target RV, $X \sim p(x)$, the transformed RV follows an unnormalized PDF $\rho(z) = p(f^{-1}(z)) |\dot{f}^{-1}(z)|$, and, given a sample $z'$ from $\rho(z)$, then $x' = f^{-1}(z')$ is clearly distributed as $p_0(x) \propto p(x)$.
On the other hand, if the invertible transformation $\widetilde{U} = h(Y)$ is applied instead to the inverse target RV, $Y \sim p^{-1}(y)$, then the resulting RV $\widetilde{U}$ follows an unnormalized PDF $q(\tilde{u}) = p^{-1}(h^{-1}(\tilde{u})) |\dot{h}^{-1}(\tilde{u})|$.
Unfortunately, the relationship between samples $\tilde{u}'$ from $q(\tilde{u})$ and samples $x'$ from $p_0(x) \propto p(x)$ is not trivial, but can still be found and exploited to obtain samples from the target PDF, as shown in Section \ref{sec:ExtIoD}.

\subsection{Transformation of the target random variable $X$}
\label{sec:TRS_X}

In this section we look for suitable transformations, $f(x)$, applied directly to the target RV, $X \sim p_0(x) \propto p(x)$, such that the resulting RV, $Z = f(X) \sim \rho(z)$ is bounded with bounded support.
In the sequel we will consider, without loss of generality, that $f(x)$ is a class $C^1$ monotonic (either increasing or decreasing) function inside the range of interest (i.e. inside the domain of the target RV $X$, $\mathcal{D}_X$).\footnote{A function $f(x)$ is said to be of class $C^1$ if it is \emph{continuously differentiable}, i.e. if $f(x)$ is continuous, differentiable, and its derivative, $\dot{f}(x)$, is also a continuous function.}
This implies that $f(x)$ is invertible, and its inverse, $f^{-1}(z)$, is also a class $C^1$ monotonic function inside the range of interest (the domain of the transformed RV $Z$, $\mathcal{D}_Z = \langle 0,1 \rangle$, in this case).
Finally, regarding the unnormalized target PDF, $p(x)$, we do not make any assumption (e.g. we do not require that $p(x)$ is neither monotonic nor continuous) and consider a domain $\mathcal{D}_X = \mathbb{R}$ for PDFs with unbounded support (cases 1 and 3) and $\mathcal{D}_X = \langle a,b \rangle$, with $a,b \in \mathbb{R}$ and $a < b$, for PDFs with bounded support (case 2).

\subsubsection{Bounded target PDF $p(x)$ with unbounded support}
\label{sec:TRS_X_boundedPDF}

When the target PDF is bounded with unbounded domain and the transformation is applied directly to the target RV $X$, the sampling technique obtained is known in the literature as the {\em transformed rejection method}, due to \citep{Botts11,Hormann93,Hormann94,Wallace76}.
However, this approach is also called the {\em almost exact inversion method} in \citep[Chapters 3]{Devroye86} and the {\em exact approximation method} in \citep{Marsaglia84}, remarking its close relationship with the inversion method \citep[Chapter 2]{Devroye86}, as explained later.

Let $p(x)$ be a bounded density with unbounded support, $\mathcal{D}_X = \mathbb{R}$, and let us consider a class $C^1$ monotonic transformation, $f: \mathbb{R} \rightarrow \langle 0,1 \rangle$.
If $X$ is an RV with unnormalized PDF $p(x)$, then the transformed random variable $Z = f(X)$ has an unnormalized density
\begin{equation}
	\rho(z) = p\big(f^{-1}(z)\big) \Bigg| \frac{\textrm{d} f^{-1}(z)}{\textrm{d}z}\Bigg| = p\big(f^{-1}(z)\big) 
		|\dot{f}^{-1}(z)|, \qquad \textrm{for} \quad z \in \langle 0,1 \rangle,
\label{eq:pdfZ}
\end{equation}
where $f^{-1}(z)$ is the inverse function of $f(x)$.
Thus, the key idea in \citep{Wallace76} is using an RS algorithm to draw samples from $\rho(z)$ and then generating samples from the target PDF by inverting the transformation $f(x)$, i.e. drawing $z' \sim \rho(z)$ and then taking $x'=f^{-1}(z')$.
By choosing an adequate transformation $f(x)$, such that $\rho(z)$ is also bounded, this strategy allows the proposal PDF, $\pi(x)$, to be a uniform density, as in Figure \ref{fig:3cases}(c).

\begin{figure}[htb]
\centerline{
	\subfigure[]{\includegraphics[width=4.5cm]{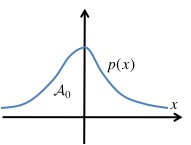}}
	\subfigure[]{\includegraphics[width=4.5cm]{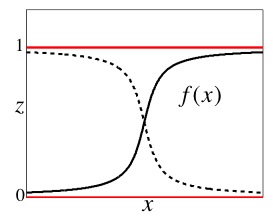}}	
	\subfigure[]{\includegraphics[width=4.5cm]{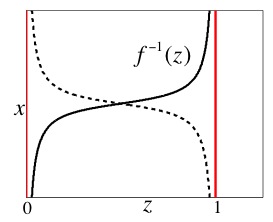}}
}	
\caption{\textbf{(a)} A bounded target PDF $ p(x)$ with an unbounded domain $\mathcal{D}_X = \mathbb{R}$. \textbf{(b)} Two possible examples, monotonically increasing (solid line) and monotonically decreasing (dashed line), of the transformation $f(x)$ with horizontal asymptotes at $x=0$ and $x=1$. \textbf{(c)} The corresponding inverse transformations $f^{-1}(z)$ with vertical asymptotes at $z=0$ and $z=1$.}
\label{TRSfigura}
\end{figure}

Obviously, the domain of $\rho(z)$, $\mathcal{D}_Z = \langle 0,1 \rangle$, is bounded. However, in general the density $\rho(z)$ can be unbounded, i.e. it may have vertical asymptotes, depending on the choice of the transformation $f(x)$.
Indeed, taking a closer look at \eqref{eq:pdfZ} we notice that, although the first term $p\big(f^{-1}(z)\big)$ is bounded (since $p_0(x)$ is assumed to be bounded), the second term, $|\dot{f}^{-1}(z)|$, is unbounded in general, since
\begin{equation}
	\lim_{z\rightarrow 0}  \left|\frac{d f^{-1}(z)}{dz}\right| = \lim_{z\rightarrow 1} \left|\frac{d f^{-1}(z)}{dz}\right| = \infty.
\label{eq:limitTRSA1}
\end{equation}
This is due to the fact that $f(x)$ must have horizontal asymptotes, since it is a monotonic continuous function that converts the infinite support of $p(x)$, $\mathcal{D}_X$, into a finite domain, $\mathcal{D}_Z = \langle 0,1 \rangle$.
Consequently, $f^{-1}(z)$ must have vertical asymptotes at the extreme points of $\mathcal{D}_Z$, implying that the limits in \eqref{eq:limitTRSA1} diverge to infinity.
Figure \ref{TRSfigura} illustrates this situation, showing an example of a non-monotonic unnormalized target PDF with support $\mathcal{D}_X = \mathbb{R}$ and two examples of possible transformations $f(x)$ and $f^{-1}(z)$ (strictly increasing and decreasing respectively), where the asymptotes can be clearly appreciated.

Hence, as a conclusion, it is clear from \eqref{eq:pdfZ} and \eqref{eq:limitTRSA1} that the unnormalized density $\rho(z)$ resulting from the transformation $f(x)$ remains bounded when the tails of $p(x)$ decay to zero quickly enough, namely, faster than the derivative $\frac{d f^{-1}(z)}{dz} = \left(\frac{d f(x)}{dx }\right)^{-1}$ diverges when $z \rightarrow z^* \in \{0,1\}$ (or equivalently, when $x \rightarrow f^{-1}(z^*) = \pm \infty$).
More formally, let us note that the limit of interest can be expressed as
\begin{equation}
	L_1 = \lim_{z\rightarrow z^*} \rho(z) = \lim_{z\rightarrow z^*} p\big(f^{-1}(z)\big) \big|\dot{f}^{-1}(z)\big|
		= \lim_{z \rightarrow z^*} \frac{p\big(f^{-1}(z)\big)}{\big|\dot{f}(x)\big|_{x=f^{-1}(z)}}
		= \lim_{x \rightarrow f^{-1}(z^*)} \frac{p(x)}{|\dot{f}(x)|},
\end{equation}
for $z^* \in \{0,1\}$, with both $ p\big(f^{-1}(z)\big)$ and $\big|\dot{f}(x)\big|_{x=f^{-1}(z)}$ tending to zero as $z \to z^*$.
Hence, this limit will be finite if and only if $ p\big(f^{-1}(z)\big)$ is an infinitesimal of the same or higher order than $\big|\dot{f}(x)\big|_{x=f^{-1}(z)}$ at $z = z^*$.
Alternatively, using the last expression of the limit, $L_1$ will be finite if and only if $ p(x)$ is an infinitesimal of the same or higher order than $|\dot{f}(x)|$ at $x = f^{-1}(z^*) = \pm \infty$.

We also remark that $f^{-1}(z)$ has vertical asymptotes at both extreme points of $\mathcal{D}_Z$ because the support considered for the target RV $X$, $\mathcal{D}_X$, is a bi-infinite interval (i.e. it extends towards infinity in both directions). If the support of $X$ is a semi-infinite interval in $\mathbb{R}$ (i.e. an interval that extends towards infinity only in one direction, e.g. $\mathcal{D}_X=\mathbb{R}^+$ or $\mathcal{D}_X=\mathbb{R}^-$), then $f^{-1}(z)$ only has one vertical asymptote either at $z=0$ or at $z=1$, depending on the open end of the interval and on whether $f(x)$ is increasing or decreasing. However, by focusing on the single asymptote of $f^{-1}(z)$, the discussion performed above remains valid.

Finally, it is also important to realize that better acceptance rates can be obtained by a suitable choice of the transformation function $f(x)$. Indeed, when $f(x)$ is similar to the unnormalized CDF, $F_X(x)$, the PDF $\rho(z)$ becomes flatter and closer to a uniform density, so that the acceptance rate using a uniform proposal, $\pi(z) = \mathcal{U}(\mathcal{D}_Z)$, is improved. In fact, if $f(x)$ is exactly equal to the unnormalized CDF of $X$, i.e. $f(x) = F_X(x)$, then $\rho(z)$ is the uniform density inside the interval $\mathcal{D}_Z = [0,1]$. For this reason, this technique is also termed \emph{almost exact inversion method} by some authors (see e.g. \citep{Devroye86}).


\subsubsection{Unbounded target PDF $p(x)$ with bounded support}
\label{sec:TRMup}

A similar methodology can also be applied when the target PDF, $p(x)$, is unbounded but has a bounded support, $\mathcal{D}_X = \langle a,b \rangle$.
In this case, using again a class $C^1$ monotonic transformation, $f: \langle a,b \rangle \rightarrow \langle 0,1 \rangle$, we can also transform $p(x)$ into a bounded density with bounded domain, $\mathcal{D}_Z = \langle 0,1 \rangle$.
For ease of exposition, and without loss of generality, let us assume that $p(x)$ has only one vertical asymptote at $x = x^*$, i.e. $0 \le p(x) < \infty$ for $x \ne x^*$ and $\lim_{x\rightarrow x^*}  p(x) = \infty$, as illustrated in Figure \ref{fig:3cases}(b), where $x^*=a$.\footnote{In many cases, the vertical asymptote of $p(x)$ is located at one of the extreme points of the support. Hence, for a monotonically decreasing target PDF with support $\mathcal{D}_X=(a,b]$ we will typically have $x^* = a$, as shown in Figure \ref{fig:3cases}(b).}
Now, let us consider a target RV $X$ with PDF $p(x)$ and $Z = f(X)$. We already know that the unnormalized density of $Z$ is given by
\begin{equation}
	\rho(z) =  p\big(f^{-1}(z)\big) \big|\dot{f}^{-1}(z)\big|, \qquad \textrm{for} \quad  z \in \langle 0,1 \rangle.
\label{EQq2}
\end{equation}
Unfortunately, although $\big|\dot{f}^{-1}(z)\big|$ is bounded (since $f^{-1}(z)$ is a class $C^1$ function), $\rho(z)$ is unbounded in general, as the first term diverges, i.e.
\begin{equation}
\lim_{z\rightarrow f(x^*)}  p\big(f^{-1}(z)\big)= \lim_{x \rightarrow x^*}  p\big(x\big) = \infty.
\end{equation}
Following a similar line of reasoning as in the previous section, we notice that now the limit of interest is given by
\begin{equation}
	L_2 = \lim_{z\rightarrow f(x^*)} \rho(z) = \lim_{z\rightarrow f(x^*)} p\big(f^{-1}(z)\big) \big|\dot{f}^{-1}(z)\big|
		= \lim_{x \rightarrow x^*}  p(x) \big|\dot{f}^{-1}(z)\big|_{z=f(x)}
		= \lim_{x \rightarrow x^*} \frac{ p(x)}{\big|\dot{f}(x)\big|}.
\label{eq:limitTRSA2}
\end{equation}
Thus, since $ p(x) \to \infty$ when $x \to x^*$, a necessary condition for obtaining $L_2 < \infty$ is having $\big|\dot{f}^{-1}(z)\big|_{z=f(x)} \to 0$, or equivalently $\big|\dot{f}(x)\big| \to \infty$, when $x \to x^*$.
However, this condition is not sufficient for ensuring that \eqref{eq:limitTRSA2} is bounded.
Focusing on the last expression of this limit, we notice that $L_2$ will be finite if and only if $1/\big|\dot{f}(x)\big|$ is an infinitesimal of equal or higher order than $1/ p(x)$ at $x = x^*$.

\begin{figure}[htb]
\centering
\centerline{ 
\subfigure[]{\includegraphics[width=5cm]{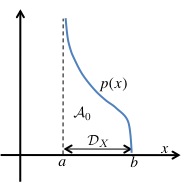}} 
\subfigure[]{\includegraphics[width=5cm]{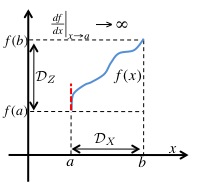}} 
 }
\caption{{\bf a)} Example of an unbounded target PDF, $p(x)$, with a bounded support $\mathcal{D}_X = (a,b]$. {\bf b)}  Example of a suitable increasing transformation $f(x)$ fulfilling that $\dot{f}(x) \to \infty$ when $x \to x^* = a$ faster than $p(x)$.}
\label{figurasTRunboundedpdf}
\end{figure}

Figure \ref{figurasTRunboundedpdf} shows an example of an unbounded target PDF, $ p(x)$, with a bounded support, $\mathcal{D}_X = (a,b]$, as well as an adequate transformation $f(x)$ that allows us to achieve a bounded unnormalized transformed PDF, $\rho(z)$.
Finally, let us remark that, for a more general unnormalized target PDF, $p(x)$, with several vertical asymptotes located at $x^* \in \mathcal{X}^*$, the same restrictions apply.
Indeed, $\rho(z)$ will be finite if and only if $1/\big|\dot{f}(x)\big|$ is an infinitesimal of equal or higher order than $1/p(x)$ at all $x = x^* \in \mathcal{X}^*$.

\subsubsection{Unbounded target PDF $p(x)$ with unbounded support}
\label{sec:TRSA3}

%
%
Combining the reasoning followed in the previous two subsections, it is straightforward to see that the necessary and sufficient conditions for obtaining a bounded PDF, $\rho(z)$, when the target PDF, $p(x)$, is unbounded and has an unbounded support, are:
\begin{enumerate}
	\item The target PDF, $p(x)$, must be an infinitesimal of the same or higher order than $\big|\dot{f}(x)\big|$ at $x = f^{-1}(z^*)$
		for all $z^* \in \mathcal{Z}^*$ and $\mathcal{Z}^*$ denoting the set of vertical asymptotes of $f^{-1}(z)$.
	\item $1/\big|\dot{f}(x)\big|$ must be an infinitesimal of the same or higher order than $1/p(x)$ at $x = x^*$ for all
		$x^* \in \mathcal{X}^*$ and $\mathcal{X}^*$ denoting the set of vertical asymptotes of $p(x)$.
\end{enumerate}

\subsection{Transformation of the inverse target random variable $Y$}
\label{sec:TRS_Y}

In this section we perform the complementary study of the one shown in Section \ref{sec:TRS_X}, analyzing suitable transformations, $h(y)$, applied to the unnormalized inverse target RV, $Y \sim p^{-1}(y)$, such that the resulting RV, $\widetilde{U} = h(Y) \sim q(\tilde{u})$ is bounded with bounded support.
With respect to the transformations we will consider the same restrictions as in the previous section: $h(y)$ belongs to the set of class $C^1$ monotonic functions inside the range of interest (i.e. inside the domain of $Y$, $\mathcal{D}_Y$).
Once more, this means that $h(y)$ is invertible, and its inverse, $h^{-1}(\tilde{u})$, is also a class $C^1$ monotonic function inside the range of interest: the domain of the transformed RV $\widetilde{U}$, $\mathcal{D}_{\widetilde{U}} = \langle 0,1 \rangle$.
Finally, regarding the unnormalized target PDF, $p(x)$, now we assume, without loss of generality, that it is monotonic and strictly decreasing with a domain $\mathcal{D}_X = \mathbb{R}^+$ for target PDFs with unbounded support (cases 1 and 3) and $\mathcal{D}_X = (0,b]$ for PDFs with bounded support (case 2).\footnote{The same conclusions can be obtained using a monotonically decreasing target PDF. However, since monotonically decreasing PDFs are more frequently used, we have chosen to work with this class of PDFs.}
This ensures that $p(x)$ is invertible and $p^{-1}(y)$ is also a well-defined monotonic and strictly decreasing PDF,\footnote{The discussion performed in the sequel can be extended to non-monotonic PDFs. However, in this case we must work with the generalized inverse PDF, $p_G^{-1}(y)$, which may be difficult to define in some cases. Thus, for the sake of simplicity we focus on monotonic PDFs here, leaving the discussions related to non-monotonic PDFs for Section \ref{sec:GenPdfs}.} with a domain $\mathcal{D}_Y = \mathbb{R}^+$ for unbounded target PDFs (cases 2 and 3) and $\mathcal{D}_Y = (0,1]$ for bounded target PDFs (case 1).\footnote{Note that using $\mathcal{D}_Y = (0,1]$ implies assuming that the normalization constant, $K$, is chosen in such a way that $p(0)=1$ and $p(x) \to 0$ when $x \to \infty$.}

\subsubsection{Bounded target PDF $p(x)$ with unbounded support}
\label{sec:OurCase}

Let us consider a monotonically decreasing and bounded unnormalized target PDF, $p(x)$, with unbounded support, $\mathcal{D}_X = \mathbb{R}^{+}$, such that $p(0)=1$ and $p(x) \to 0$ when $x \to \infty$.
This implies that the unnormalized inverse target PDF, $p^{-1}(y)$, is unbounded, but with bounded support, $\mathcal{D}_Y = (0, 1]$.
Let us consider another RV, $\widetilde{U} = h(Y)$ with $Y \sim p^{-1}(y)$, obtained applying a monotonic transformation, $h(y)$, bounded inside the domain of $Y$, $h: (0, 1] \rightarrow \langle 0,1 \rangle$.
The unnormalized density of $\widetilde{U}$ is then given by
\begin{equation}
	q(\tilde{u}) = p^{-1}\big(h^{-1}(\tilde{u})\big) \big|\dot{h}^{-1}(\tilde{u})\big|
		\qquad \textrm{for} \quad \tilde{u} \in \rightarrow \langle 0,1 \rangle.
\label{eq:pdfQ1}
\end{equation}
Now, since $p^{-1}(y)$ is unbounded when $y \to y^* = 0$, in order to obtain a bounded PDF, $q(\tilde{u})$, a necessary condition is
\begin{equation}
	\lim_{\tilde{u} \to h(y^*)} \big|\dot{h}^{-1}(\tilde{u})\big| = \lim_{y \to y^* = 0} \big|\dot{h}(y)\big|^{-1} = 0.
\end{equation}
However, as it happened in Section \ref{sec:TRMup}, this is not a sufficient condition. Once more, focusing on the limit of interest in this case,
\begin{equation}
	L_3 = \lim_{\tilde{u} \to h(y^*)} q(\tilde{u})
		= \lim_{\tilde{u} \to h(y^*)} p^{-1}\big(h^{-1}(\tilde{u})\big) \big|\dot{h}^{-1}(\tilde{u})\big|
		= \lim_{\tilde{u} \to h(y^*)} \frac{p^{-1}\big(h^{-1}(\tilde{u})\big)}{\big|\dot{h}(y)\big|_{y=h^{-1}(\tilde{u})}}
		= \lim_{y \to y^* = 0} \frac{p^{-1}(y)}{\big|\dot{h}(y)\big|},
\end{equation}
we realize that a necessary and sufficient condition is that $1/\big|\dot{h}(y)\big|$ is an infinitesimal of equal or higher order than $1/p^{-1}(y)$ at $y = y^* = 0$.

\subsubsection{Unbounded target PDF $p(x)$ with bounded support}
\label{sec:TRSB2}

Here we consider the complementary case of the one discussed in the previous section: an unbounded monotonically decreasing unnormalized target PDF, $p(x)$, with a vertical asymptote at $x = x^* = 0$ (i.e. $\lim_{x \to x^* = 0} p(x) = \infty$), but bounded support, $\mathcal{D}_X = (0,b]$.
Hence, the unnormalized inverse target PDF, $p^{-1}(y)$, is monotonically decreasing and bounded ($0 < p^{-1}(y) \le b$), but has an unbounded support, $\mathcal{D}_Y = \mathbb{R}^+$.
Now, considering another RV, $\widetilde{U} = h(Y)$ with $Y \sim p^{-1}(y)$, obtained applying a continuous and monotonic (either increasing or decreasing) transformation, $h: \mathbb{R}^{+} \rightarrow \langle 0, 1 \rangle$, to the unnormalized inverse target RV, $\widetilde{U}$ has an unnormalized PDF
\begin{equation}
	q(\tilde{u}) = p^{-1}\big(h^{-1}(\tilde{u})\big) \big|\dot{h}^{-1}(\tilde{u})\big|
		\qquad \textrm{for} \quad \tilde{u} \in \langle 0, 1 \rangle.
\label{eq:pdfQ2}
\end{equation}
Again, although the first term, $p^{-1}\big(h^{-1}(\tilde{u})\big)$, is bounded (as $p^{-1}(y)$ is bounded), this PDF may be unbounded, since the second term will be unbounded in general.
Indeed, since $h(y)$ transforms an infinite domain, $\mathcal{D}_Y = \mathbb{R}^+$, into a finite domain, $\mathbb{D}_{\widetilde{U}} = \langle 0, 1 \rangle$, it must reach a horizontal asymptote when $y \to \infty$.
This results in a vertical asymptote for $h^{-1}(\tilde{u})$ either at $\tilde{u} = \tilde{u}^* = 1$ (when $h^{-1}(\tilde{u})$ is increasing) or at $\tilde{u} = \tilde{u}^* = 0$ (when $h^{-1}(\tilde{u})$ is decreasing), implying that
\begin{equation}
	\lim_{\tilde{u} \to \tilde{u}^*} \big|\dot{h}^{-1}(\tilde{u})\big|
		= \lim_{y \to h^{-1}(\tilde{u}^*)} \big|\dot{h}(y)\big|^{-1} = \infty.
\end{equation}
The limit of interest in this case is given by
\begin{equation}
	L_ 4 = \lim_{\tilde{u} \to \tilde{u}^*} q(\tilde{u})
		= \lim_{\tilde{u} \to \tilde{u}^*} p^{-1}\big(h^{-1}(\tilde{u})\big) \big|\dot{h}^{-1}(\tilde{u})\big|
		= \lim_{\tilde{u} \to \tilde{u}^*} \frac{p^{-1}\big(h^{-1}(\tilde{u})\big)}{\big|\dot{h}(y)\big|_{y=h^{-1}(\tilde{u})}}
		= \lim_{y \to h^{-1}(\tilde{u}^*)} \frac{p^{-1}(y)}{\big|\dot{h}(y)\big|},
\end{equation}
Therefore, a necessary and sufficient condition for having $L_4 < \infty$ is that $p^{-1}(y)$ is an infinitesimal of equal or higher order than $\big|\dot{h}(y)\big|$ at $y = h^{-1}(\tilde{u}^*)$.

\subsubsection{Unbounded target PDF $p(x)$ with unbounded support}
\label{sec:TRSB3}

In the more general case ($p(x)$ unbounded and with infinite support, $\mathcal{D}_X = \mathbb{R}^+$), combining the results obtained in the previous two subsections, it can be easily demonstrated that the PDF of the transformed RV $\widetilde{U}$, $q(\tilde{u})$, will be bounded if and only if:
\begin{enumerate}
	\item The unnormalized inverse target PDF, $p^{-1}(y)$, is an infinitesimal of the same or higher order than $\big|\dot{h}(y)\big|$
		at $y = h^{-1}(\tilde{u}^*)$ for all $\tilde{u}^* \in \widetilde{\mathcal{U}}^*$ and $\widetilde{\mathcal{U}}^*$ denoting the set
		of vertical asymptotes of $h^{-1}(\tilde{u})$.
	\item $1/\big|\dot{h}(y)\big|$ is an infinitesimal of the same or higher order than $1/p^{-1}(y)$ at $y = y^*$ for all $y^* \in
		\mathcal{Y}^*$ and $\mathcal{Y}^*$ denoting the set of vertical asymptotes of $p^{-1}(y)$.
\end{enumerate}

\subsection{Summary of the conditions for all the possible cases}
\label{sec:summaryCases}

Table \ref{summaryTRS} summarizes all the possible cases considered in the previous subsections, showing the restrictions imposed both on the transformation ($f(x)$ or $h(y)$) and the target PDF ($p(x)$ or $p^{-1}(y)$), the vertical asymptotes (again both for the transformation and the target PDF) and the conditions required for attaining a bounded transformed PDF, either $\rho(z)$ as given by \eqref{eq:pdfZ} or \eqref{EQq2} for $Z=f(X)$ or $q(\tilde{u})$ as given by \eqref{eq:pdfQ1} or \eqref{eq:pdfQ2} for $\widetilde{U}=h(Y)$.
\begin{table}[!hbt]
	\begin{center}
	\caption{Summary of the conditions required for attaining a bounded transformed PDF}
	\label{summaryTRS}
	\begin{tabular}{|c|c|c|c|c|}
		\hline
		\hline
		\multicolumn{2}{|c|}{\bf conditions} & \multicolumn{2}{|c|}{\bf vertical asymptotes} & {\bf conditions for}\\
		\cline{1-4}
		{\bf transformation} & {\bf PDF} & {\bf transformation} & {\bf PDF} & {\bf  bounded PDF}\\
		\hline
		$Z = f(X)$ & $p(x)$ bounded & $\big|\dot{f}^{-1}(z)\big| \to \infty$ & None & $p(x) \to 0$ faster than\\
		$f: \mathbb{R} \to \langle 0, 1 \rangle$ & $\mathcal{D}_X = \mathbb{R}$ & when $z \to z^* \in \{0,1\}$ & & $\big|\dot{f}(x)\big| \to 0$ at $x = f^{-1}(z^*)$\\
		$f$ class $C^1$ monotonic & & & & with $z^* \in \{0,1\}$.\\
		\hline
		$Z = f(X)$ & $p(x)$ unbounded & None & $p(x) \to \infty$ & $1/\big|\dot{f}(x)\big| \to 0$ faster than\\
		$f: \langle a,b \rangle \to \langle 0, 1 \rangle$ & $\mathcal{D}_X = \langle a,b \rangle$ &  & when $x \to x^*$ & $1/p(x) \to 0$
			at $x = x^*$.\\
		$f$ class $C^1$ monotonic &  $a,b \in \mathbb{R}$, $a<b$ & & & \\
		\hline
		$Z = f(X)$ & $p(x)$  unbounded & $\big|\dot{f}^{-1}(z)\big| \to \infty$ & $p(x) \to \infty$ & $p(x) \to 0$ faster than\\
		$f: \mathbb{R} \to \langle 0, 1 \rangle$ & $\mathcal{D}_X = \mathbb{R}$ & when $z \to z^*$ & when $x \to x^*$ & $\big|\dot{f}(x)\big| \to 0$ at
			$x = f^{-1}(z^*)$.\\
		$f$ class $C^1$ monotonic & & & & \\
		 & & & & $1/\big|\dot{f}(x)\big| \to 0$ faster than\\
		 & & & & $1/p(x) \to 0$ at $x = x^*$.\\
		\hline
		\hline
		$\widetilde{U} = h(Y)$ & $p^{-1}(y)$ monotonically & None & $p^{-1}(y) \to \infty$ & $1/\big|\dot{h}(y)\big| \to 0$ faster than\\
		$h: (0,1] \to \langle 0, 1 \rangle$ & decreasing & & when $y \to y^* = 0$ & $1/p^{-1}(y) \to 0$ at $y = y^* = 0$.\\
		$h$ class $C^1$ monotonic & $\mathcal{D}_Y = (0,1]$ & & & \\
		 & unbounded & & & \\
		\hline
		$\widetilde{U} = h(Y)$ & $p^{-1}(y)$ monotonically & $\big|\dot{h}^{-1}(\tilde{u})\big| \to \infty$ & None &
			$p^{-1}(y) \to 0$ faster than\\
		$h: \mathbb{R}^+ \to \langle 0, 1 \rangle$ & decreasing & when $\tilde{u} \to \tilde{u}^*$ & &
			$\big|\dot{h}(y)\big| \to 0$ at $y = h^{-1}(\tilde{u}^*)$.\\
		$h$ class $C^1$ monotonic & $\mathcal{D}_Y = \mathbb{R}^+$ & with $\tilde{u}^* = 0$ or $\tilde{u}^* = 1$ & & \\
		& bounded & & & \\
		\hline
		$\widetilde{U} = h(Y)$ & $p^{-1}(y)$ monotonically & $\big|\dot{h}^{-1}(\tilde{u})\big| \to \infty$ & $p^{-1}(y) \to \infty$ & 
			$1/\big|\dot{h}(y)\big| \to 0$ faster than\\
		$h: \mathbb{R}^+ \to \langle 0, 1 \rangle$ & decreasing & when $\tilde{u} \to \tilde{u}^*$ & when $y \to y^*$ & 
			$1/p^{-1}(y) \to 0$ at $y = y^*$.\\
		$h$ class $C^1$ monotonic & $\mathcal{D}_Y = \mathbb{R}^+$ & & & \\
		 & unbounded & & & $p^{-1}(y) \to 0$ faster than\\
		 & & & & $\big|\dot{h}(y)\big| \to 0$ at $y = h^{-1}(\tilde{u}^*)$.\\
		\hline
		\hline
\end{tabular}
\end{center}
\end{table} 

\section{Generalized ratio of uniforms method (GRoU)}
\label{sec:GRoU}

A general version of the standard ratio of uniforms (RoU) method, proposed in \citep{Kinderman77},  can be established with the following theorem \citep{Wakefield91}.   

\begin{Teorema}
\label{GRoU_Theorem}
Let $g(u): \mathbb{R}^+\rightarrow \mathbb{R}^+$ be a strictly increasing (in $\mathbb{R}^+ \backslash \{0\}=(0,+\infty)$) differentiable function such that $g(0)=0$ and let $p(x)\geq 0$ be a PDF known only up to a proportionality constant. Assume that $(v,u)\in \mathbb{R}^2$ is a sample drawn from the uniform distribution on the set 
\begin{equation}
\label{regionAdefgen1}
\mathcal{A}_g= \Bigg\{(v,u)\in \mathbb{R}^2: 0\leq u \leq g^{-1}\Bigg[c\mbox{ }p\Bigg(\frac{v}{\dot{g}(u)}\Bigg)\Bigg]\Bigg\}, 
\end{equation}
where $c>0$ is a positive constant and $\dot{g}=\frac{dg}{du}$. Then $x=\frac{v}{\dot{g}(u)}$ is a sample from $p_0(x)$. 
\end{Teorema}
The proof can be found in the Appendix \ref{App_a_ver}. Choosing $g(u)=\frac{1}{2}u^2$, we come back to the standard RoU method.
Other generalizations of the RoU method can be found in the literature \citep{Chung97,Jones96}. Moreover, in Appendices \ref{App_a_ver2} and \ref{App_DecrG} we provide extensions of the GRoU relaxing some assumptions.   
Further developments involving ratio of RV's can be found in \citep{Barbu82,Curtiss41,Dieter89,Marsaglia65,Perez08,Stefanescu87,Vaduva82}. In literature, the GRoU is also combined with MCMC techniques \citep{Groendyke08}.

The theorem above provides a way to generate samples from $p_0(x)$. Indeed, if we are able to draw uniformly a point $(v',u')$ from $\mathcal{A}_g$, then the sample $x'=v'/\dot{g}(u')$ is distributed according to $p_0(x)\propto p(x)$. Therefore, the efficiency of the (standard or generalized) RoU methods depend on the ease with which we can generate points uniformly within the region $\mathcal{A}_g$. For this reason, the cases of practical interest are those in which the region $\mathcal{A}_g$ is bounded.
Moreover, observe that if $g(u)=u$ and $c=1$ we come back to {\it the fundamental theorem of simulation} described in Section \ref{sec:FToS}, since $\mathcal{A}_g$ becomes exactly $\mathcal{A}_0$.

\begin{figure*}[htb]
\centering
\centerline{ 
\includegraphics[width=6cm]{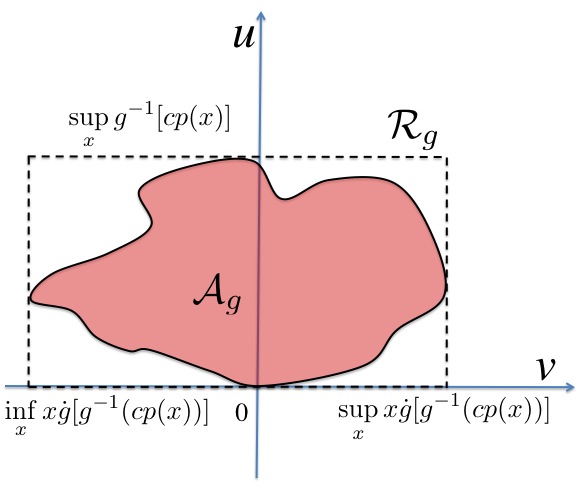} 
 }
\caption{ Example of a bounded region $\mathcal{A}_g$, embedded  
in the rectangular region $\mathcal{R}_g$.  
}
\label{figuraRoUrect}
\end{figure*}

Note that in the boundary of the region $\mathcal{A}_g$ we have $u= g^{-1}[cp(x)]$ and, since $v=x\dot{g}(u)$, we also have  
$v=x\dot{g}[g^{-1}(cp(x))]$. The contour of  $\mathcal{A}_g$ is described parametrically by the following two equations 
\begin{gather}
\label{EqBoundaryAg}
\left\{
\begin{split}
&u= g^{-1}[cp(x)], \\
&v=x\dot{g}[g^{-1}(cp(x))],\\
\end{split}
\right.
\end{gather}
where $x$ plays the role of a parameter. 
Hence, if the two functions $g^{-1}[cp(x)]$ and $x\dot{g}[g^{-1}(cp(x))]$ are bounded, the region $\mathcal{A}_g$ is embedded  
in the rectangular region  
\begin{gather}
\label{EqRect_g}
\begin{split}
\mathcal{R}_g=\Big\{(v,u)\in \mathbb{R}^2: \mbox{ }&0 \leq u \leq \sup_x g^{-1}[cp(x)],  \\
&\inf_x x\dot{g}[g^{-1}(cp(x))]\leq  v \leq  \sup_x x\dot{g}[g^{-1}(cp(x))]\Big\} .
\end{split}
\end{gather}
Figure \ref{figuraRoUrect} depicts a generic bounded region $\mathcal{A}_g$, embedded  
in the rectangular region $\mathcal{R}_g$, defined above.

Once, for instance,  the rectangle $\mathcal{R}_g$ is constructed, it is straightforward to draw uniformly from $\mathcal{A}_g$ by rejection sampling: simply draw uniformly from $\mathcal{R}_g$ and then check whether the candidate point belongs to $\mathcal{A}_g$. Note that to use this rejection procedure we do not need to know the analytical expression of the boundary of the region $\mathcal{A}_g$, i.e., it is not necessary to know the analytical relationship between the variables $v$ and $u$ that describes the contour of the $\mathcal{A}_g$.  Indeed, Eq. (\ref{regionAdefgen1}) provides a way to check whether a point $(v,u)\in \mathbb{R}^2$ falls inside $\mathcal{A}_g$ or not and this is enough to apply a RS scheme. 

\begin{figure*}[htb]
\centering
\centerline{ 
\subfigure[]{\includegraphics[width=4.5cm]{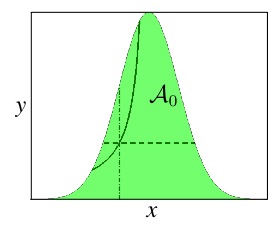}} 
 \subfigure[]{\includegraphics[width=4.5cm]{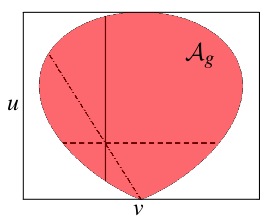}}  
 }
\caption{Examples of the regions $\mathcal{A}_g$ obtained applying the standard RoU transformation to standard Gaussian PDF. \textbf{(a)} A standard Gaussian density $p_0(x)\propto p(x)=\exp\{-x^2/2\}$.  \textbf{(b)} The region $\mathcal{A}_g$ corresponding to a standard Gaussian PDF, obtained using $g(u)=\frac{1}{2}u^2$ and $c=1$.  
}
\label{figurasRoU2}
\end{figure*}

Figure \ref{figurasRoU2}(b) provides an example in which the region $\mathcal{A}_g$ (obtained with $g(u)=\frac{1}{2}u^2$ and $c=1/2$, i.e., with the standard RoU method) corresponds to standard Gaussian density (shown in Figure \ref{figurasRoU2}(a)). The pictures also illustrate different lines corresponding to $x$ constant (dotted line), $y$ constant (dashed line), $v$ constant (solid line) in the domain $x-y$ and in the transformed domain $v-u$. 

In the next section, we obtain the conditions that the function $g(u)$ has to satisfy in order that $g^{-1}[cp(x)]$ and $x\dot{g}[g^{-1}(cp(x))]$ be bounded.

\subsection{Conditions to obtain a bounded $\mathcal{A}_g$}
\label{SectCondAg}

The region $\mathcal{A}_g$ is bounded if the two functions $g^{-1}[cp(x)]$ and $x\dot{g}[g^{-1}(cp(x))]$ are bounded. Now, we study the conditions that the functions $g(u)$ and $p(x)$ have to fulfill in order to obtain $u=g^{-1}[cp(x)]$ and $v=x\dot{g}[g^{-1}(cp(x))]$ bounded.  
\begin{enumerate}
\item {\it First function $u=g^{-1}[cp(x)]$:} since $g$ is an increasing ($\dot{g}\geq 0$) and continuous function,  $g^{-1}$ is also increasing so that the function $g^{-1}[cp(x)]$ is bounded if, and only if,  $p(x)$ is bounded, i.e.,
 \begin{equation}
p(x)\leq M, 
\end{equation}
for all $x\in \mathcal{D}$, where $M$ is a constant.  
\item {\it Second function $v=x\dot{g}[g^{-1}(cp(x))]$:} since $\dot{g}\geq 0$ and hence $g^{-1}$ is also increasing,  the function $x\dot{g}[g^{-1}(cp(x))]$ is bounded if:
\begin{enumerate}
\item $p(x)$ is bounded, i.e., $p(x)\leq M$,
\item  and the limits 
\begin{equation}
\label{LimXgg1eq}
\lim_{x\rightarrow + \infty} x\dot{g}[g^{-1}(cp(x))]=L_1\leq +\infty,  
\end{equation}
\begin{equation}
\label{LimXgg2eq}
\lim_{x\rightarrow - \infty} x\dot{g}[g^{-1}(cp(x))]=L_2\leq +\infty,  
\end{equation}
are both finite. 
The attainment of Eqs. (\ref{LimXgg1eq})-(\ref{LimXgg2eq}) entails the following conditions:
\begin{itemize}
\item Since the first factor in $x \cdot \dot{g}[g^{-1}(cp(x))]$ is  $x$,  we need that $\dot{g}\circ g^{-1}\circ cp$ vanishes as $x\rightarrow \pm \infty$, i.e.,
\begin{equation}
\label{limiteancora}
 \lim_{x\rightarrow \pm \infty} \dot{g}[g^{-1}(cp(x))]=0.
\end{equation}
Hence, since $\lim_{x\rightarrow \pm \infty} p(x)=0$ ($x\in \mathcal{D}=\mathbb{R}$), $g^{-1}(0)=0$ (we have assumed $g(0)=0$ in the GRoU), and $u=g^{-1}[cp(x)]$ (see Eq. \eqref{EqBoundaryAg}),
 the limit in Eq. (\ref{limiteancora}) becomes
\begin{equation}
\label{LimitCondEqnose}
\lim_{x\rightarrow \pm \infty} \dot{g}[\underbrace{g^{-1}(cp(x))}_{u\rightarrow 0}]=\lim_{u\rightarrow 0} \dot{g}(u)=\lim_{u\rightarrow 0} \frac{dg}{du} =0. 
\end{equation}
\item  Moreover, since we desire Eqs. \eqref{LimXgg1eq}-\eqref{LimXgg2eq}, it is also necessary that this factor $\dot{g}[g^{-1}(cp(x))]=\dot{g}(u)$, for $u\rightarrow 0$, must decay to zero equal or faster than $1/x\rightarrow 0$ when $x\rightarrow \pm \infty$. 

This condition can be rewritten in other forms. For instance, setting $y=p(x)$ (and assuming now $p(x)$ invertible, for instance, monotonic decreasing) we can rewrite $\dot{g}[g^{-1}(cp(x))]$ as
 \begin{equation}
\label{LimitCondEqnose2}
\dot{g}[g^{-1}(cp(x))]=\left. \frac{dg}{du}\right|_{u=g^{-1}(cp(x))}=\left. \frac{dg}{du}\right|_{u=g^{-1}(cy)}= \frac{1}{ \left.\frac{dg^{-1}}{dy}\right|_{cy}},
\end{equation}
(recall that $c$ is just a constant) hence when $x\rightarrow + \infty$, $y=p(x)\rightarrow 0$, we need that $\frac{dg^{-1}}{dy}\rightarrow \infty$,  for $y\rightarrow 0$, must diverge equal or faster than $x=p^{-1}( y)\rightarrow +\infty$. That is equivalent to assert $\frac{1}{\dot{g}^{-1}(y)}\rightarrow 0$ when $y\rightarrow 0$ equal or  faster than $1/x\rightarrow 0$ for $x\rightarrow +\infty$. Since  $x=p^{-1}(y)$ we can rewrite it as $\frac{1}{\dot{g}^{-1}(y)}\rightarrow 0$ vanishes equal or faster than $\frac{1}{p^{-1}(y)} \rightarrow 0$, both for $y\rightarrow 0$ .
\end{itemize}
\end{enumerate}
\end{enumerate}

\subsection{Summary of conditions}
\label{SectCondRoU}
The region $\mathcal{A}_g$ generated by GRoU is bounded if: 
\begin{enumerate}
\item The function $y=p(x)$ is bounded (i.e., if $p(x)$ is monotonic, $x=p^{-1}(y)$ has finite support).
\item the limit $\lim_{u\rightarrow 0} \frac{dg}{du}=0,$ is verified. Since $u=g^{-1}[cp(x)]$ and we set $y=p(x)$, this limit is equivalent to  
 $\lim_{y\rightarrow 0} \frac{dg^{-1}}{dy}=\infty$,
 as written in Eq. (\ref{LimitCondEqnose2}).
\item The derivative $ \frac{dg}{du} \rightarrow 0$ when $u \rightarrow 0$, has to vanish to zero equal or faster than $1/x\rightarrow 0$ for $x\rightarrow \pm \infty $. Setting $y=p(x)$ and consider a monotonic $p(x)$ (so that we can write $x=p^{-1}(y)$), it is equivalent to assert that $\frac{1}{\dot{g}^{-1}(y)}\rightarrow 0$ vanishes equal or faster than $\frac{1}{p^{-1}(y)} \rightarrow 0$, for $y\rightarrow 0$.
\end{enumerate}
Moreover, we recall that in the GRoU theorem also assumes other conditions over the function $g(u)$:
\begin{enumerate}
\item[4)] $g(u)$ must be increasing,
\item[5)] $g(u):\mathbb{R}^{+}\rightarrow\mathbb{R}^{+}$,
\item[6)] $g(0)=0$.
\end{enumerate}
We will show that these $3$ last conditions can be relaxed. Indeed, they are used to prove the GRoU (see Appendix \ref{App_a_ver}), however they are not conditions needed to obtain a bounded region $\mathcal{A}_g$.

\section{Extended Inverse of density method}
\label{sec:ExtIoD}

The standard inverse-of-density (IoD) method of Section \ref{sec:IoD} provides the relationship between a RV $Y$ distributed as a PDF proportional to $p^{-1}(y)$ and the RV $X$ with a PDF  proportional to $p(x)$.\footnote{Recall that we refer to $p(x)$ and $p^{-1}(y)$ as densities although they are unnormalized.} In this section, we study the connection between a transformed random variable $U=h(Y)$, where $Y$ is distributed according to $p^{-1}(y)$, and the random variable $X$ with PDF $p(x)$.

\begin{figure}[htb]
    \label{App2Fig2}
       \centering 
       \subfigure[]{\includegraphics[width=6.7cm]{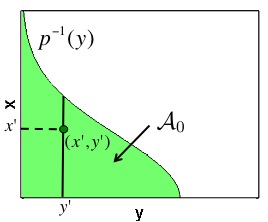}}
       \subfigure[]{\includegraphics[width=6.7cm]{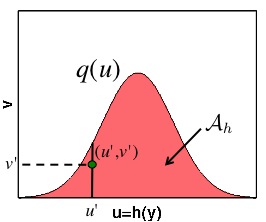}}
          \caption{\textbf{(a)} Given a point $(x',y')$ uniformly distributed on $\mathcal{A}_0$, $y'$ has PDF $p^{-1}(y)$ while  $x'$ is distributed as $p(x)$, as affirmed by fundamental theorem of simulation and the inverse-of-density-method. \textbf{(b)} Given a transformation of RV $U=h(Y)$ with PDF $q(u)$, and a point $(u',v')$ uniformly distributed on the area $\mathcal{A}_h$ below $q(u)$, then the sample $x'=v'\dot{h}(h^{-1}(u'))$ has density $p(x)$.}
\end{figure}

Given a random variable $Y$ with PDF $p^{-1}(y)$ and  $\tilde{U}=h(Y)$, where $h$ is a monotonic function, we know that the density of $\tilde{U}$ is 
\begin{equation}
\label{ancoraBastaEq}
q(\tilde{u})=p^{-1}(h^{-1}(\tilde{u}))\bigg|\frac{d h^{-1}}{d\tilde{u}}\bigg|.
\end{equation}
Denoting as $\mathcal{A}_h$ the area below $q(\tilde{u})$ (see Figure \ref{App2Fig2}(b)), our goal is now to find the relationship between the pair $(U,V)$ uniformly distributed on $\mathcal{A}_h$ and the RV $X$ with density $p(x)$. 

It is important to observe that $(U,V)=(\tilde{U},V)$ since, for the fundamental theorem of simulation (see Section \ref{sec:FToS}),  if $(U,V)$ is uniformly distributed on $\mathcal{A}_h$ then $U$ has pdf $q(\tilde{u})$ so that $U=\tilde{U}$. Hence, for lack of simplicity, in the sequel we use $u$ instead of $\tilde{u}$, and  $U$ instead of $\tilde{U}$.
Obviously, if we are able to draw a sample $u'$ from $q(u)$, we can easily generate a sample $y'$ from $p^{-1}(y)$ as
\begin{equation}
\label{UtoXrelEQ2}
y'=h^{-1}(u').
\end{equation}
Therefore, using the inverse-of-density method, we can obtain a sample $x'$ from $p(x)$ as 
\begin{equation}
\label{EqX1}
x'=z'p^{-1}(y')=z'p^{-1}(h^{-1}(u')),
\end{equation}
where $z'\sim \mathcal{U}([0,1])$, $u'\sim q(u)$ and $y'=h^{-1}(u')$ from Eq. (\ref{UtoXrelEQ2}). Equation \eqref{EqX1} above connects he RV's $U$ and $X$. However, we are looking for a relationship involving also the random variable $V$.

Moreover, we denote as $\mathcal{A}_h$ the region delimited by the curve $v=q(u)$ and the axis $u$. Figure \ref{App2Fig2}(b) illustrates the PDF $q(u)$, the area $\mathcal{A}_h$ and a point $(u',v')$ drawn uniformly from $\mathcal{A}_h$. To draw a point $(u',v')$ uniformly from $\mathcal{A}_h$, we can first draw a sample $u'$ from $q(u)$ and then $v'$ uniformly the interval $[0,q(u')]$, i.e., $v'\sim \mathcal{U}([0,q(u')])$. Therefore, the sample $v'$ can be also expressed as 
\begin{equation}
\label{EqV2}
v'=z'q(u'),
\end{equation}
where $z'\sim \mathcal{U}([0,1])$. Substituting $q(u)$ in Eq. (\ref{ancoraBastaEq}) into Eq. (\ref{EqV2}), we obtain
\begin{equation}
\label{EqV22}
v'=z'\underbrace{p^{-1}(h^{-1}(u'))\bigg|\frac{d h^{-1}}{du}\bigg|_{u'}}_{q(u')}.
\end{equation}
Furthermore, recalling Eq. (\ref{EqX1}) we can see that 
\begin{equation}
\label{EqV222}
v'=\underbrace{z'p^{-1}(h^{-1}(u'))}_{x'}\bigg|\frac{d h^{-1}}{du}\bigg|_{u'}, 
\end{equation}
hence 
\begin{equation}
\label{EqV2222}
v'=x'\bigg|\frac{d h^{-1}}{du}\bigg|_{u'}.
\end{equation}
Then, finally we can also write 
\begin{equation}
\label{EqX2}
x'=\frac{v'}{\big|\frac{d h^{-1}}{du}\big|_{u'}}=v'|\dot{h}(h^{-1}(u'))|,
\end{equation}
that is a sample from $p(x)$. We indicate with $\dot{h}=\frac{dh}{dx}$ the first derivative of $h(x)$. Eq.  \eqref{EqX2} can be also seen as an extension of the {\em fundamental theorem of simulation} (Section \ref{sec:FToS}). 

Figure \ref{App2Fig2}(a) depicts the area $\mathcal{A}_0$ delimited by $p(x)$ and a point $(y',x')$ drawn uniformly from $\mathcal{A}_0$. As explained in Section \ref{sec:IoD}, $x'$ is distributed as $p(x)$ and  $y'$ is distributed as $p^{-1}(y)$. For the standard IoD method we know that $X=Zp^{-1}(Y)$ where $Z\sim \mathcal{U}([0,1])$ and $Y\sim p^{-1}(y)$. 

Moreover, Equation (\ref{EqX2}) connects a uniform random point $(U,V)\in \mathcal{A}_h$, as illustrated in Figure \ref{App2Fig2}(b), and the RV $X$. Therefore, if we are able to draw points $(u',v')$ uniformly from $\mathcal{A}_h$ we can generate sample $x'$ from the density $p(x)$ using Eq. (\ref{EqX2}), as formalized by the following proposition. 
\begin{Proposicion}
\label{Prop1}
Let $Y$ be a RV with a monotonic PDF $p^{-1}(y)$, and let $U=h(Y)$ be another (transformed) RV, where $h(y)$ is a monotonic transformation. Let us denote with $q(u)$ the density of $U$ and let $\mathcal{A}_h$ be the area below $q(u)$. If we are able to draw a point $(u',v')$ uniformly from the region $\mathcal{A}_h$, then  
\[x'=\frac{v'}{\big|\frac{d h^{-1}}{du}\big|_{u'}}=v'|\dot{h}(h^{-1}(u'))|,\]
is a sample from the PDF $p(x)$.  
\end{Proposicion} 
Below, we provide two interesting special cases.
\begin{itemize}
\item Choosing $h(y)=y$ (hence $\dot{h}=1$), we have $U=Y$ and as a consequence $q(u)=q(y)=p^{-1}(y)$ and the region $\mathcal{A}_h$ is exactly $\mathcal{A}_0$, so that Eq. (\ref{EqX2}) becomes 
\begin{equation}
x'=v',
\end{equation}
i.e., we come back to the fundamental theorem of simulation. Indeed, if we are able to draw a point $(u'=x',v'=y')$ uniformly from $\mathcal{A}_0\equiv \mathcal{A}_h$, for the fundamental theorem of simulation, it yields that $x'=u'$ has PDF $p(x)$ while, clearly  $y'=v'$ has distributed as the inverse PDF $p^{-1}(y)$ (consideration used in the standard IoD method).
\item Moreover, if we take $h(y)=\sqrt{2y}$, $y\geq 0$, since $h^{-1}(u)=\frac{1}{2}u^2$, we have
\begin{equation}
x'=\frac{v'}{u'},
\end{equation}
that corresponds to the standard RoU method.
\end{itemize}

\section{Relationship between the GRoU, transformed rejection and IoD methods}
\label{sec:RoUvsTRS}

This section is devoted to expose the following proposition. 
\begin{Proposicion}
\label{RoUasTodo}
The {\it generalized RoU method} can be seen as a combination of the {\it transformed rejection method} applied to random variable $Y$ distributed according to the inverse density $p^{-1}(y)$, described in Section \ref{sec:TRMup}, with the {\it extended inverse-of-density method} explained in Section \ref{sec:ExtIoD}.  
\end{Proposicion}
We first investigate the connection between GRoU and transformed rejection, and then the connection between GRoU and the inverse-of-density.

\subsection{Connection between GRoU and transformed rejection}
Let us recall the region defined by the GRoU in the Eq. (\ref{regionAdefgen1}) (for simplicity in the treatment we set $c=1$)
\begin{equation}
\label{regionAdefgen1_2}
\mathcal{A}_g= \left\{(v,u)\in \mathbb{R}^2: 0\leq u \leq g^{-1}\left[p\left(\frac{v}{\dot{g}(u)}\right)\right]\right\}, 
\end{equation}
where, $p(x)\propto p_0(x)$, $g(u)$ is a {\em increasing} function and $g(0)=0$. 
 
\begin{enumerate}
\item  We consider first for lack of simplicity a monotonic {\em decreasing} bounded target density $y=p(x)\propto p_0(x)$ with an unbounded support $\mathcal{D}=[0,+\infty)$ (hence the mode is at $x=0$). 
Since that $g$ is increasing (then $g^{-1}$ is also increasing) we can write
\[ 
g(u) \leq p\left(\frac{v}{\dot{g}(u)}\right).
\] 
Moreover, recalling that $p(x)=p_{dec}(x)$ is decreasing (hence also $p^{-1}(y)=p_{dec}^{-1}(y)$ is decreasing), we have 
\[ 
p_{dec}^{-1}(g(u)) \geq \frac{v}{\dot{g}(u)},
\] 
and since $\dot{g}(u)\geq 0$ ($g$ is increasing), we obtain
\[ 
v\leq  p_{dec}^{-1}(g(u))\dot{g}(u).
\] 
Finally, since   $x\in[0,+\infty)$, then $x=p^{-1}(y)\geq 0$ and $p^{-1}(g(u))\dot{g}(u)\geq 0$ (we recall $\dot{g}(u)\geq 0$).  Therefore, we can write 
\[ 
0 \leq v\leq  p_{dec}^{-1}(g(u))\dot{g}(u).
\]
Then these trivial calculations lead us to express the set $\mathcal{A}_g$ as
\begin{equation}
\label{secondDefAg}
\mathcal{A}_{g,p_{dec}}= \left\{(v,u)\in \mathbb{R}^2: 0\leq v \leq p_{dec}^{-1}\left(g(u)\right)\dot{g}(u) \right\},
\end{equation}
where $p_{dec}^{-1}(y)$ is the inverse of the target density. 
It is important to remark that the inequalities depend on the sign of the first derivative of $g$ (increasing) and $p$ (decreasing).

\item Similar considerations can be developed for monotonic {\em increasing} PDF $p(x)=p_{inc}(x)$ with an unbounded support $x\in \mathcal{D}=(-\infty,0]$ (i.e., $x\leq 0$).  Indeed, in this case we can rewrite $\mathcal{A}_g$ as
\begin{equation}
\label{secondDefAg_inc}
\mathcal{A}_{g,p_{inc}}= \left\{(v,u)\in \mathbb{R}^2: p_{inc}^{-1}\left(g(u)\right)\dot{g}(u)\leq v \leq 0  \right\},
\end{equation}
Note that since $x\in (-\infty,0]$, i.e. $x\leq 0$, then $x=p_{inc}^{-1}(y)\leq 0$ and $p_{inc}^{-1}\left(g(u)\right)\leq 0$ then finally  $p_{inc}^{-1}\left(g(u)\right)\dot{g}(u)\leq 0$. The inequalities are different because here $p(x)=p_{inc}(x)$ is increasing.

\item Similar arguments can be also extended for non-monotonic PDFs. 
See for instance Figure \ref{figurasRoU2}(b) where we have $p(x)$ is increasing in $\mathcal{D}_1=(-\infty,0]$ and  $p(x)$ is decreasing in $\mathcal{D}_2=[0,+\infty)$, i.e., $p(x)$ is non-monotonic with mode located at $0$.
Moreover, if the mode of the PDF are not located in zero o there are several modes, then more but similar considerations are needed. In Section \ref{SectModenotzero} we discuss these more general cases. 
\end{enumerate}


Consider now an increasing differentiable transformation $u=h(y)$ and consider the random variable $Y$ with a decreasing PDF $p^{-1}(y)$ and the transformed variable $U=h(Y)$ with density $q(u)=p^{-1}(h^{-1}(u))\dot{h}^{-1}(u)$. The region below $q(u)$ is 
\begin{equation}
\label{defAh}
\mathcal{A}_h= \left\{(v,u)\in \mathbb{R}^2: 0\leq v \leq p^{-1}(h^{-1}(u))\dot{h}^{-1}(u) \right\},
\end{equation}
and we can note that Eq. (\ref{secondDefAg}) is equivalent to Eq. (\ref{defAh}) when 
\begin{equation}
y=g(u)=h^{-1}(u). 
\end{equation}
Moreover, clearly, the cases of interest are those in which the region $\mathcal{A}_g$ and $\mathcal{A}_h$ are bounded, as seen in Sections \ref{sec:OurCase} and \ref{sec:GRoU}. Specifically, in Section \ref{sec:OurCase} we have  discussed the properties that a transformation $h(y)$ have to fulfill in order to obtain a bounded region $\mathcal{A}_h$, while in  Section \ref{SectCondRoU} we have described  the conditions to obtain a bounded set  $\mathcal{A}_g$. 

\begin{figure*}[htb]
\centering
\centerline{                         
\subfigure[]{\includegraphics[width=4.6cm]{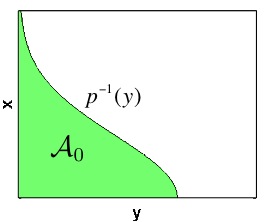}}
\subfigure[]{\includegraphics[width=4.6cm]{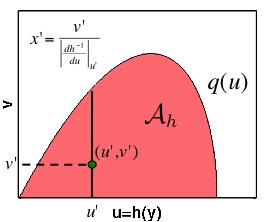}}                         
\subfigure[]{\includegraphics[width=4.6cm]{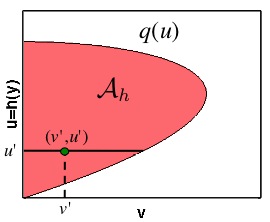}}                          
}
\caption{\textbf{(a)} Example of region $\mathcal{A}_0$ defined by the inverse density $p^{-1}(y)$. \textbf{(b)} The density $q(u)=\left|\frac{dh^{-1}}{du}\right|p^{-1}(h^{-1}(u))$ obtained transforming the RV $Y$, i.e., $U=h(Y)$. Generating uniformly the point $(u',v')$ in the area $\mathcal{A}_h$  we can obtain samples $x'$ from $p(x)$ using Eq. (\ref{RoUcompara}). \textbf{(c)} The region $\mathcal{A}_h$ rotated $90^\circ$ in order to show it how appears when we apply the GRoU technique.} 
\label{figRoUcompara}
\end{figure*}

It is important to remark that these conditions coincides if we choose $g(u)=h^{-1}(u)$ (with $h(0)=0$, see Section \ref{SectModenotzero} and Appendix \ref{App_a_ver2} about this assumption).
Namely, the conditions that the function $g(u)$  in Section \ref{sec:GRoU} must satisfy in order to guarantee the the region $\mathcal{A}_g$ be bounded are exactly the same conditions that have to be imposed on the function $h^{-1}(u)$ of Section \ref{sec:TRMup} in order to apply the transformed rejection method.
Therefore, we can state the following result.
\begin{Proposicion}
\label{RoUTRM}
 The region $\mathcal{A}_g$ can be obtained as a transformation $h=g^{-1}$ of a random variable $Y$ distributed according to the inverse PDF $p^{-1}(y)$. Specifically, given a RV $U=h(Y)=g^{-1}(Y)$ with PDF indicated as $q(u)$, the region $\mathcal{A}_g$ coincides with the area $\mathcal{A}_h$ below the curve $q(u)$.  
\end{Proposicion}
This proposition means that the set $\mathcal{A}_g$ defined by Eq. (\ref{regionAdefgen1}) or  (\ref{secondDefAg}) is obtained by applying the transformed rejection idea for unbounded PDF's to the inverse density $p^{-1}(y)$ (see Section \ref{sec:TRMup}). Figure \ref{figRoUcompara}(b) displays the region $\mathcal{A}_h$ (that coincides with $\mathcal{A}_g$ if $g=h^{-1}$) defined in Eq. (\ref{defAh}). Figure \ref{figRoUcompara}(c) depicts the same region $\mathcal{A}_h$ rotated $90^\circ$. 
Proposition \ref{RoUTRM} can also be deduced as shown in Appendix \ref{App_FantasticoSect}.

Moreover, Proposition \ref{RoUTRM} yields the following corollary about the two marginal densities of the random variables $U$ and $V$ with uniform joint pdf on the region $\mathcal{A}_g$ provided by the GRoU. 
\begin{Corolario}
\label{RoUTRM2}
Consider a random vector $(V,U)$ uniformly distributed on the region $\mathcal{A}_g$ provided by the GRoU. We already know that the RV $X=\frac{V}{\dot{g}(U)}$ has pdf $p(x)$ as proven by the GRoU. Moreover, we can assert that $U$ is distributed as $q(u)=p^{-1}(g(u))\frac{dg}{du}$ (since $U=g^{-1}(Y)$) and $V$ is distributed as the {\it generalized inverse density} $q^{-1}(v)$ of $q(u)$ (see Section \ref{sec:GenPdfs}, for the definition of the generalized inverse pdf).  
\end{Corolario}


\subsection{Connection among GRoU, extended IoD and fundamental theorem}
Moreover, In Section \ref{sec:ExtIoD} we have analyzed the relationship between $X$ with PDF $p(x)$  and the RV $U=h(Y)$, where $Y$ is distributed as $p^{-1}(y)$. Hence, given two samples $v'$ and $u'$ uniformly distributed on the set $\mathcal{A}_h$, the area below the PDF $q(u)$,  we can assert that the sample
\begin{equation}
\label{RoUcompara}
x'=\frac{v'}{\big|\frac{d h^{-1}}{du}\big|_{u'}}=\frac{v'}{\big| \dot{h}^{-1}(u')\big|},
\end{equation}
is distributed as $p_0(x)\propto p(x)$, as we prove in Section \ref{sec:ExtIoD} for the extended IoD and extended fundamental theorem of simulation. Note that, if we set $h(y)=g^{-1}(y)$, we obtain $x'=v'/\dot{g}(u')$ that is exactly equivalent to the GRoU technique in Section \ref{sec:GRoU}. Therefore,  we can also assert the following two propositions.
\begin{Proposicion}
\label{RoUInvofD}
The GRoU extends the underlying idea of the {\em classical inverse-of-density} approach, described in Section \ref{sec:IoD}. Indeed, the classical IoD method uses a random variable $Y$ distributed as the inverse density $p^{-1}(y)$ to draw samples from $p(x)$, whereas the GRoU uses a transformation of the random variable $Y$, $U=g^{-1}(Y)$, to generate samples from $p(x)$.
\end{Proposicion}
\begin{Proposicion}
\label{RoUInvofD2}
The GRoU can be also seen as an extension of the {\em fundamental theorem of simulation}, described in Section \ref{sec:FToS}. Indeed, the fundamental theorem links the coordinates of a random point $(X,Y)\in \mathcal{A}_0$ with the PDFs $p(x)$ and $p^{-1}(y)$, i.e., $X\sim p(x)$, $Y\sim p^{-1}(y)$, whereas GRoU links the coordinates  a random point $(U,V)\in \mathcal{A}_g$ with the same PDFs $p(x)$ and $p^{-1}(y)$, i.e., $\frac{V}{\dot{g}(U)}\sim p(x)$, $g(U)\sim p^{-1}(y)$. 
\end{Proposicion}

Clearly, Propositions \ref{RoUTRM}, \ref{RoUInvofD} and \ref{RoUInvofD2} entail the Proposition \ref{RoUasTodo}.
Note that the GRoU can also be extended for  a decreasing $g(u)$, as we will show in Appendix \ref{App_DecrG} (see also Appendix \ref{App_Increible}). In this case, we have $x=-\frac{v}{\dot{g}(u)}$ then in general for the GRoU we can write
\[
x=\frac{v}{\big| \dot{g}(u)\big|},
\]
exactly as in Eq. (\ref{RoUcompara}).


\subsection{Function $g(u)$ to obtain a rectangular region $\mathcal{A}_g$ and first formulation of the IoD}
\label{OptimalGbutnonoptimal}

Clearly, the easiest case to perform {\em exact} sampling with GRoU is that $\mathcal{A}_g$ be a rectangular region\footnote{Clearly it is just one possibility, there are other situations where we can perform exact sampling (for instance, if $\mathcal{A}_g$ is a circle or a triangle).}.
The considerations in Section \ref{sec:RoUvsTRS} are very useful to clarify which $g(u)$ produces a rectangular region $\mathcal{A}_g$.
More specifically, Proposition \ref{RoUasTodo} allows us to infer which is the optimal (theoretical) choice of the function $g(u)$. 

Indeed, since the GRoU is a transformation of a RV $Y$ with pdf $p^{-1}(y)$ (considering, for instance, a decreasing $p(x)$), specifically $U=g^{-1}(Y)$ with $c=1$, the well-known {\it inversion method} \citep{Devroye86} asserts that if the function $g^{-1}(y)$ is the cumulative distribution function (CDF)\footnote{The CDF $F_{Y}(y)$ of RV $Y$ can be easily expressed as function of $F_{X}(x)$ (the CDF of $X$) for monotonic decreasing target pdfs $p_0(x) \propto p(x)$ as we show in Appendix \ref{App_Fy_Fx}.} of $Y$ then the transformation produces a uniform RV $U$. Hence, the set $\mathcal{A}_g$  is a rectangular region if we use 
\begin{gather}
\begin{split}
\label{GoptimalEq}
g^{-1}(y)=F_{Y}(y) \Rightarrow g(u)=F_{Y}^{-1}(u), 
\end{split}
\end{gather}
where $F_{Y}(y)$ is the CDF of RV $Y$, i.e.,
\begin{equation}
F_Y(y)=\int_{-\infty}^{y} p^{-1}(y)dy.
\end{equation} 
Since $p^{-1}(y)$ is unnormalized, note that $F_Y(y)\rightarrow 1/K$ with $y\rightarrow +\infty$ (instead of $F_Y(y)\rightarrow 1$) where
$$\frac{1}{K}=\int_{\mathcal{D}_Y} p^{-1}(y)dy=\int_{\mathcal{D}_X} p(x)dx.$$
Therefore, if $g^{-1}(y)=F_{Y}(y)$ then $U=F_{Y}(Y)$ is a uniform RV in $[0,1/K]$, and $\mathcal{A}_g$ is a rectangle $0\leq u\leq \frac{1}{K}$ and $0 \leq v \leq 1$ as we show below in Eq. (\ref{regionAdefNosecuantas3}).
Indeed, when $g^{-1}(y)=F_{Y}(y)$, the region $\mathcal{A}_g$ is defined ($c=1$) as  
\begin{gather}
\begin{split}
\label{regionAdefNosecuantas}
\mathcal{A}_g= \left\{(v,u)\in \mathbb{R}^2: 0\leq u \leq F_Y\left[p\left(\frac{v}{\dot{F}_Y^{-1}(u)}\right)\right]\right\}, 
\end{split}
\end{gather}
and since $\dot{F}_Y^{-1}(u)=\frac{1}{\dot{F}_Y(F_Y^{-1}(u))}=\frac{1}{p^{-1}(F_Y^{-1}(u))}$, 
\begin{gather}
\begin{split}
\label{regionAdefNosecuantas2}
\mathcal{A}_g= \left\{(v,u)\in \mathbb{R}^2: 0\leq u \leq F_Y\left[p\Big(vp^{-1}(F_Y^{-1}(u) ) \Big)\right]\right\},
\end{split}
\end{gather}
and $x=vp^{-1}(F_Y^{-1}(u))$ is distributed as $p_0(x)\propto p(x)$.  Since $0 \leq F_Y(y) \leq 1/K$, then the suitable values of the variable $u$ are contained in $[0,1/K]$.
The variable $v$ is contained in $[0,1]$ independently of the values of $u$, because inverting the inequalities in Eq. (\ref{regionAdefNosecuantas2}) we obtain
$$ 0\leq v \leq \frac{p^{-1}(F_Y^{-1}(u))}{p^{-1}(F_Y^{-1}(u))}=1,$$
so that $\mathcal{A}_g$ is completely described by the inequalities 
\begin{gather}
\begin{split}
\label{regionAdefNosecuantas3}
\mathcal{A}_g= \left\{(v,u)\in \mathbb{R}^2:   \mbox{  } 0\leq v \leq 1, \mbox{  }\mbox{  } 0\leq u \leq \frac{1}{K} \right\}.
\end{split}
\end{gather}
Moreover, observe that the expression $x=vp^{-1}(F_Y^{-1}(u))$ is exactly the same of Eq. (\ref{eq:IoD1}) obtained with the first formulation of IoD method. 
Indeed, in this case $\mathcal{A}_g$ is a rectangle $(v,u)\in [0,1]\times[0,1/K]$  then $v\sim\mathcal{U}([0,1])$ and the sample $y=g(u)=F_Y^{-1}(u)$ is distributed as $p^{-1}(y)$, so that  $x=vp^{-1}(y)$ is equivalent to Eq. (\ref{eq:IoD1}).  
\subsection{The second formulation of the IoD (Khintchine's theorem) as special case of the GRoU }
\label{SecFormasSpecialCase}

Here we show that the second formulation of the inverse of density method described at the end of Section \ref{sec:IoD} is contained by the generalized ratio of uniforms technique.
 Assuming an increasing target PDF $p(x)$,  if we set $g(u)=p(u)$, namely we use as function $g$ exactly our target PDF $p$,  and $c=1$ then
\begin{equation}
\mathcal{A}_{g=p}= \left\{(v,u)\in \mathbb{R}^2: 0\leq u \leq p^{-1}\left[p\left(\frac{v}{\dot{p}(u)}\right)\right]\right\}, 
\end{equation}
\begin{equation}
\mathcal{A}_{g=p}= \left\{(v,u)\in \mathbb{R}^2: 0\leq u \leq \frac{v}{\dot{p}(u)}\right\}, 
\end{equation}
or using the alternative definition of the area $\mathcal{A}_g$ in Eq. (\ref{secondDefAg}), we have
\begin{equation}
\label{InvasSpecialCase}
\mathcal{A}_{g=p}= \left\{(v,u)\in \mathbb{R}^2: 0\leq v \leq u\dot{p}(u)\right\}.
\end{equation}
Hence, the region $\mathcal{A}_{g=p}$ represents the area below the vertical density (see Eq. \ref{eq:vertPDF}) \citep{Jones02,Khintchine38, Troutt04} corresponding to $p^{-1}(y)$ used in the second formulation of the inverse of density method (see  Section \ref{sec:IoD}). The ratio of uniforms approach assert that 
\begin{equation}
\label{miracoloITA}
x'=\frac{v'}{\dot{p}(u')},
\end{equation}
is a sample from $p(x)$ if $u'$ is distributed as the vertical density $u\dot{p}(u)$ and $v'\sim\mathcal{U}([0,u'\dot{p}(u')])$. Now, we want to certify if this statement is also true using the second formulation of the inverse of density technique.  

Consider a sample $z'\sim\mathcal{U}([0,1])$ then $z'u'\dot{p}(u')\sim\mathcal{U}([0,u'\dot{p}(u')])$, hence we can write 
\begin{equation}
v'=z'u'\dot{p}(u').
\end{equation}
Replacing the relationship above in Eq. (\ref{miracoloITA}) we obtain 
\begin{equation}
\label{miracoloITA2}
x'=z'w',
\end{equation} 
where $z'\sim\mathcal{U}([0,1])$ and $w'$ is drawn from $u\dot{p}(u)$. Note that Equations (\ref{SecondFormIoD}) and (\ref{miracoloITA2}) coincide since, $u\dot{p}(u)$ is exactly the vertical density of $p^{-1}(y)$ in Eq. (\ref{eq:vertPDF}).
Therefore, we can assert that the second formulation of the standard inverse of density in Section \ref{sec:IoD} can be found choosing $g(u)=p(u)$ in the GRoU (where $p(u)$ is our target PDF).

\begin{figure*}[htb]
\centering
\centerline{                         
\includegraphics[width=11cm]{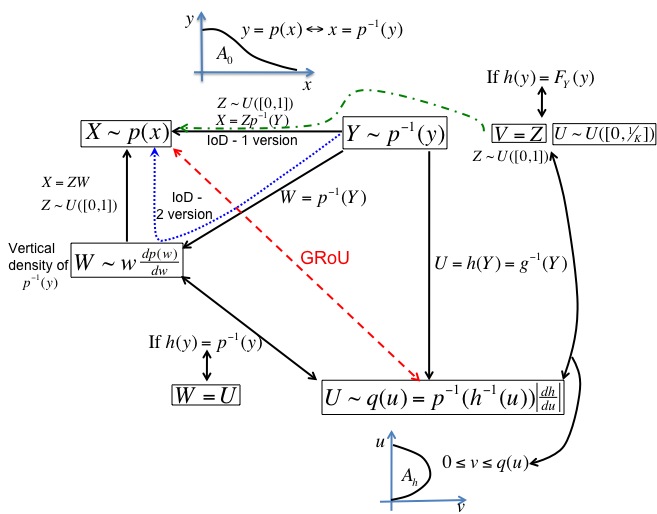}
}
\caption{Relationships among densities, random variables and sampling methods for a decreasing target PDF $y=p(x)$. Dashed line shows the connection produced by the GRoU algorithm whereas dotted line describes the second formulation of the inverse of density (IoD) method. The random variable $W\sim w\frac{dp}{dw}$ is the vertical density corresponding to the pdf $p^{-1}(y)$.  Finally, dashed-dotted line depicts the relationship between the first formulation of the IoD and GRoU when $h(y)=g^{-1}(y)=F_{Y}(y)$ (see Section \ref{OptimalGbutnonoptimal}).} 
\label{figResumen}
\end{figure*}

 Figure \ref{figResumen} summarizes the relationships among densities, random variables and sampling methods (the two versions of the IoD, the VDR and the GRoU) for a decreasing target PDF $p(x)$. 

\subsection{Effect of the constant $c$}
\label{Effect_C}
So far, for simplicity we have set $c=1$. However, all the previous considerations and remarks remain valid. Indeed, assuming a decreasing $p(x)=p_{dec}(x)$ and $c>0$, for instance Eq. \eqref{secondDefAg} becomes    
 \begin{equation}
\label{secondDefAgConC}
\mathcal{A}_{g,p_{dec}}= \left\{(v,u)\in \mathbb{R}^2: 0\leq v \leq  p_{dec}^{-1}\left(\frac{g(u)}{c}\right)\dot{g}(u) \right\}.
\end{equation}
Since, as we show in Section \ref{sec:background}, all this techniques work with unnormalized PDF, we can also multiply both inequalities for a positive constant $1/c$ obtaining    
 \begin{gather}
 \label{secondDefAgConC2}
 \begin{split}
\mathcal{A}_{g,p_{dec}}= \left\{(v,u)\in \mathbb{R}^2: 0\leq v \leq  \frac{1}{c}p_{dec}^{-1}\left(\frac{g(u)}{c}\right)\dot{g}(u) \right\}, \\
\mathcal{A}_{g,p_{dec}}= \left\{(v,u)\in \mathbb{R}^2: 0\leq v \leq  p_{dec}^{-1}\left(\frac{g(u)}{c}\right) \frac{\dot{g}(u)}{c} \right\}, \\
\end{split}
\end{gather}
then $\mathcal{A}_{g,p_{dec}}$ represents the area below $q(u)=p_{dec}^{-1}\left(\frac{g(u)}{c}\right) \frac{\dot{g}(u)}{c}$ that is the (unnormalized) PDF of the RV $U= g^{-1}(cY)$ where $Y$ is distributed according to (unnormalized) PDF $p_{dec}^{-1}(y)$. See also Appendix \ref{App_FantasticoSect}.


\section{Further Considerations}
\label{sec:further}

In this section, we provide further observations about the connection among the GRoU and the transformed rejection sampling, and about some assumptions over the function $g(u)$.    

\subsection{Minimal rectangular region}
Here, we show that the minimal rectangle $\mathcal{R}_g$ such that $\mathcal{A}_g\subseteq  \mathcal{R}_g$ is equivalent to a rectangular region $\mathcal{R}_h$ embedding a set $\mathcal{A}_h$ obtained with a trasformation of a random variable $Y$ with density $p^{-1}(y)$.

We have seen that the minimal rectangular region embedding the region $\mathcal{A}_g$ ($\mathcal{A}_g\subseteq  \mathcal{R}_g$) of the GRoU is defined as
\begin{gather}
\begin{split}
\mathcal{R}_g=\Big\{(v,u)\in \mathbb{R}^2: \mbox{ }&0 \leq u \leq \sup_x g^{-1}[cp(x)],  \\
&\inf_x x\dot{g}[g^{-1}(cp(x))]\leq  v \leq  \sup_x x\dot{g}[g^{-1}(cp(x))]\Big\} .
\end{split}
\end{gather}
For lack of simplicity,  in the following we consider a bounded decreasing PDF $p(x)$ defined for all $x \in \mathbb{R}^+$  with mode localized at $x=0$. 
Recalling also that $g$ is a positive increasing function, the first important observation is that $x\dot{g}[g^{-1}(cp(x))]$ is also positive so that 
$$\inf_{x\in \mathbb{R}^+} x\dot{g}[g^{-1}(cp(x))]=0.$$
Moreover, since $g$ is increasing then also $g^{-1}$ is increasing, a second observation is that
$$\sup_{x\in \mathbb{R}^+} g^{-1}(cp(x))=g^{-1}(cp(0)),$$
where $x=0$ is location of the mode of $p(x)$ (namely $p(0)=\sup_{x\in \mathbb{R}^+} p(x)$). 
Therefore for a bounded $p(x)$ defined in $\mathbb{R}^+$ and a mode at $0$, we can rewrite $\mathcal{R}_g$ as 
\begin{gather}
\label{EqRect_g_otraVez}
\begin{split}
\mathcal{R}_g=\Big\{(v,u)\in \mathbb{R}^2: \mbox{ }&0 \leq u \leq \sup_x g^{-1}[cp(0)],  \\
&0 \leq  v \leq  \sup_x x\dot{g}[g^{-1}(cp(x))]\Big\} .
\end{split}
\end{gather}

Now, let us consider a increasing transformation $h(y)$ and a random variable $Y$ distributed according to the inverse PDF $p^{-1}(y)$. Note that, since we assume a  $p(x)$ is decreasing, bounded with mode at $x=0$, $p^{-1}(y)$ has bounded domain $0 \leq y \leq p(0)$ but it is unbounded with a vertical asymptote at  $0$. 
Then, we consider the RV $U=h(Y)$ with PDF 
\[
q(u)=p^{-1}(h^{-1}(u))\frac{dh^{-1}(u)}{du}, \mbox{ }\mbox{ }\mbox{ with }\mbox{ }\mbox{ } h(0) \leq u \leq h(p(0))
\]
and indicate with $\mathcal{A}_h$ the area below $q(u)$. We also assume that $h(y)$ is chosen adequately such that $\mathcal{A}_h$ is bounded. In this case, a  minimal rectangle $\mathcal{R}_h$ embedding  $\mathcal{A}_h$ exists and clearly it is 
\begin{gather}
\begin{split}
\mathcal{R}_h=\Bigg\{(v,u)\in \mathbb{R}^2: \mbox{ }& h(0) \leq u \leq h(p(0)),  \\
& 0 \leq  v \leq  \sup_u p^{-1}(h^{-1}(u))\frac{dh^{-1}(u)}{du} \Bigg\}.
\end{split}
\end{gather}
Now, we desire to express $q(u)$ first as a function of $y$, obtaining $q(y)$, and later as a function of $x$, obtaining $q(x)$. Recall that $Y$ a RV with PDF $p^{-1}(y)$ and $U=h(Y)$. Then, we have $u=h(y)$ and $y=h^{-1}(u)$ and we can write 
\[
q(y)= p^{-1}(y)\frac{dh^{-1}(h(y))}{du}, \mbox{ }\mbox{ }\mbox{ with }\mbox{ }\mbox{ } 0 \leq y \leq p(0).
\]
Moreover, since $x=p^{-1}(y)$ and $y=p(x)$ we can also write
\[
q(x)= x\cdot \frac{dh^{-1}(h(p(x)))}{du}, \mbox{ }\mbox{ }\mbox{ with }\mbox{ }\mbox{ } 0\leq x \leq p^{-1}(0) \rightarrow +\infty,
\]
i.e.,
\[
q(x)= x\cdot {\dot h}^{-1}[h(p(x))], \mbox{ }\mbox{ }\mbox{ with }\mbox{ }\mbox{ } 0\leq x \leq  +\infty.
\]
Then, we can rewrite the minimal rectangle $\mathcal{R}_h$ as
\begin{gather}
\begin{split}
\mathcal{R}_h=\Big\{(v,u)\in \mathbb{R}^2: \mbox{ }& h(0) \leq u \leq h(p(0)),  \\
&0 \leq  v \leq  \sup_x x\cdot {\dot h}^{-1}[h(p(x))], \Big\} .
\end{split}
\end{gather}
and if we choose  $h=g^{-1}$ with $h(0)=0$ and $c=1$  then
\begin{gather}
\label{EqR_h}
\begin{split}
\mathcal{R}_h=\Big\{(v,u)\in \mathbb{R}^2: \mbox{ }& 0 \leq u \leq g^{-1}(p(0)),  \\
&0 \leq  v \leq  \sup_x x\cdot {\dot g}[g^{-1}(p(x))], \Big\} .
\end{split}
\end{gather}
Note that $\mathcal{R}_h$ in Eq. (\ref{EqR_h}) is exactly the same rectangle $\mathcal{R}_g$ in Eq. (\ref{EqRect_g_otraVez}) when $a=0$ and $c=1$.

\subsection{About the condition $g(0)=0$}
\label{Sectveryimp}

One assumption of the GRoU is that $g(0)=0$: is it strictly necessary? can this condition be relaxed? 
We can disclose that the condition $g(0)=0$ is needed with the version of the GRoU that we have tackled so far, in Section \ref{sec:GRoU}. 
However, note that it is possible to propose different versions GRoU as we show in the Appendices \ref{App_Increible}, \ref{App_a_ver2} and  \ref{App_DecrG}.  

To relax the assumption $g(0)=0$ we can study two different possibilities: $g(0)=c\neq 0$ and $g(b)=0$.
 
\begin{figure*}[htb]
\centering
\centerline{                         
\subfigure[]{\includegraphics[width=4.5cm]{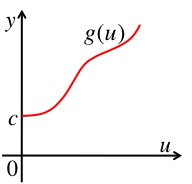}}
\subfigure[]{\includegraphics[width=4.5cm]{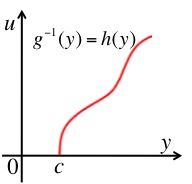}}
\subfigure[]{\includegraphics[width=4.5cm]{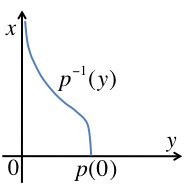}}                                    
}
\caption{Explication why $g(0)=c \neq 0$ is not possible in the GRoU. \textbf{(a)} Example of (increasing) function $g(u)$ with $g(0)=c \neq 0$. \textbf{(b)} The inverse function $g^{-1}(y)=h(y)$ corresponding to the $g(u)$ in Figure (a). \textbf{(c)} Example of inverse PDF $p^{-1}(y)$.} 
\label{figSect7folliapura}
\end{figure*}

The condition $g(0)=c \neq 0$, depicted in Figure \ref{figSect7folliapura}(a), is impossible (at least, in the classical formulation of the GRoU of Section \ref{sec:GRoU}). Indeed, we have that
\begin{itemize}
\item in the standard GRoU the function $g(u)$ has to be increasing,  
\item and we know that the transformation $h(y)=g^{-1}(y)$ is applied to a RV $Y$ with PDF $p^{-1}(y)$ that is defined in $(0,p(0)]$.  
\end{itemize}
In this situation, the inverse function $h(y)=g^{-1}(y)$ is shown in Figure \ref{figSect7folliapura}(b). We know that this transformation $U=h(Y)$ has to be apply to a RV $Y$ with PDF $p^{-1}(y)$ of type in Figure \ref{figSect7folliapura}(c). Then, the RV $Y$ takes values in $(0,p(0)]$. Therefore, if $c\neq 0$ in the interval $(0,c]$ $h(y)$ is not defined but $Y$ can take values there, then the transformation $U=h(Y)=g^{-1}(Y)$ is not possible.


However, the second case $g(b)=0$ with $b\neq 0$ is possible with a slight extension of the GRoU that we show in the Appendix \ref{App_a_ver2}.
Indeed, in this case we have $g(u): [b,+\infty)\rightarrow \mathbb{R}^+$ and then $h(0)=g^{-1}(0)=b$.

\begin{figure*}[htb]
\centering
\centerline{                         
\subfigure[]{\includegraphics[width=4.5cm]{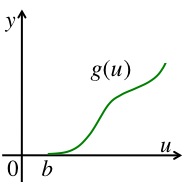}}                                    
\subfigure[]{\includegraphics[width=4.5cm]{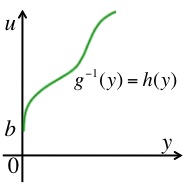}}                                    
\subfigure[]{\includegraphics[width=6.cm]{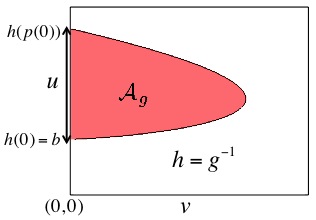}}                                     
}
\caption{Explication why $g(b)=0$ with $b\neq 0$ is possible in an extended version of the GRoU (see Appendix \ref{App_a_ver2}).  \textbf{(a)}  Example of (increasing) function $g(u)$ with $g(b)=0$ with $b \neq 0$. \textbf{(b)} The inverse function $g^{-1}(y)=h(y)$ corresponding to the $g(u)$ in Figure (a). \textbf{(c)} Example of region $\mathcal{A}_g$ when $g(b)=0$ with $b\neq0$.} 
\label{figGcondotro}
\end{figure*}

Figure \ref{figGcondotro}(a) illustrates an example of function $g(u)$ with $g(b)=0$ and $b\neq 0$. Figure \ref{figGcondotro}(b) shows the corresponding inverse function $g^{-1}(y)=h(y)$. Finally, Figure \ref{figGcondotro}(c) depicts the area $\mathcal{A}_g$  this case when $g(b)=0$ then $h(0)=b$.

Another way to understand this issue is the following: in the definition of $\mathcal{A}_g$ we need to combine $g^{-1}$ and $cp(x)$, i.e.,
\[g^{-1}\circ cp(x)=g^{-1}[cp(x)].\]
Since $p(x)\in [0,M]$ where $M=\sup_{x\in\mathbb{R}} p(x)$ ($M=p(0)$, we are assuming the mode is localized at $0$) then $g^{-1}$ has to be defined in $[0,M]$.  




\subsection{Constant $c$ and image of $g(u)$}
\label{SectConstCcazzo}
The sign of the constant $c$ is related to the domain of $g^{-1}(y)$, namely, the image of $g(u)$. 
Indeed, if $c>0$, it is straightforward to see the $g^{-1}(y)$ must be defined in $\mathbb{R}^+$ since we have the composition of functions $g^{-1}\circ cp(x)$ where $p(x)\geq 0$. 
Indeed, since we have assumed $c>0$, so far we have considered always functions $g(u): \mathbb{R} \rightarrow \mathbb{R}^{+}$, i.e., the domain of $g^{-1}$ is $ \mathbb{R}^{+}$.

Moreover, as we have seen in Section \ref{Effect_C}, for general values of $c$, the GRoU is equivalent to the transformation of RVs $U=g^{-1}(c Y)$ where $Y$ has PDF $p^{-1}(y)$, then another time we can deduce that the RV $cY$ must take values into the domain of $g^{-1}(y)$.
Therefore, if we consider a function $g(u): \mathbb{R} \rightarrow \mathbb{R}^{-}$ then we must use a negative $c$, i.e., $c<0$.

\section{General PDFs}
\label{sec:GenPdfs}

In this section, we investigate the connection between GRoU and Inverse-of-density ( Khintchine's theorem) for generic densities. 

\subsection{Inverse-of-density (and Khintchine's theorem) for generic PDFs}
\label{sec:GenericINVofDEN}
Before to analyze the GRoU applied for generic non-monotonic PDFs, first of all we discuss and recall how it is possible apply the inverse-of-density approach in Section \ref{sec:IoD} for generic PDFs.  
Let us define the set of points
\begin{equation}
\mathcal{A}_{0|y}=\{(x,z)\in\mathcal{A}_0, \mbox{ } \mbox{ } z=y\},
\end{equation}
i.e.,  all the points in $\mathcal{A}_0$ such that $z=y$ for all $y\in \mathbb{R}^+$.  Then we can define the {\it generalized} inverse PDF as
\begin{equation}
p_{G}^{-1}(y)=|\mathcal{A}_{0|y}|,
\end{equation}
where $|\mathcal{A}_{0|y}|$ is the Lebesgue measure of $\mathcal{A}_{0|y}$.

Then the inverse-of-density approach (and all the extended versions of Khintchine's theorem \citep{Bryson82,Devroye84,DeSilva78,Chaubey10,Olshen70,Shepp62}) can be summarized in this way: we can draw samples from $p(x)$ if we are able  
\begin{itemize}
\item to generate a sample $y'$ from $p_{G}^{-1}(y)$, 
\item and then generate uniformly a point $(x',y')$ on $\mathcal{A}_{0|y'}$.  Then $x'$ is distributed according to $p(x)$.
\end{itemize}
Note that this approach is strictly related to the {\it slice sampling} algorithm. Clearly, this general approach can be expressed in different ways in different specific cases (as symmetric unimodal PDF with mode at $0$ \citep{Shepp62}), yielding different versions of Khintchine's theorem \citep{Chaubey10,Shepp62}.

\begin{figure*}[htb]
\centering
\centerline{                         
\subfigure[]{\includegraphics[width=5cm]{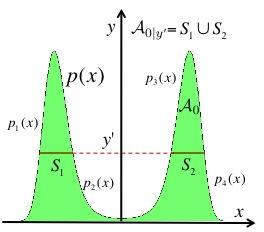}}                                    
\subfigure[]{\includegraphics[width=5cm]{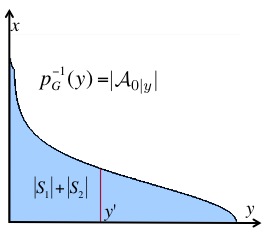}}                                    
}
\caption{{\bf (a)} A bimodal PDF $p(x)$. Monotonic parts of $p(x)$ are indicated by $p_i(x)$, $i=1,...,4$.  {\bf (b)} The corresponding generalized inverse PDF $p_{G}^{-1}(y)$.} 
\label{figinvrsepdfgeneric}
\end{figure*}

It is interesting to observe that (a) is monotone non-increasing \citep{Damien01,Jones02}, (b) it has an vertical asymptote at $0$ (if the domain of $p(x)$ is unbounded) and minimum at $\sup_{x} p(x)$.  Figure \ref{figinvrsepdfgeneric} shows an example of bimodal PDF and the corresponding generalized inverse PDF $p_{G}^{-1}(y)$. Observe that, for instance, in case the set $\mathcal{A}_{0|y}$ can be formed by two disjoint segments (as in Figure \ref{figinvrsepdfgeneric}(a), $S_1$ and $S_2$) or just one depending on the value of $y$.  Clearly, the length of the sets $S_1$ and $S_2$ depend on the $4$ monotonic pieces $p_i(x)$, $i=1,...,4$, that form $p(x)$.  


\subsection{GRoU for unimodal PDF with mode at $0$}
In Section \ref{sec:RoUvsTRS} we have already seen  the definition of $\mathcal{A}_g$ when $p(x)$ is increasing or decreasing with mode at $0$. If the target PDF $p(x)$ is unimodal with mode at $0$, we can divide the domain $\mathcal{D}=\mathcal{D}_1\cup \mathcal{D}_2$ where $\mathcal{D}_1=[0,+\infty)$ with $p(x)=p_{dec}(x)$ is decreasing and $\mathcal{D}_2=(-\infty,0]$ with $p(x)=p_{inc}(x)$ is increasing.   
Hence, in this case, the complete set $\mathcal{A}_g$ can be also written as (combining Eq. \ref{secondDefAg} and Eq. \ref{secondDefAg_inc})
\begin{equation}
\mathcal{A}_{g}= \left\{(v,u)\in \mathbb{R}^2: p_{inc}^{-1}\left(g(u)\right)\dot{g}(u)\leq v \leq p_{dec}^{-1}\left(g(u)\right)\dot{g}(u)  \right\}.
\end{equation}
Note that the inequalities depend on the sign of the first derivative of $p(x)$ (i.e., where $p(x)$ is increasing or decreasing). Then, we could interpret it as if the GRoU applies a transformation $U=g^{-1}(Y)$ over two random variables, $Y_1$ with PDF $p_{dec}^{-1}(y)$ and $Y_2$ with PDF $-p_{inc}^{-1}(y)$. 

\begin{figure*}[htb]
\centering
\centerline{                         
\subfigure[]{\includegraphics[width=5cm]{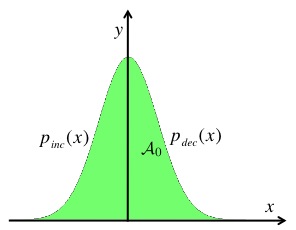}}                                   
\subfigure[]{\includegraphics[width=4.7cm]{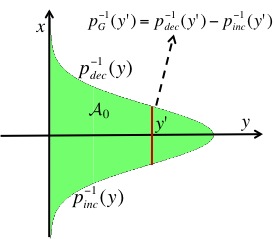}}                                    
\subfigure[]{\includegraphics[width=6cm]{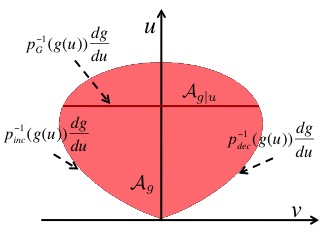}}                                    
}
\caption{{\bf (a)} A unimodal PDF $p(x)$ with mode localized at zero. {\bf (b)} The region $\mathcal{A}_0$ rotated $90^\circ$. In this case, the generalized inverse PDF $p_G^{-1}(y)$ can be written as $p_G^{-1}(y)=p^{-1}_{dec}(y)-p^{-1}_{inc}(y)$. If the target PDF $p(x)$ is also symmetric we have  $p^{-1}_{dec}(y)=-p^{-1}_{inc}(y)$, hence $p_G^{-1}(y)=2p^{-1}_{dec}(y)$. {\bf (c)} The corresponding region $\mathcal{A}_g= \left\{(v,u)\in \mathbb{R}^2: p_{inc}^{-1}\left(g(u)\right)\dot{g}(u)\leq v \leq p_{dec}^{-1}\left(g(u)\right)\dot{g}(u)  \right\}$ obtained by the GRoU using $g(u)=u^2/2$.} 
\label{figUnimodalpdfMode0}
\end{figure*}

Figure \ref{figUnimodalpdfMode0}(a) shows an example of unimodal PDF with mode localized at zero.  Figure \ref{figUnimodalpdfMode0}(b) illustrates the same region  $\mathcal{A}_0$ rotated $90^\circ$ (i.e., switching the axes $x$ and $y$). It is possible to figure out that $p_G^{-1}(y)=p^{-1}_{dec}(y)-p^{-1}_{inc}(y)$ where $p_G^{-1}(y)$ is the generalized inverse density associated to $p(x)$. Finally, Figure \ref{figUnimodalpdfMode0}(c) depicts the corresponding region $\mathcal{A}_g$ using $g(u)=u^2/2$. We can also observe that for a given value $u$ and defining the subset $\mathcal{A}_{g|u}\subset \mathcal{A}_{g}$
\begin{equation}
\mathcal{A}_{g|u}\dfn \{(v,z)\in\mathcal{A}_g, z=u\},
\end{equation}
then we can write the expression 
\begin{equation}
p_G^{-1}(g(u)) \frac{dg}{du}=|\mathcal{A}_{g|u}|,
\end{equation}
where in the first side we have the PDF of a transformed RV $U=g^{-1}(Y)$ where $Y$ is distributed as $p_G^{-1}(y)$ and $|\mathcal{A}_{g|u}|$ is the Lebesgue measure of the subset $\mathcal{A}_{g|u}$. 

\subsection{Unimodal PDF with mode at $a\neq0$}
\label{SectModenotzero}
\begin{figure*}[htb]
\centering
\centerline{                         
\subfigure[]{\includegraphics[width=5cm]{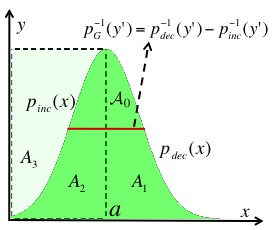}}                                   
\subfigure[]{\includegraphics[width=5cm]{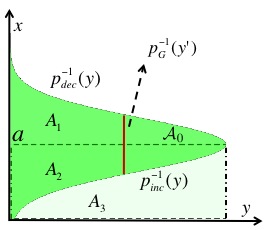}}                                    
}
\centerline{ 
\subfigure[]{\includegraphics[width=5cm]{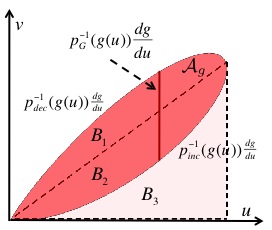}}                                   
\subfigure[]{\includegraphics[width=5cm]{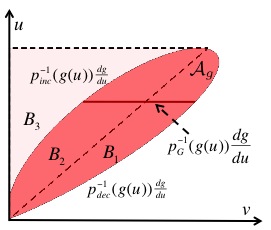}}                                    
}
\caption{{\bf (a)} An example of unimodal density $p(x)$. {\bf (b)} The region $\mathcal{A}_0$ represented switching the axes $x-y$. {\bf (c)} The region $\mathcal{A}_g$ obtained with the GRoU technique with $g(u)=u^2/2$. {\bf (d)} The same region $\mathcal{A}_g$ represented switching the axes $u-v$ in the previous picture (it is typical representation of the GRoU regions).} 
\label{figBnoterminanunca}
\end{figure*}

In this section, we consider the application of GRoU method to a unimodal PDF $p(x)$ ($x\in\mathbb{R}^+$ without loss of generality) with mode at $a\neq0$. We can see an example in Figure \ref{figBnoterminanunca}(a).

In Figure \ref{figBnoterminanunca}(b) is depicted the region $\mathcal{A}_0$ below $p(x)$ with the axis $x-y$ switched (as rotated $90^\circ$). In this case the region $\mathcal{A}_0$ can be described as 
\begin{equation}
\label{Eq:A0defotra}
\mathcal{A}_0=\{(x,y)\in\mathbb{R}^2: p_{inc}^{-1}(y)\leq x \leq p_{dec}^{-1}(y)\}.
\end{equation}
Moreover, observing Figure \ref{figBnoterminanunca}(b) we can individuate and define $5$ random variables: a RV $Y_1$ with PDF $p^{-1}_{dec}(y)-a$ (associated to the region $A_1$),  RV $Y_2$ with PDF $a-p^{-1}_{inc}(y)$ (associated to the region $A_2$), $Y_3$ with PDF $p^{-1}_{inc}(y)$ (associated to the region $A_2$), $Y_4$ with PDF $p_{dec}^{-1}(y)$ (associated to the regions $A_1$, $A_2$ and $A_2$) and finally $Y_5$ with PDF the generalized inverse density $p_{G}^{-1}(y)=p^{-1}_{dec}(y)-p^{-1}_{inc}(y)$. 
Note that $\mathcal{A}_0$ is only composed by $A_1$ and $A_2$, i.e., $\mathcal{A}_0=A_1\cup A_2$.\footnote{If we desire to draw uniformly on $\mathcal{A}_0$ defined as in Figure \ref{figBnoterminanunca}(b), we should to be able to simulate a RV $Y_5$ with PDF $p_{G}^{-1}(y)$. Indeed,  to do it we could simulate a r.v. $Y_4$, i.e., generate a sample $y'$ according to a PDF proportional to  $p^{-1}_{dec}(y)$, then draw $u'\sim \mathcal{U}([0,1])$, finally calculate $x'=u' p^{-1}_{dec}(y')$ and accept $x'$ if $x' \geq p^{-1}_{inc}(y')$ (hence, $(x',y')$ is uniformly distributed on $\mathcal{A}_0$, $x'$ is distributed according $p(x)$ and $y'$ as $p_G^{-1}(y)$).} 

Now, we consider the transformation of random variables $U_1=g^{-1}(Y_3)$ and $U_2=g^{-1}(Y_4)$ (with $g^{-1}$ an increasing function) and plot together the two PDFs $q_1(u)\propto p^{-1}_{inc}(g(u))\frac{dg}{du}$ and $q_2(u)\propto p^{-1}_{dec}(g(u))\frac{dg}{du}$ obtaining the regions $B_1$, $B_2$ and $B_3$ as represented in Figure \ref{figBnoterminanunca}(c). The region attained with the GRoU method is exactly $\mathcal{A}_g=B_{1}\cup B_{2}$ (in Figure \ref{figBnoterminanunca}(c) we use $g(u)=u^2/2$) and  we can write it as 
\begin{equation}
\label{Eq:Agdefotra}
\mathcal{A}_g=\left\{(v,u)\in\mathbb{R}^2: p_{inc}^{-1}(g(u))\frac{dg}{du} \leq v \leq p_{dec}^{-1}(g(u))\frac{dg}{du}\right\}.
\end{equation}
Note that we can interpret that the boundary of $\mathcal{A}_g$ can be obtained through a ``transformation'' of the contour of the region $\mathcal{A}_0$ (see Eqs. (\ref{Eq:A0defotra}) and (\ref{Eq:Agdefotra})).   
Finally, recalling the subset $\mathcal{A}_{g|u}= \{(v,z)\in\mathcal{A}_g, z=u\}$ then note that in this case we also have 
$p_G^{-1}(g(u)) \frac{dg}{du}=|\mathcal{A}_{g|u}|.$


\subsection{Generic PDF}

Let us assume that we can divide the domain $\mathcal{D}_{X}$ of the PDF $p(x)$ with a partition formed by $N$ disjoint sets, i.e.,  $\mathcal{D}_X=\mathcal{D}_1\cup \mathcal{D}_2\cup ...\mathcal{D}_N$, where $p(x)$ is monotonic increasing or decreasing, i.e.,
\begin{equation}
p(x)=p_{j}(x) \mbox{  }\mbox{  }\mbox{ and }\mbox{  }\mbox{  } x\in \mathcal{D}_j,
\end{equation}
where $p_j(x)$ is an increasing or decreasing function. 

\begin{figure*}[htb]
\centering
\centerline{                         
\subfigure[]{\includegraphics[width=6cm]{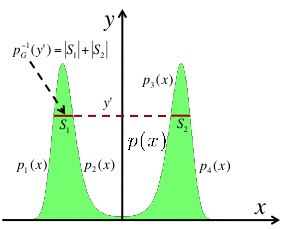}}                                 
\subfigure[]{\includegraphics[width=6cm]{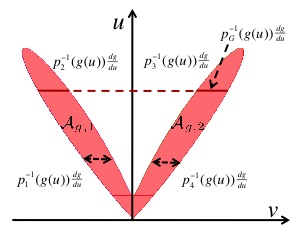}}                                     
}
\caption{{\bf (a)} A bimodal density $p_0(x)\propto p(x)=\exp\{-(x^2-4)^2/4\}$ formed by $4$ monotonic pieces $p_i(x)$, $i=1,...,4$. {\bf (b)} The the corresponding region $\mathcal{A}_g=\mathcal{A}_{g,1}\cup \mathcal{A}_{g,2}$ obtained by the GRoU using $g(u)=u^2/2$.} 
\label{figBimodalpdfAg}
\end{figure*}

Let us assume, moreover, that $p(x)$ is a continuous function with $\mathcal{D}_X=\mathbb{R}$.
Since $\int_{\mathcal{D}_X}p(x)dx<+\infty$, then $N$ is even and
$p_{2i-1}(x)$, with $i=1,...,N/2$, are increasing functions whereas $p_{2i}(x)$, with $i=1,...,N/2$, are decreasing functions. Then, the region $\mathcal{A}_g$ generated by the GRoU can be expressed as 
\begin{equation}
\mathcal{A}_{g}=\mathcal{A}_{g,1}\cup \mathcal{A}_{g,2} \cup \ldots \cup \mathcal{A}_{g,N/2}, 
\end{equation}
where
\begin{equation}
\mathcal{A}_{g,i}= \left\{(v,u)\in \mathbb{R}^2: p_{2i-1}^{-1}\left(g(u)\right)\dot{g}(u)\leq v \leq p_{2i}^{-1}\left(g(u)\right)\dot{g}(u)  \right\},
\end{equation}
for $i=1,...,N/2$.
Figure \ref{figBimodalpdfAg}(a) shows the bimodal PDF $p_0(x)\propto p(x)=\exp\{-(x^2-4)^2/4\}$ and the corresponding region $\mathcal{A}_g$ obtained by the GRoU with $g(u)=\frac{1}{2}u^2$ is illustrated in Figure \ref{figBimodalpdfAg}(b).
We recall that, as illustrated in Figure \ref{figBimodalpdfAg}(a), we can define 
\begin{equation}
\mathcal{A}_{0|y}= \{(x,z)\in\mathcal{A}_0, z=y\},
\end{equation}
and then we can write
\begin{equation}
p_G^{-1}(y)=|\mathcal{A}_{0|y}|.
\end{equation}
Since in this case it is composed by two segments, $\mathcal{A}_{0|y}=S_1 \cup S_2$, we have $p_G^{-1}(y)=|S_1|+|S_2|$. Then, recalling the definition of the subset $\mathcal{A}_{g|u}= \{(v,z)\in\mathcal{A}_g, z=u\}$, hence note that we have again that 
$$p_G^{-1}(g(u)) \frac{dg}{du}=|\mathcal{A}_{g|u}|,$$
as depicted in Figure \ref{figBimodalpdfAg}(b). 



\subsection{Discussion about GRoU}
The RoU and GRoU techniques were introduced \citep{Kinderman77, Wakefield91} as a bivariate transformation of the bidimensional region $\mathcal{A}_0$ below the target PDF $p(x)$. To be specific, the RoU techniques were presented as a transformation of a  bidimensional {\it uniform random variable} defined over $\mathcal{A}_0$. This bivariate transformation follows the equations $x=\frac{v}{\dot{g}(u)}$ and $y=u$ (see Appendix \ref{App_a_ver}). These relationships describe all the points {\it within} the transformed region $\mathcal{A}_g$.

In this work (and, specially, in this section)  we have also seen that the GRoU can be interpreted as transformations of random variables $Y_i$ with PDFs the monotonic pieces $p_i^{-1}(y)$, $i=1,..,N$ (where the monotonic functions $p_i(x)$, $i=1,..,N$  compose the target density $p(x)$). These transformed densities describe disjoint parts of the {\it boundary} of region $\mathcal{A}_g$ obtained with the GRoU.

Furthermore, given the random vector $(V,U)$ uniformly distributed on $\mathcal{A}_g$, we have seen that the second random coordinate $U$ is distributed according to $q(u)\propto p_G^{-1}(g(u)) \frac{dg}{du}$. Namely, we can  write the RV $U$ as a transformation of a RV $Y$, i.e., exactly as $U=g^{-1}(Y)$, where $Y$ is distributed according to the generalized inverse PDF $p_{G}^{-1}(y)$.



\section{GRoU for unbounded PDFs}
\label{sec:GRoUforUnbounded}

Another assumption used on the Theorem \ref{GRoU_Theorem} of the GRoU is that $p_0(x)\propto p(x)$ must be bounded.
In this section, we discuss as to design a GRoU technique for unbounded PDFs (with bounded support, for simplicity) using the observations in Section \ref{sec:RoUvsTRS}.  We will refer to this technique as {\it unbounded GRoU} (U-GRoU).
Then consider, for instance, a {\it decreasing} target PDF $p_0(x)\propto p(x)$, where
$$p(x): \mathcal{D}_X=(0,b]\rightarrow \mathbb{R}^{+},$$
with an vertical asymptote at $x^*=0$. In this case, to apply a kind of GRoU approach to draw samples from $p_0(x)\propto p(x)$, we have two possibilities:
\begin{enumerate}
\item the first option is to apply the standard GRoU for bound PDFs of Section \ref{sec:GRoU} to the inverse PDF $p^{-1}(y)$ (that is clearly bounded with unbounded domain, in this case), in order to produce  a sample $y'$ from $p^{-1}(y)$. Then,  samples distributed according to $p_0(x)$ can be obtained using the IoD method in Section \ref{sec:IoD}, i.e., $x'=z'y'$ where $z' \sim \mathcal{U}([0,1])$. 
However, we need to be able to evaluate $p^{-1}(y)$, namely to invert $p(x)$, and it could be difficult or impossible, in general. 
\item A more general approach is to design a GRoU technique to tackle directly this kind of unbounded target PDFs. To do that, we can use the observations and discussions about the GRoU provided in the previous Section \ref{sec:RoUvsTRS}.
\end{enumerate}
In Section \ref{sec:RoUvsTRS}, we have emphasized that the GRoU is equivalent to a transformation $g^{-1}(y)$ of a RV $Y\sim p^{-1}(y)$ (with $c=1$ and $p(x)$ monotonic). If the transformation $U=g^{-1}(Y)$ is adequately chosen the region $\mathcal{A}_g$ defined by the GRoU is bounded. 

In this situation, $p^{-1}(y) :\mathbb{R}^{+}\rightarrow (0,b]$ is bounded with unbounded support. In Section \ref{sec:TRSB2} we have described the conditions that an increasing transformation $\varphi(y): \mathbb{R}^{+}\rightarrow [d_1,d_2)$ (where $d_1<d_2$ are generic constant) has to fulfill in order to obtain bounded PDFs with bounded support. The random variable $U=\varphi(Y)$ has PDF
\begin{equation}
q(u)=p^{-1}(\varphi^{-1}(u))\frac{d\varphi^{-1}}{du} \quad\mbox{ with } \quad d_1 \leq u\leq d_2. 
\end{equation}
The PDF $q(u)$ is bounded if $p^{-1}(y)$ is is an infinitesimal of the same or higher order than $\dot{\varphi}(y)$ at $y\rightarrow+\infty$, as we have shown in Section \ref{sec:TRSB2}.
Therefore, with this suitable function $\varphi(y)$ and  the observations in Section \ref{sec:RoUvsTRS} we can define the corresponding suitable region $\mathcal{A}_\varphi$ as 
\begin{equation}
\mathcal{A}_\varphi = \left\{(v,u)\in \mathbb{R}^{2}: \quad 0\leq u \leq \varphi\left[ p\left( \frac{v}{\dot{\varphi}^{-1}(u)} \right)\right] \right\},
\end{equation}
so that the sample
$$x=\frac{v}{\dot{\varphi}^{-1}(u)}$$
is  distributed as $p_0(x)\propto p(x)$ if $(v,u)$ are uniformly distributed on $\mathcal{A}_\varphi$. In the sequel, we provide two examples of suitable transformations $\varphi(y)$.
\begin{Ex}    
Consider the unbounded target pdf
\begin{equation}
 p_0(x)\propto p(x)=\sqrt{-2\log(x)} \quad \mbox{ with } \quad x\in (0,1].
\end{equation}
 In this case, a first U-GRoU scheme can be found using 
 \begin{gather}
 \left\{
 \begin{split}
  \varphi(y)&=\arctan(y): \mathbb{R}^+\rightarrow \left[0,\frac{\pi}{2}\right),\\
  \varphi^{-1}(u)&=\tan(u):  \left[0,\frac{\pi}{2}\right) \rightarrow \mathbb{R}^+,  \\
  \dot{\varphi}^{-1}(u)&=\tan(u)^2+1,
 \end{split}
 \right.
 \end{gather}
 i.e., if $(v,u)$ is uniformly distributed on $\mathcal{A}_\varphi$
\begin{equation}
\mathcal{A}_\varphi= \left\{(v,u)\in \mathbb{R}^{2}: \quad 0\leq u \leq \arctan\left[p\left( \frac{v}{\tan(u)^2+1} \right)\right] \right\},
\end{equation} 
then $x= \frac{v}{\tan(u)^2+1}$ is distributed as $p_0(x)$. Figure \ref{UNB_1fig} depicts the region $\mathcal{A}_\varphi$ for this choice of  $\varphi(y)$. The acceptance rate with $\varphi(y)=\arctan(y)$,  using the optimal overbounding rectangle, is $\approx 65 \%$.  
\begin{figure}[htb]
\centering 
\centerline{
  \subfigure[]{\includegraphics[width=5.6cm]{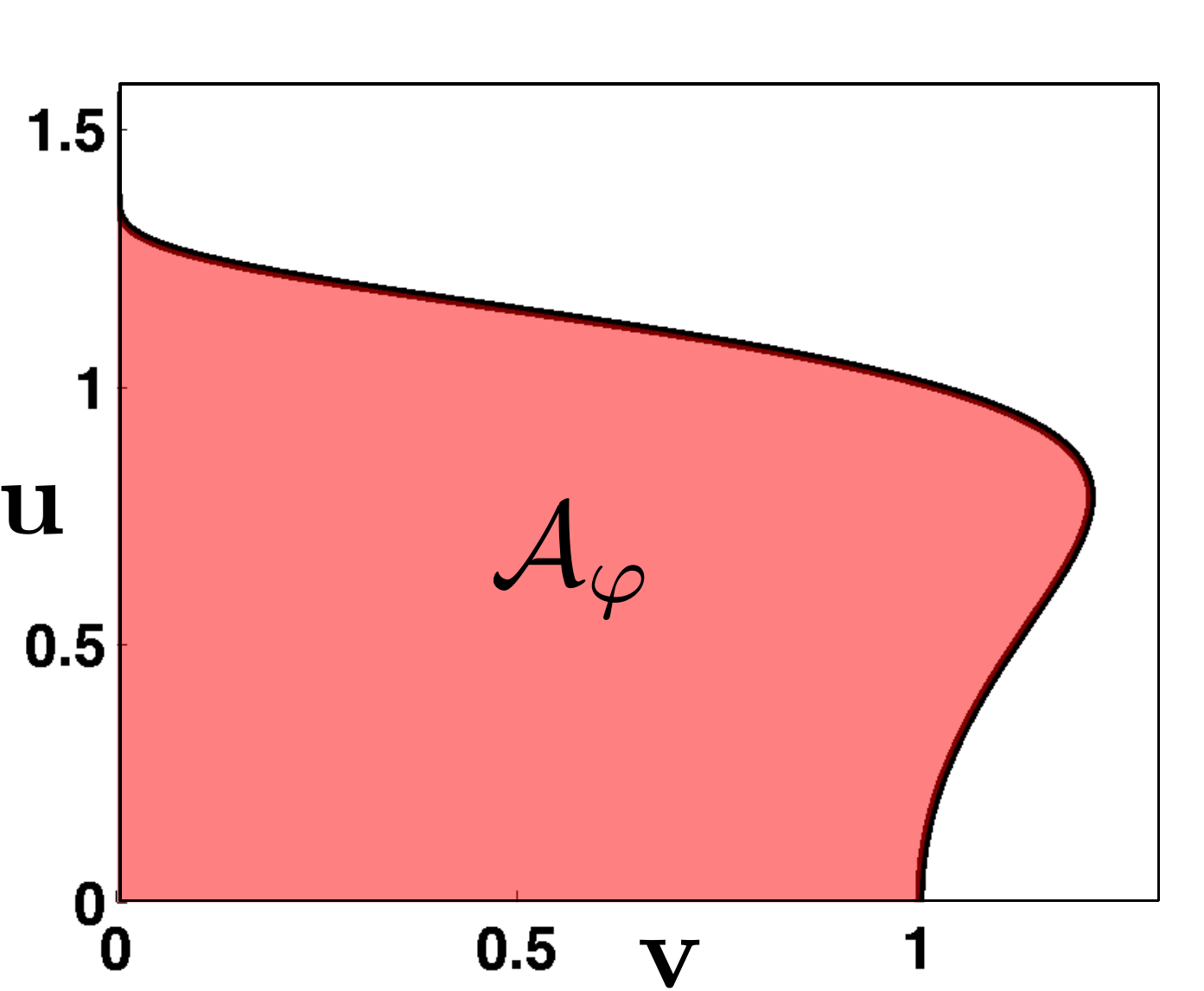}} 
  \subfigure[]{\includegraphics[width=6cm]{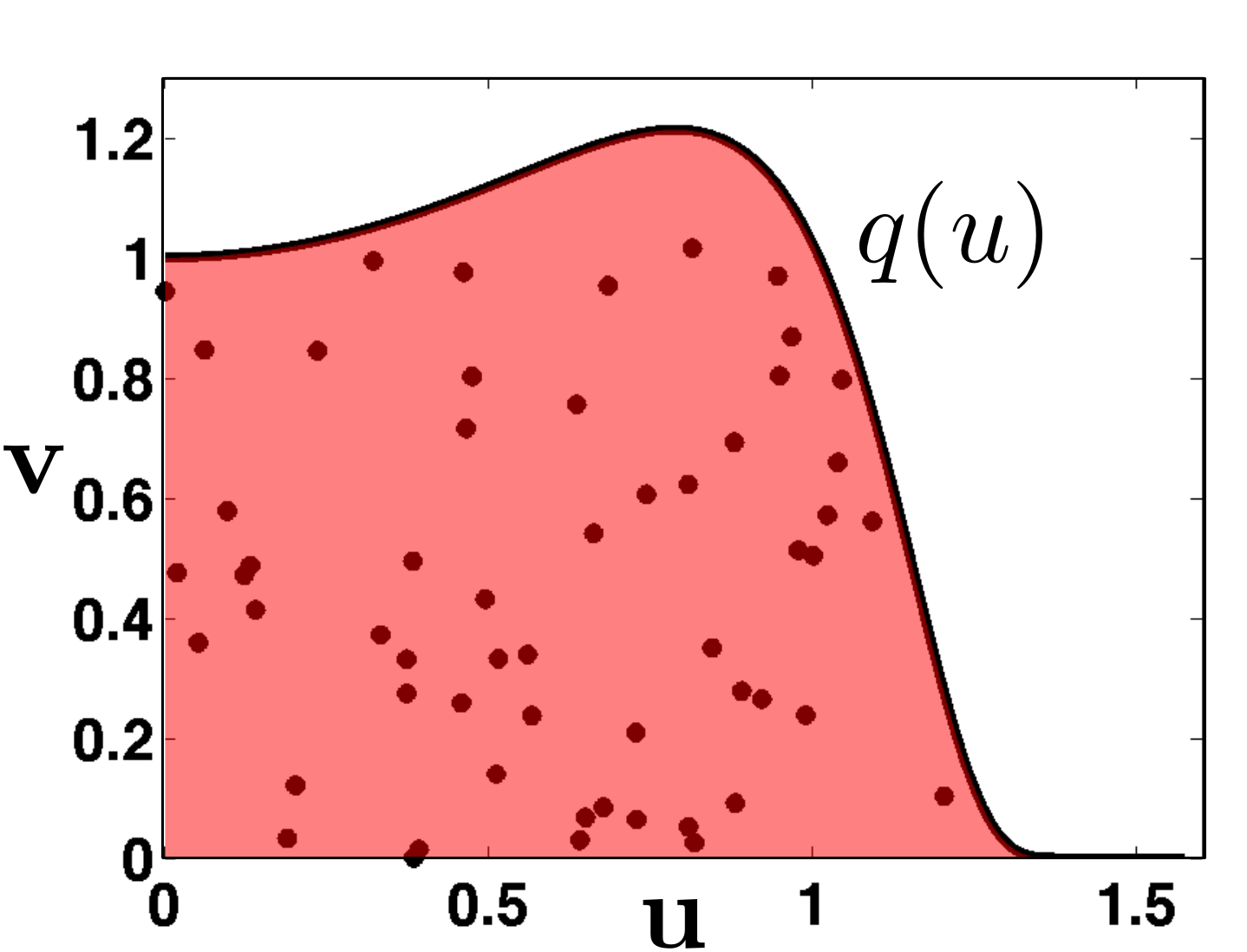}}
      \subfigure[]{\includegraphics[width=6cm]{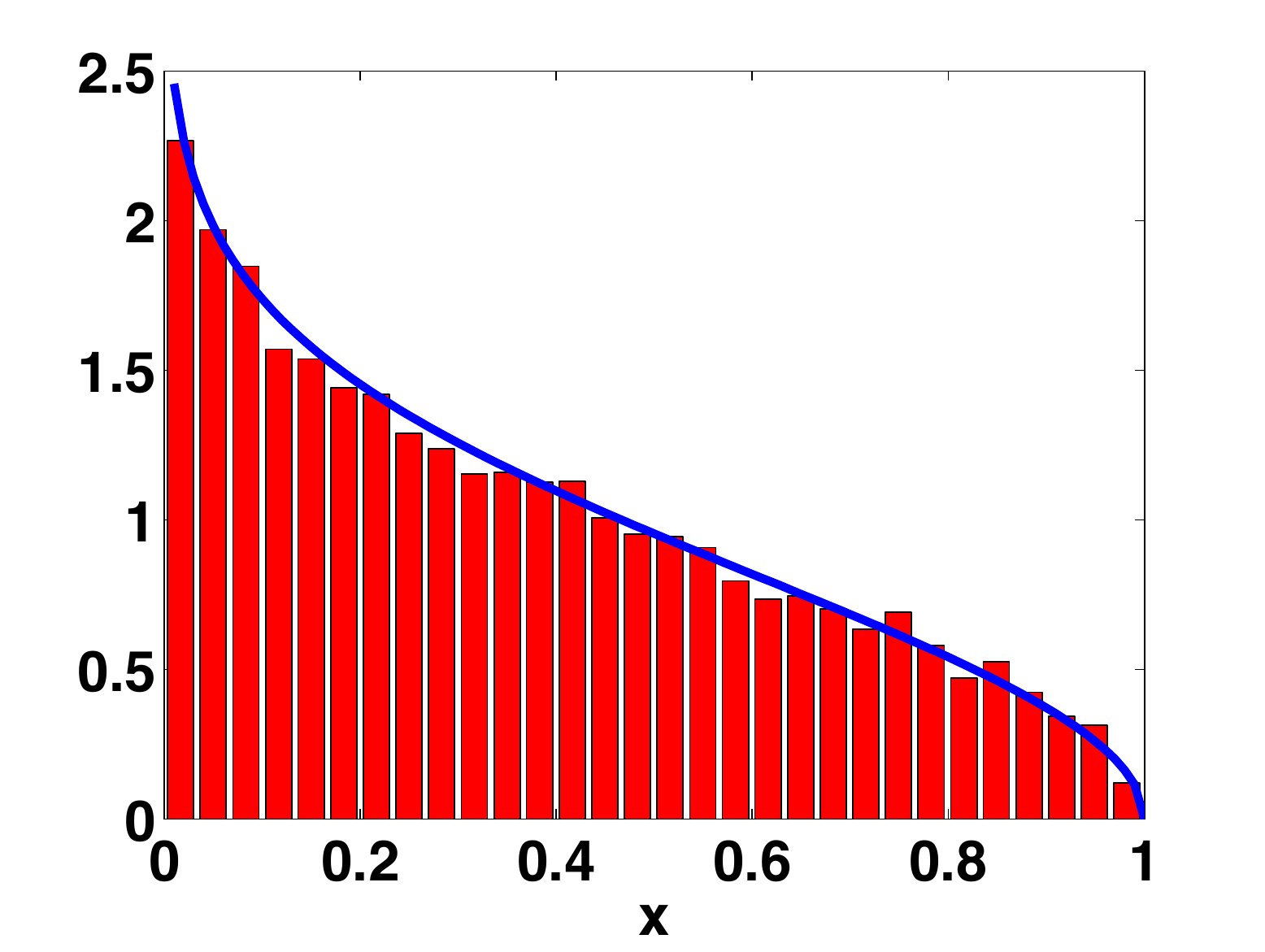}}
  }
  \caption{{\bf (a)} Region $\mathcal{A}_\varphi$ for $p_0(x)\propto p(x)=\sqrt{-2\log(x)}$ ($x\in [0,1)$) corresponding to the choice $\varphi(y)=\arctan(y)$. {\bf (b)} The same region $\mathcal{A}_\varphi$ with rotated axes and $55$ samples uniformly distributed on $\mathcal{A}_\varphi$. The boundary is described by the function $q(u)=p^{-1}(\varphi^{-1}(u))\frac{d\varphi^{-1}}{du}$. {\bf (c)} Normalized histogram of  $10000$ generated and accepted samples via GRoU.}
  \label{UNB_1fig}
  \end{figure}

 A second possibility is given using, for instance,
\begin{gather}
 \left\{
 \begin{split}
 \varphi(y) & =\frac{y}{y+1} : \mathbb{R}^+\rightarrow [0, 1), \\
  \varphi^{-1}(u) & =-\frac{u}{u-1} :  [0, 1) \rightarrow \mathbb{R}^+, \\
\dot{\varphi}^{-1}(u)&=\frac{u}{(u - 1)^2} - \frac{1}{u - 1}, \\
 \end{split}
 \right.
 \end{gather}
 i.e., if $(v,u)$ is uniformly distributed on $\mathcal{A}_\varphi$
\begin{equation}
\mathcal{A}_\varphi\dfn \left\{(v,u)\in \mathbb{R}^{2}: \quad 0\leq u \leq \frac{p\left(\frac{v}{u/(u - 1)^2 - 1/(u - 1)}\right)}{p\left( \frac{v}{u/(u - 1)^2 - 1/(u - 1)}\right)+1} \right\},
\end{equation}   

then $x=\frac{v}{u/(u - 1)^2 - 1/(u - 1)}$ is distributed according to $p_0(x)$.\footnote{Clearly, any proposed sample $x'$ such that $x'<0$ or $x'>1$ is inadmissible since the target $p_0(x)$ is defined in $[0,1)$.}  Figure \ref{UNB_2fig} illustrates the region $\mathcal{A}_\varphi$ for this other choice of $\varphi(y)$. The acceptance rate in this case,  using the optimal overbounding rectangle, is $\approx 51 \%$.  
\begin{figure}[htb]
\centering 
\centerline{
  \subfigure[]{\includegraphics[width=5.7cm]{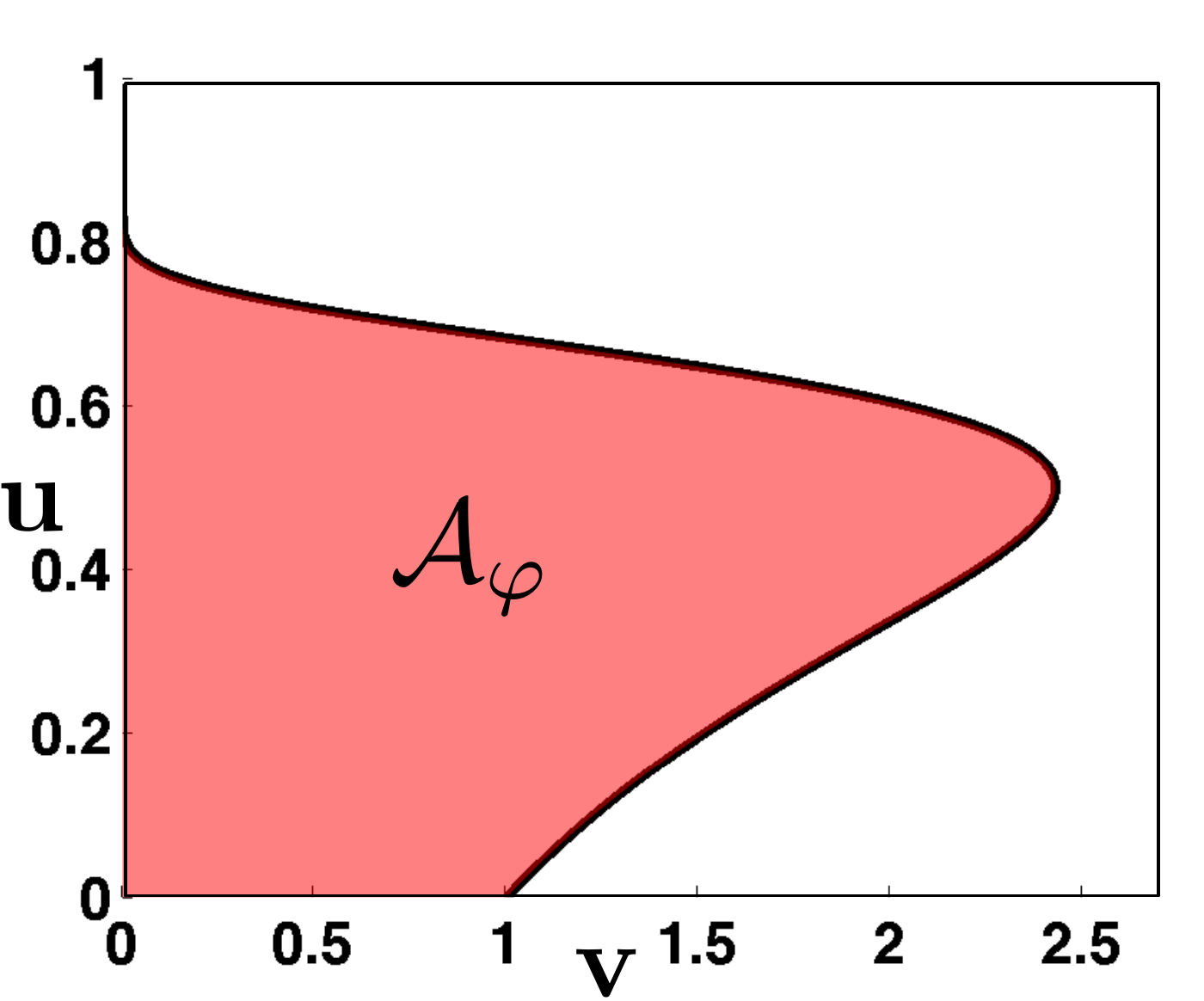}} 
  \subfigure[]{\includegraphics[width=5.7cm]{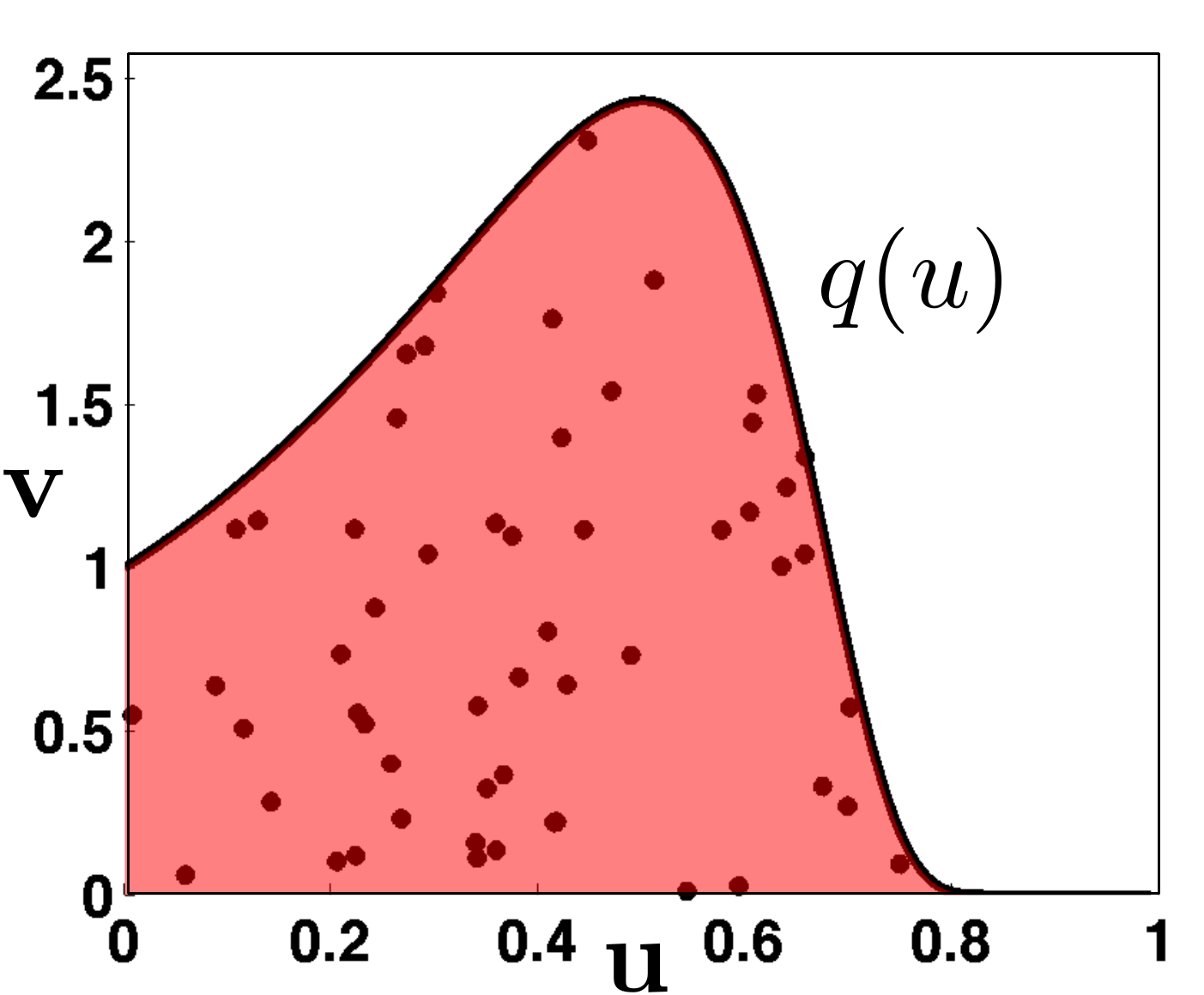}}
    \subfigure[]{\includegraphics[width=6cm]{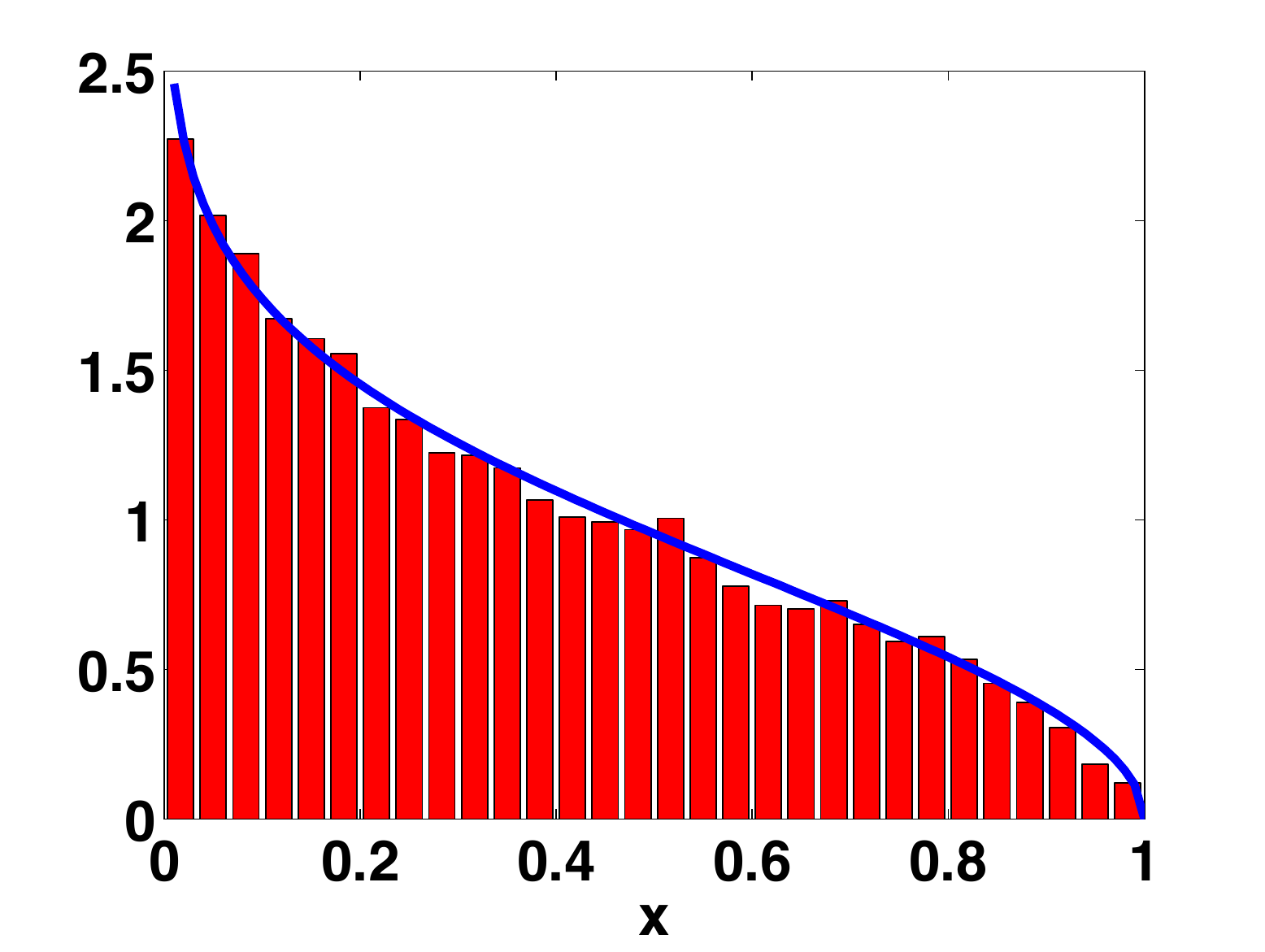}}
  }
  \caption{{\bf (a)} Region $\mathcal{A}_\varphi$ for $p_0(x)\propto p(x)=\sqrt{-2\log(x)}$ ($x\in [0,1)$) corresponding to the choice $\varphi(y)=\frac{y}{y+1}$. {\bf (b)} The same region $\mathcal{A}_\varphi$ with rotated axes and $55$ samples uniformly distributed on $\mathcal{A}_\varphi$. The boundary is described by the function $q(u)=p^{-1}(\varphi^{-1}(u))\frac{d\varphi^{-1}}{du}$. {\bf (c)} Normalized histogram of $10000$ generated and accepted samples via GRoU. }
  \label{UNB_2fig}
  \end{figure}

\end{Ex}

\section{Conclusions}
\label{sec:conclusions}

In this work, we have shown that the generalized ratio of uniforms (GRoU) algorithm \citep{Kinderman77,Wakefield91} can be seen as combination of other classical sampling strategies: an extension of inverse-of-density (IoD) method \citep{Bryson82,Chaubey10,Devroye86,Jones02,Khintchine38,Shepp62} (and, at the same time, of the fundamental theorem of simulation approach \citep{Devroye86,Robert04}) and the transformed rejection sampling (TRS) \citep{Devroye86,Hormann94,Marsaglia84,Wallace76}.


Specifically, for {\it monotonic} densities with mode at zero, the GRoU is {\it completely equivalent} to a combination of the TRS approach applied to the inverse PDF $p^{-1}(y)$ (Section \ref{sec:OurCase}), an extension of the IoD strategy (Section \ref{sec:IoD}) and an extension of the fundamental theorem idea (Section \ref{sec:FToS}). The classical IoD technique and also the fundamental theorem of simulation provide the relationship between random variates from $p^{-1}(y)$ and samples drawn from $p(x)$ whereas the GRoU links the realizations of a RV $U=g^{-1}(Y)$, where $Y$ has PDF $p^{-1}(y)$, to the samples distributed as the target PDF $p(x)$.  Moreover, we have exhibited that the conditions about the function $g(u)$ required in the GRoU \citep{Wakefield91} to obtain a {\it bounded} region $\mathcal{A}_g$ are {\it exactly} the same assumptions needed to the transformation $U=h(Y)=g^{-1}(Y)$ (where $Y$ has density $p^{-1}(y)$) in the TRS approach, in order to achieve {\it bounded} transformed PDF $q(u)=p(h^{-1}(u))\frac{dh^{-1}}{du}$ with bounded domain.
We have also seen that the TRS approach can be considered lightly more general than the GRoU approach in the sense that some conditions required by the GRoU can be relaxed as we show, for instance, in Sections \ref{Sectveryimp}-\ref{SectConstCcazzo} and the Appendices \ref{App_a_ver2}-\ref{App_DecrG}.

For generic non-monotonic densities, formed by $N$ monotonic pieces $p_i(x)$, $i=1,..,N$, the boundary of the region $\mathcal{A}_g$ of the GRoU can be expressed as
transformations of random variables $Y_i$ with PDFs $p^{-1}_i(y)$, the inverse functions of the monotonic pieces $p_i(x)$, $i = 1,..,N$.  
Moreover, the RV $U$ of the random vector $(V,U)$ uniformly distributed on $\mathcal{A}_g$ is distributed according to the PDF $q(u)\propto p_G^{-1}(g(u)) \frac{dg}{du}$ where $p_{G}^{-1}(y)$ is the generalized inverse PDF. Namely, we can  write the  RV $U$ as a transformation of a RV $Y$, i.e., exactly as $U=g^{-1}(Y)$, where $Y$ is distributed according to $p_{G}^{-1}(y)$.

Therefore, in this work we have illustrated the close relationships among GRoU, IoD and TRS approaches in Section \ref{sec:RoUvsTRS}. 
Using the considerations in Section \ref{sec:RoUvsTRS}, we have also relaxed different assumptions of the GRoU (see Sections \ref{Sectveryimp}-\ref{SectConstCcazzo} and Appendices \ref{App_a_ver2}, \ref{App_DecrG}). Moreover, the discussions in Section \ref{sec:RoUvsTRS} allow us to design of a GRoU technique to deal with unbounded target PDFs, in Section \ref{sec:GRoUforUnbounded}.
Finally, we use the considerations and remarks in Section \ref{sec:RoUvsTRS} to clarify certain aspects about the optimality on the choice of the functions $g(u)$. Indeed, we have deduced which is the function $g(u)$ to obtain a rectangular region $\mathcal{A}_g$ (see Section \ref{OptimalGbutnonoptimal}).

\section{Acknowledgment}
This work has been partially supported by the Ministry of Science and Innovation of Spain (MONIN project, ref. TEC-2006-13514-C02-01/TCM,  DEIPRO project, ref. TEC-2009-14504-C02-01 and Consolider-Ingenio program 2010 ref. CSD2008- 00010 COMONSENS) and the Autonomous Community of Madrid (project PROMULTIDIS-CM, ref. S-0505/TIC/0233).
 
\appendix


\subsection{Proof of the GRoU}
\label{App_a_ver}

Given the transformation $(v,u)\in \mathbb{R}^2\rightarrow (x,z)$
  \begin{gather}
  \label{queraronolabel}
  \left\{
  \begin{split}
 & x=\frac{v}{\dot{g}(u)} \\
 & z=u          
   \end{split}  
  \longrightarrow 
   \right.
     \left\{
       \begin{split}
 & v=x\dot{g}(z) \\
 & u=z         
   \end{split}
    \right. ,
   \end{gather}
and a pair of RV's  $(V,U)$ uniformly distributed on $\mathcal{A}_g$, we can write the joint PDF $q(x,y)$ of the transformed RV's $(X,Z)$ as 
\begin{equation}
\label{RoUqEqGen}
 q(x,z)=\frac{1}{|\mathcal{A}_g|} |J^{-1}| \  \  \ \mbox{for all } \  \  \ 0\leq z \leq g^{-1}[cp(x)],
\end{equation}
where $|\mathcal{A}_g|$ denotes the area of $\mathcal{A}_g$, and $J^{-1}$ is the Jacobian of the inverse transformation, namely, 
 \begin{equation}
 \label{JacoProof2}
 J^{-1}=\det \left[
   \begin{array}{cc} 
     \dot{g}(z) & x\ddot{g}(z) \\
      0 & 1 \\
   \end{array}
   \right]=\dot{g}(z). 
   \end{equation} 
  Since we assume $\dot{g}(z)\geq 0$ (i.e., $g$ increasing), then $|J^{-1}|=|\dot{g}(z)|=\dot{g}(z)$ and substituting (\ref{JacoProof2}) into (\ref{RoUqEqGen}) yields
   \begin{gather}
   \label{jointQ2}
     q(x,z)=\left\{
     \begin{split}
        &\frac{1}{|\mathcal{A}_g|}\dot{g}(z)  \  \  \mbox{for} \  \  \ 0\leq z \leq g^{-1}[cp(x)], \\
        & 0,  \  \    \mbox{ }\mbox{ }\mbox{ }\mbox{ }\mbox{ }\mbox{ }\mbox{ }\mbox{ } \mbox{ }\mbox{otherwise}.
      \end{split}
      \right.
   \end{gather}
   Hence, integrating $q(x,z)$ w.r.t. $z$ yields the marginal PDF of the RV $X$,
 \begin{gather}
 \label{EqfundProof}
 \begin{split}
q(x)=\int_{-\infty}^{+\infty} q(x,z) dz=& \int_{0}^{{g^{-1}[cp(x)]}} \frac{1}{|\mathcal{A}_g|}\dot{g}(z)dz= \\
= \frac{1}{|\mathcal{A}_g|}&\Big[ g(z)\Big]_{0}^{g^{-1}[cp(x)]}= \frac{c}{|\mathcal{A}_g|} p(x)- \frac{1}{|\mathcal{A}_g|}g(0) \\ 
\end{split}
\end{gather}
where the first equality follows from  Eq. (\ref{jointQ2}) and the remaining calculations are trivial. Since we have also assumed $g(0)=0$, it turns out that
\[q(x)=\frac{c}{|\mathcal{A}_g|}p(x)=p_0(x).  \]
Hence, we have proved that a marginal PDF is exactly $p_0(x)\propto p(x)$.  $\quad\Box$

\subsection{Important observation}
\label{App_FantasticoSect}

In the proof above, we have integrated the bidimensional pdf $q(x,z)$ in Eq. (\ref{jointQ2}) w.r.t. $z$ finding the marginal pdf $q(x)$ that is exactly our target $p_0(x)$. Note that the set
$$\mathcal{A}_g=\left\{(v,u)\in \mathbb{R}^2: 0\leq u \leq g^{-1}\left[c\mbox{ }p\left(\frac{v}{\dot{g}(u)}\right)\right]\right\},$$
can be expressed in terms of $(x,z)$, i.e.,
$$\mathcal{A}_g=\left\{(x,z)\in \mathbb{R}^2: 0\leq z \leq g^{-1}\left[c\mbox{ }p(x)\right]\right\},$$
and since $g^{-1}$ is increasing and assuming $p(x)$ decreasing ($c$ is positive), we can rewrite it as 
$$\mathcal{A}_g=\left\{(x,z)\in \mathbb{R}^2: 0\leq x \leq p^{-1}\left[\frac{1}{c}\mbox{ }g(z)\right]\right\}.$$
Then, if we integrate $q(x,z)$ w.r.t. $x$ we obtain (setting $k=1/c$)
 \begin{gather}
 \label{EqfundProof2}
 \begin{split}
q(z)=\int_{-\infty}^{+\infty} q(x,z) dx=& \int_{0}^{p^{-1}\left[k\mbox{ }g(z)\right]} \frac{1}{|\mathcal{A}_g|}\dot{g}(z)dx= \\
= \frac{1}{|\mathcal{A}_g|}&\Big[ \dot{g}(z)x\Big]_{0}^{p^{-1}\left[k\mbox{ }g(z)\right]}=\frac{1}{|\mathcal{A}_g|}\dot{g}(z)p^{-1}\left[k\mbox{ }g(z)\right]-\frac{1}{|\mathcal{A}_g|}\dot{g}(z)0, \\
q(z)=\frac{1}{|\mathcal{A}_g|} \dot{g}(z)p^{-1}&\left[k\mbox{ }g(z)\right].
\end{split}
\end{gather}
Namely, the RV $Z=U$ (see Eq. \eqref{queraronolabel}) is obtained as a transformation $Z=U=g^{-1}(\frac{1}{k}Y)=g^{-1}(cY)$ of the RV $Y$ with pdf $p^{-1}(y)$ that is exactly what we anticipate in Section \ref{sec:RoUvsTRS}.

\subsection{Other interesting observations}
\label{App_Increible}

It is interesting to notice that:
\begin{itemize}
\item If we consider the same $\mathcal{A}_g= \left\{(v,u)\in \mathbb{R}^2: 0\leq u \leq g^{-1}\left[c\mbox{ }p\left(\frac{v}{\dot{g}(u)}\right)\right]\right\}$ (i.e., the set $\mathcal{A}_g$ is defined in the same way) but we take $x=-\frac{v}{\dot{g}(u)}$,  then we draw samples from $p(-x)$.
\item If we consider another definition of the set $\mathcal{A}_g'= \left\{(v,u)\in \mathbb{R}^2: 0\leq u \leq g^{-1}\left[c\mbox{ }p\left(-\frac{v}{\dot{g}(u)}\right)\right]\right\}$ and later we take $x=-\frac{v}{\dot{g}(u)}$,  the set $\mathcal{A}_g'$ is  a symmetric version of $\mathcal{A}_g$ with respect the axis $u$ and we draw samples from $p(x)$.
\end{itemize}
These considerations can be easily inferred from the proof above.


\subsection{Extension of the GRoU}
\label{App_a_ver2}
We present here a light extension of the GRoU.

\begin{Teorema}
Let $g(u): [b,+\infty)\rightarrow \mathbb{R}^+$ be a strictly increasing (in $(b,+\infty)$) differentiable function such that $g(b)=0$ and let $p(x)\geq 0$ be a PDF known only up to a proportionality constant. Assume that $(v,u)\in \mathbb{R}^2$ is a sample drawn from the uniform distribution on the set 
\begin{equation}
\label{regionAdefgen2}
\mathcal{A}_g= \Bigg\{(v,u)\in \mathbb{R}^2: b\leq u \leq g^{-1}\Bigg[c\mbox{ }p\Bigg(\frac{v}{\dot{g}(u)}\Bigg)\Bigg]\Bigg\}, 
\end{equation}
where $c>0$ is a positive constant and $\dot{g}=\frac{dg}{du}$. Then $x=\frac{v}{\dot{g}(u)}$ is a sample from $p_0(x)$. 
\end{Teorema}
The proof is straightforward. Indeed, in this case, the development of the proof is identical yielding a expression similar to the Eq. (\ref{EqfundProof}) that becomes
\begin{gather}
 \begin{split}
 \nonumber
  \int_{b}^{{g^{-1}[cp(x)]}} \frac{1}{|\mathcal{A}_g|}\dot{g}(z)dz=
 \frac{1}{|\mathcal{A}_g|}&\Big[ g(z)\Big]_{b}^{g^{-1}[c_2p(x)]}= \frac{c}{|\mathcal{A}_g|} p(x)- \frac{1}{|\mathcal{A}_g|}g(b)=\frac{c}{|\mathcal{A}_g|} p(x), 
 \end{split}
\end{gather}
that is proportional to the target PDF (since $p(x)\propto p_0(x)$, as well).


\subsection{Other extension of the GRoU with a decreasing function $g(u)$}
\label{App_DecrG}
We present another light extension of the GRoU where $g(u)$ is decreasing.

\begin{Teorema}
Let $g(u): \mathbb{R}^-\rightarrow \mathbb{R}^+$ (i.e. $u\leq 0$) be a strictly {\it decreasing} (in $\mathbb{R}^- \backslash \{0\}=(-\infty,0)$) differentiable function such that $g(0)=0$ and let $p(x)\geq 0$ be a PDF known only up to a proportionality constant. Assume that $(v,u)\in \mathbb{R}^2$ is a sample drawn from the uniform distribution on the set 
\begin{equation}
\label{regionAdefgenNosequanto}
\mathcal{A}_{g_{dec}}= \Bigg\{(v,u)\in \mathbb{R}^2: g^{-1}\Bigg[c\mbox{ }p\Bigg(\frac{v}{\dot{g}(u)}\Bigg)\Bigg] \leq u \leq 0 \Bigg\}, 
\end{equation}
where $c>0$ is a positive constant and $\dot{g}=\frac{dg}{du} <0$. Then $x=\frac{v}{\dot{g}(u)}$ is a sample from $p_0(x)$. Or,  another possibility is to define   
\begin{equation}
\label{regionAbastanonnepossopiu}
\mathcal{A}_{g_{dec}}'= \Bigg\{(v,u)\in \mathbb{R}^2: g^{-1}\Bigg[c\mbox{ }p\Bigg(-\frac{v}{\dot{g}(u)}\Bigg)\Bigg] \leq u \leq 0 \Bigg\}, 
\end{equation}
and then take $x=-\frac{v}{\dot{g}(u)}$.
\end{Teorema}
It is important to note that $g^{-1}(y): \mathbb{R}^+\rightarrow \mathbb{R}^-$ then $g^{-1}(y)\leq 0$. For instance, it is possible to consider $g(u)=u^2/2$ with $u\leq 0$, the region $\mathcal{A}_{g_{dec}}$ have the same form of $\mathcal{A}_{g}$ (when we use $g(u)=u^2/2$ with $u\geq 0$) but  it is symmetric the originof the axes $(0,0)$ ,
and $\mathcal{A}_{g_{dec}}'$ is symmetric to $\mathcal{A}_{g}$ w.r.t. the axis $v$.

Finally, consider again a decreasing bounded PDF $p(x)$ with $x\in \mathbb{R}^+$. Then $\mathcal{A}_{g_{dec}}$ can be rewritten as (with $c=1$) 
\begin{equation}
\label{regionAdefgenNosequanto2}
\mathcal{A}_{g_{dec}}=\left\{(v,u)\in \mathbb{R}^2: p^{-1}(g(u))\dot{g}(u)  \leq v \leq 0\right\}, 
\end{equation}
whereas $\mathcal{A}_{g_{dec}}'$ can be rewritten as
\begin{equation}
\label{regionAdefgenNosequanto3}
\mathcal{A}_{g_{dec}}'=\left\{(v,u)\in \mathbb{R}^2: 0  \leq v \leq -p^{-1}(g(u))\dot{g}(u)\right\}.
\end{equation}
Note that above $-\dot{g}(u)>0$. Moreover, for instance we can consider jointly  Eq.  (\ref{regionAdefgenNosequanto3}) above and Eq. (\ref{secondDefAg}), and then we can write
\begin{equation}
\label{regionAgenericlocura}
\mathcal{A}_{g}=\left\{(v,u)\in \mathbb{R}^2: 0  \leq v \leq p^{-1}(g(u))|\dot{g}(u)|\right\},
\end{equation}
where $g(u)$ can be increasing or decreasing. Eq. (\ref{regionAgenericlocura}) is clearly the expression of a transformation of a random variable $Y$ with PDF $p^{-1}(y)$. Clearly, we can come back and obtain Eq. (\ref{regionAbastanonnepossopiu}) or Eq. (\ref{regionAdefgen1_2}) depending if  $g(u)$ is decreasing or increasing, respectively (note that, although we have $|\dot{g}(u)|$, however we also have to invert $g(u)$; in one case the region  is defined for $u\leq 0$ and in the other for $u\geq 0$ but both for $v\geq0$). If $g(u)$ is decreasing we obtain Eq.  (\ref{regionAbastanonnepossopiu}) and we need to set $x=-\frac{v}{\dot{g}(u)}$, whereas if $g(u)$ is increasing we obtain Eq.  (\ref{regionAdefgen1_2}) and we need to set $x=+\frac{v}{\dot{g}(u)}$, hence finally in both we can summarize both cases using the following expression 
\begin{equation}
x=\frac{v}{|\dot{g}(u)|}.
\end{equation}

\subsection{Relationship between $F_Y(y)$ and $F_X(x)$}
\label{App_Fy_Fx}
\begin{figure*}[htb]
\centering
\centerline{                         
\includegraphics[width=5cm]{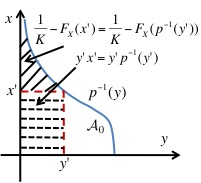}
}
\caption{Relationships between the CDFs $F_{Y}(y)$ of a RV $Y$ with pdf $p^{-1}(y)$ and $F_{X}(x)$ of $X$ with a monotonic decreasing unnormalized target pdf $p(x)$ ($1/K=\int_{\mathcal{D}_X} p(x)dx.$). The area $F_{Y}(y')=\Prob\{Y\leq y'\}=\int_{0}^{y'} p^{-1}(y)dy$ can be obtained as sum of the area $\frac{1}{K}-F_{X}(x')=\frac{1}{K}-F_{X}(p^{-1}(y'))$ (depicted with oblique solid lines) and the rectangular area $y'x'=y'p^{-1}(y')$ (indicated with dashed lines). } 
\label{figOptimalG}
\end{figure*}

The CDF $F_{Y}(y)$ of RV $Y\sim p^{-1}(y)$ can be easily expressed as function of $F_{X}(x)$ (the CDF of $X\sim p(x)$) for monotonic decreasing target pdfs $p_0(x) \propto p(x)$. Indeed, we can write 
\begin{equation}
\label{LocuraaaaMonoPDF}
F_{Y}(y)=\frac{1}{K}-F_{X}(p^{-1}(y))+p^{-1}(y)y,
\end{equation}
that can be easily deduced observing Figure \ref{figOptimalG}. Indeed, in Figure \ref{figOptimalG} we can see that the area $F_{Y}(y')=\Prob\{Y\leq y'\}=\int_{0}^{y'} p^{-1}(y)dy$ can be obtained as sum of the area $\frac{1}{K}-F_{X}(x')=\frac{1}{K}-F_{X}(p^{-1}(y'))$ and the rectangular area $y'x'=y'p^{-1}(y')$ (where we have use the relationship $x'=p^{-1}(y')$).

Observe that if we calculate the first derivative of $F_Y(y)$ in Eq. (\ref{LocuraaaaMonoPDF}), we obtain
\begin{gather}
\begin{split}
\label{LocuraaaaMonoPDF2}
&\frac{dF_{Y}(y)}{dy}=0-\frac{dF_{x}(p^{-1}(y))}{dx}\frac{p^{-1}(y)}{dy}+\frac{p^{-1}(y)}{dy}y+p^{-1}(y), \\
&\frac{dF_{Y}(y)}{dy}=-\underbrace{p(p^{-1}(y))}_{y}\frac{p^{-1}(y)}{dy}+\frac{p^{-1}(y)}{dy}y+p^{-1}(y), \\
&\frac{dF_{Y}(y)}{dy}=p^{-1}(y), \\
\end{split}
\end{gather}
that is exactly the foreseen relationship between CDF and the corresponding PDF.


\subsection{Considerations about the Eq. (\ref{LimitCondEqnose2})}
\label{App_BASTAA}
By definition of inverse function we have
$$g(g^{-1}(cy))=cy,$$
and then we can calculate the derivative with respect to $y$ of both sides (using the chain rule)
$$\left. \frac{dg}{du}\right|_{g^{-1}(cy)} \left.\frac{dg^{-1}}{dy}\right|_{cy} c=c, $$
and finally
$$\left. \frac{dg}{du}\right|_{g^{-1}(cy)} \left.\frac{dg^{-1}}{dy}\right|_{cy}=1, $$
so that 
$$\left. \frac{dg}{du}\right|_{g^{-1}(cy)}=\frac{1}{ \left.\frac{dg^{-1}}{dy}\right|_{cy}}. $$

\newpage

\section*{Acronyms and Notation}
\label{sec:acronyms_notation}

\subsection{Acronyms}
\label{sec:acronyms}

\begin{center}
	\tablefirsthead{%
		\multicolumn{2}{c}{\textbf{List of acronyms used in the text}}\\
		\hline}
	\tablehead{%
		\multicolumn{2}{c}{\textbf{List of acronyms used in the text (continued)}}\\
		\hline}
	\tabletail{\hline}
	\begin{supertabular}{|p{0.10\textwidth}|p{0.88\textwidth}|}	
		CDF & Cumulative Distribution Function\\
		GRoU & Generalized Ratio of Uniforms\\
		IID & Independent Identically Distributed\\
		IoD & Inverse of Density\\
		MC & Markov Chain\\
		MCMC & Markov Chain Monte Carlo\\
		PDF & Probability Density Function\\
		RoU & Ratio of Uniforms\\
		RS & Rejection Sampling\\
		RV & Random Variable\\
		SMC & Sequential Monte Carlo\\
		TR & Transformed Rejection\\
		TRS & Transformed Rejection Sampling\\
		VDR & Vertical Density Representation \\
		U$-$GRoU &  Generalized Ratio of Uniforms for Unbounded densities \\
		\hline
	\end{supertabular}
\end{center}

\newpage

\subsection{Notation}
\label{sec:notation}

\begin{center}
	\tablefirsthead{%
		\multicolumn{2}{c}{\textbf{Summary of the notation used in the text}}\\
		\hline}
	\tablehead{%
		\multicolumn{2}{c}{\textbf{Summary of the notation used in the text (continued)}}\\
		\hline}
	\tabletail{\hline}
	\begin{supertabular}{|p{0.07\textwidth}|p{0.88\textwidth}|}
		$X$ & Target random variable.\\
		$\sim$ & Symbol used to indicate that a certain RV $X$ is distributed according to a given proper normalized or unnormalized PDF.
			For instance, $X \sim p_0(x)$ indicates that the PDF of $X$ is exactly $p_0(x)$ (since $p_0(x)$ is a normalized PDF), whereas
			$Y \sim p^{-1}(y)$ indicates that the PDF of $Y$ is proportional to $p^{-1}(y)$ (since $p^{-1}(y)$ is an unnormalized PDF, and
			the normalized PDF of $Y$ is actually $Kp^{-1}(y)$).\\
		$x$, $x'$ & Particular values taken by a single realization of the target RV $X$.\\
		$p_0(x)$ & Proper normalized PDF of the target RV $X$, indicated as $X \sim p_0(x)$, meaning that $\textrm{Pr}\{X=x\} = p_0(x)$.\\
		$p(x)$ & Proper but unnormalized PDF of the target RV $X$, such that $p_0(x) = Kp(x)$ for some $K>0$. Alternatively, the
			relationship between $p_0(x)$ and $p(x)$ is frequently indicated as $p_0(x) \propto p(x)$, omitting the proportionality constant,
			$K$. The notation $X \sim p(x)$ will also be used to indicate that the PDF of $X$ is proportional to $p(x)$, i.e. that $X$ is
			distributed according to $p(x)$ up to a proportionality constant $K>0$, meaning that $\textrm{Pr}\{X=x\} \propto p(x)$.\\
		$K$ & Proportionality or normalization constant, $K>0$, for the target PDF. This constant is independent from the value taken by
			the RV, $x$, and can be formally obtained as
			\[
				K = \left[\int_{-\infty}^{\infty}{p(x)\ \textrm{d}x}\right]^{-1} = \left[\int_{\mathcal{D}_X}{p(x)\ \textrm{d}x}\right]^{-1}=\left[\int_{\mathcal{D}_Y}{p^{-1}(y)\ \textrm{d}y}\right]^{-1}.
			\]\\
		$\langle\cdot,\cdot\rangle$ & Symbols used to indicate that an interval may be left/right open or closed. For instance,
			$\langle a,b]$ indicates a right-closed interval which may be either left-closed or left-open. Similarly, $[a,b \rangle$
			indicates a left-closed interval which may be either right-closed or right-open. Finally, $\langle a,b \rangle$ denotes an
			interval that may be either closed or open on both sides.\\
		$\mathcal{D}_X$ & Support of the RV $X$, given by the range of values of $X$ where its PDF is strictly greater than zero:
			\[
				\mathcal{D}_X = \{x \in \mathbb{R}: p_0(x)>0\} = \{x \in \mathbb{R}: p(x)>0\}.
			\]
			The most common supports used in the paper are $\mathcal{D}_X = \mathbb{R}$, $\mathcal{D}_X = \mathbb{R}^+$ and
			$\mathcal{D}_X = \langle a,b \rangle$ for two arbitrary real numbers $a,b$.\\
		$F_X(x)$ & Unnormalized CDF of the target RV $X$:
			\[
				F_X(x) = \int_{-\infty}^{x}{p(x)\ \textrm{d}x}.
			\]
			Note that we have $F_X(\infty)=1/K$ instead of $F_X(\infty)=1$, since we use the unnormalized target PDF, $p(x)$, instead of the
			normalized target PDF, $p_0(x)$.\\
		$\mathcal{U}(\mathcal{C})$ & Uniform density with bounded support $\mathcal{C}$, which can be a unidimensional or bidimensional 
			region. In one dimension the most commonly used support is an interval starting at zero, $[0,a]$, denoted by 
			$\mathcal{U}([0,a])$, although more general uniform distributions with other types of supports, such as
			$\mathcal{U}(\mathcal{D}_X)$ when $\mathcal{D}_X$ is a bounded interval, can be considered. In two dimensions the most commonly
			used support will be the region associated to the area enclosed by certain PDF, e.g. $\mathcal{U}(\mathcal{A}_0)$, with
			$\mathcal{A}_0$ defined as below. It is important to remark that $\mathcal{U}(\mathcal{C})$ is always used to denote a proper
			PDF, implying that the Lebesgue measure of $\mathcal{C}$, $|\mathcal{C}|$, must be finite (i.e. that $\mathcal{C}$ must have a
			finite length for unidimensional regions or a finite area for bidimensional regions).\\
		$\pi(x)$ & Proposal density from which samples can be easily drawn that is used to generate random variables, typically in RS and
			TRS algorithms. It can be normalized or not, but always refers to a proper PDF.\\
		$L$ & Proportionality constant for the RS method, such that $p(x)/\pi(x) \le L$ for some $0 < L < \infty$ and any
			$x \in \mathcal{D}_X$.\\
		$Y$ & Random variable distributed according to the inverse of the target PDF. In the sequel, this RV will be called the inverse
			target RV, and may be used to denote $Y \sim p_0^{-1}(y)$ or (more often) $Y \sim p^{-1}(y)$.\\
		$y$, $y'$ & Particular values taken by a single realization of the inverse target RV $Y$.\\
		$p_0^{-1}(y)$ & For monotonic target PDFs (such as exponentials or half Gaussians), proper normalized PDF of the inverse target RV
			$Y$, indicated as $Y \sim p_0^{-1}(y)$, meaning that $\textrm{Pr}\{Y=y\} = p_0^{-1}(y)$. This PDF is given by the inverse
			function of $p_0(x)$, implying that $p_0 \circ p_0^{-1}(y) = y$ or alternatively that $p_0^{-1} \circ p_0(x) = x$, with
			$f \circ g (\cdot) = f(g(\cdot))$ denoting the functional composition of functions $f$ and $g$.\\
		$p^{-1}(y)$ & For monotonic target PDFs, inverse of the unnormalized PDF of the target variable, $p(x)$, obtained inverting $p(x)$
			in the same way as described for $p_0^{-1}(y)$. Note that $p^{-1}(y)$ is a proper but unnormalized PDF, since $p^{-1}(y) \ge 0$
			for any value of $y$ and
			\[
				\int_{-\infty}^{\infty}{p^{-1}(y)\ \textrm{d}y} = \int_{-\infty}^{\infty}{p(x)\ \textrm{d}x} = \frac{1}{K},
			\]
			for $K>0$ but $K \ne 1$ in general. However, unlike in the case of the target RV, the normalized PDF of the inverse target RV, 
			$p_0^{-1}(y)$, can no longer be obtained from $p^{-1}(y)$ simply multiplying $p_0^{-1}(y)$ by a normalization constant, as the 
			support of $p^{-1}(y)$ can be different from the support of $p_0^{-1}(y)$. Hence, a scaling operation must be performed instead
			on $p^{-1}(y)$ in order to obtain $p_0^{-1}(y)$. Formally, $Y \sim p_0^{-1}(y) = p^{-1}(y/K)$. Finally, we also note that the
			normalized version of $p^{-1}(y)$ is different from the normalized inverse target PDF, i.e. $Kp^{-1}(y) \ne p_0^{-1}(y)$.\\
		$|\cdot|$ & For real numbers it is used to indicate their absolute value (e.g. $|x|$), whereas for a unidimensional or bidimensional
			region, $\mathcal{C}$, it is used to denote its Lebesgue measure (i.e. its length for one-dimensional regions or its area for 
			two-dimensional regions), e.g. $|\mathcal{A}_0|=1/K$ is the  Lebesgue measure of $\mathcal{A}_0$, which is identical to the
			integral of $p(x)$ over its support.\\
		$p^{-1}_G(y)$ & Unnormalized generalized inverse PDF. Given the set
				\[
					\mathcal{A}_{0|y} = \{(x,z)\in\mathcal{A}_0: z=y\},
				\]
				where $\mathcal{A}_0$ is the region enclosed by $p(x)$, as defined below, then the unnormalized generalized inverse PDF is
				defined as
				\[
					p_{G}^{-1}(y) = |\mathcal{A}_{0|y}|,
				\]
				where $|\mathcal{A}_{0|y}|$ is the Lebesgue measure of $\mathcal{A}_{0|y}$, as defined above.\\
		$\mathcal{D}_Y$ & Support of the inverse target RV $Y$, given by the range of values of $Y$ where its PDF is strictly greater than
			zero. It may be referred to $Y \sim p_0^{-1}(y)$, implying that
			\[
				\mathcal{D}_Y = \{y \in \mathbb{R}: p_0^{-1}(y)>0\} = \{y \in \mathbb{R}: p^{-1}(y/K)>0\},
			\]
			or (more often) to $Y \sim p^{-1}(y)$, resulting in
			\[
				\mathcal{D}_Y = \{y \in \mathbb{R}: p^{-1}(y)>0\}.
			\]
			The most common supports used in the paper are $\mathcal{D}_Y = \mathbb{R}^+$ and $\mathcal{D}_Y = \langle 0,1 \rangle$. Note
			that, since $p_0(x) = Kp(x) \ge 0$, in this case we can never have negative values of $y$ in the support of $Y$, implying that
			$\mathbb{R}$ can never be the support of $\mathcal{D}_Y$.\\
		$f(x)$ & Invertible transformation, $f: \mathcal{D}_X \to \mathcal{D}_Z$, used by the TR method to convert the target RV, $X$,
			unbounded and/or with infinite support, into another RV, $Z$, bounded and with finite support.\\
		$Z$ & Transformed target RV obtained using the TR method by applying the invertible transformation $f(x)$ to the target RV $X$,
			i.e. $Z = f(X)$. In some cases it is also used to indicate an auxiliary RV. For instance, in certain situations $Z$ is used to
			denote a uniform RV in $[0,1]$. \\
		$z$, $z'$ & Particular values taken by a single realization of the transformed target RV $Z$. In some cases, it is also used to
			denote a single realization of $\mathcal{U}([0,1])$, as indicated above.\\
		$f^{-1}(z)$ & Inverse of the transformation used in the TR method.\\
		$\dot{f}^{-1}(z)$ & Derivative of $f^{-1}(z)$:
			\[
				\dot{f}^{-1}(z) = \frac{\textrm{d}f^{-1}(z)}{\textrm{d}z} = \left(\frac{\textrm{d}f(z)}{\textrm{d}z}\right)^{-1}.
			\]\\
		$\rho(z)$ & Unnormalized PDF of the transformed target RV $Z = f(X)$, indicated as $Z \sim \rho(z)$. Since $f(x)$ is an invertible
			function, this PDF can be expressed compactly as $\rho(z) = p(f^{-1}(z)) \times |\dot{f}^{-1}(z)|$.\\
		$\mathcal{D}_Z$ & Support of the transformed RV $Z$, given by the range of values of $Z$ where its PDF is strictly greater than
			zero:
			\[
				\mathcal{D}_Z = \{z \in \mathbb{R}: \rho(z)>0\}.
			\]
			Since the goal of the TR method is obtaining a bounded PDF with bounded support, in the paper we consider $\mathcal{D}_Z =
			\langle 0,1 \rangle$ without loss of generality.\\
		$h(y)$ & Invertible transformation, $h: \mathcal{D}_Y \to \mathcal{D}_{\widetilde{U}}$, used by the inverse TR method to convert
			the inverse target RV, $Y$, unbounded and/or with infinite support, into another RV, $\widetilde{U}$, bounded and with finite
			support.\\
		$\widetilde{U}$ & Transformed inverse target RV obtained using the inverse TR method by applying the invertible transformation,
			$h(y)$, to the inverse target RV, i.e. $\widetilde{U} = h(Y)$. In some cases, we may write $\widetilde{U}=U$ (see below) and the
			reason (due to the fundamental theorem of simulation) is given in the text.\\
		$\tilde{u}$, $\tilde{u}'$ & Particular values taken by a single realization of the transformed inverse target RV $\widetilde{U}$.\\
		$h^{-1}(y)$ & Inverse of the transformation used in the inverse TR method.\\
		$\dot{h}^{-1}(y)$ & Derivative of $h^{-1}(y)$:
			\[
				\dot{h}^{-1}(y) = \frac{\textrm{d}h^{-1}(y)}{\textrm{d}y} = \left(\frac{\textrm{d}h(y)}{\textrm{d}y}\right)^{-1}.
			\]\\
		$q(\tilde{u})$ & Unnormalized PDF of the transformed target RV $\widetilde{U} = h(Y)$, indicated as $Y \sim q(\tilde{u})$. Since
			$h(y)$ is an invertible function, this PDF can be expressed compactly as $q(\tilde{u}) = p^{-1}(h^{-1}(\tilde{u})) \times
			|\dot{h}^{-1}(\tilde{u})|$ when $Y \sim p^{-1}(y)$, which is the case of interest considered in the paper.\\
		$\mathcal{D}_{\widetilde{U}}$ & Support of the transformed RV $\widetilde{U}$, given by the range of values of $\widetilde{U}$ where
			its PDF is strictly greater than zero:
			\[
				\mathcal{D}_{\widetilde{U}} = \{\tilde{u} \in \mathbb{R}: q(\tilde{u})>0\}.
			\]
			Since the goal of the TR method is obtaining a bounded PDF with bounded support, in the paper we consider 
			$\mathcal{D}_{\widetilde{U}} = \langle 0,1 \rangle$ without loss of generality.\\
		$\mathcal{A}_0$ & Region defined by the IoD method inside which samples must be drawn uniformly in order to obtain samples from the
			target PDF, $p_o(x) \propto p(x)$, and the unnormalized inverse target PDF, $p^{-1}(y)$. This region can be defined using the
			unnormalized target PDF, $p(x)$, or the inverse target PDF, $p^{-1}(y)$, as discussed in the paper. Formally,
			\[
				\mathcal{A}_0 = \{(x,y) \in \mathbb{R}^2: \  0\leq y \leq p(x)\} = \{(y,x) \in \mathbb{R}^2: \  0\leq x \leq p^{-1}(y)\},
			\]
			and letting $(X,Y) \sim \mathcal{U}(\mathcal{A}_0)$, then $X \sim p(x)$ and $Y \sim p^{-1}(y)$.\\
		$\mathcal{A}_r$ & Region defined by the RoU method inside which samples must be drawn uniformly in order to obtain samples
			$x = v/u$ distributed according to the target PDF. Formally, this region
			can be defined using the unnormalized target PDF, $p(x)$, as
			\[
				\mathcal{A}_r = \{(v,u) \in \mathbb{R}^2:\ 0 \leq u \leq \sqrt{p(v/u)}\},
			\]
			and letting $(V,U) \sim \mathcal{U}(\mathcal{A}_r)$, then $X = V/U \sim p(x)$.\\
		$g(u)$ & Strictly increasing differentiable function on $\mathbb{R}^+$ such that $g(0)=0$ used by the GRoU method. Setting
			$g(u)=u^2/2$ and $c=1/2$ the GRoU becomes the RoU.\\
		$\dot{g}(u)$ & Derivative of the function $g(u)$ used by the GRoU method.\\
		$g^{-1}(v)$ & Inverse of the function $g(u)$ used by the GRoU method.\\
		$c$ & Constant, $c > 0$, used by the GRoU. Setting $g(u)=u^2/2$ and $c=1/2$ the GRoU becomes the RoU.\\
		$\mathcal{A}_g$ & Region defined by the GRoU method inside which samples must be drawn uniformly in order to obtain samples $x = 
			v/\dot{g}(u)$ distributed according to the target PDF. Formally, this region 
			can be defined using the unnormalized target PDF, $p(x)$, as
			\[
				\mathcal{A}_g = \{(v,u) \in \mathbb{R}^2: 0 \le u \le g^{-1}(c p(v/\dot{g}(u)))\},
			\]
			and letting $(V,U) \sim \mathcal{U}(\mathcal{A}_g)$, then $X = V/\dot{g}(U) \sim p(x)$. Note that, setting $g(u)=u^2/2$ and
			$c=1/2$ we obtain $\mathcal{A}_g = \mathcal{A}_r$ and the GRoU method becomes the RoU.\\
		$U$, $V$ & RVs used by the RoU and GRoU methods such that the pair $(U,V)$ is uniformly distributed inside $\mathcal{A}_r$ (for the
			RoU) or $\mathcal{A}_g$ (for the GRoU). In some cases, we have $U=\widetilde{U}$ (see above) due to the fundamental theorem of
			simulation.\\
		$u$, $u'$ & Particular values taken by a single realization of the RV $U$.\\
		$v$, $v'$ & Particular values taken by a single realization of the RV $V$.\\
		$W$ & Uniform RV used by the IoD and TR methods.\\
		$w$, $w'$ & Particular values taken by a single realization of the uniform RV $W$.\\
		$C^1$ & Used to denote a class $C^1$ function. A function $f(x)$ is said to be of class $C^1$ if it is continuously
			differentiable, i.e. if $f(x)$ is continuous, differentiable, and its derivative, $\dot{f}(x)$, is also a continuous function.\\
		$\mathcal{X}^*$ & Set of vertical asymptotes of $p(x)$, i.e. set of points $x \in \mathcal{D}_X$ where $p(x) \to \infty$.\\
		$x^*$ & Used to denote the vertical asymptotes of $p(x)$ (i.e. each of the points of $\mathcal{X}^*$).\\
		$\mathcal{Y}^*$ & Set of vertical asymptotes of $p^{-1}(y)$, i.e. set of points $y \in \mathcal{D}_Y$ where
			$p^{-1}(y) \to \infty$.\\
		$y^*$ & Used to denote the vertical asymptotes of $p^{-1}(y)$ (i.e. each of the points of $\mathcal{Y}^*$).\\
		$\mathcal{Z}^*$ & Set of vertical asymptotes of $f^{-1}(z)$, i.e. set of points $z \in \mathcal{D}_Z$ where
			$f^{-1}(z) \to \pm \infty$.\\		
		$z^*$ & Used to denote the vertical asymptotes of $f^{-1}(z)$ (i.e. each of the points of $\mathcal{Z}^*$).\\
		$\widetilde{\mathcal{U}}^*$ & Set of vertical asymptotes of $h^{-1}(\tilde{u})$, i.e. set of points
			$\tilde{u} \in \mathcal{D}_{\widetilde{U}}$ where $h^{-1}(\tilde{u}) \to \pm \infty$.\\		
		$\tilde{u}^*$ & Used to denote the vertical asymptotes of $h^{-1}(\tilde{u})$ (i.e. each of the points of
			$\widetilde{\mathcal{U}}^*$).\\
		\hline
	\end{supertabular}
\end{center}

\newpage

\bibliography{bibliografia} 
   
\end{document}